\documentclass{article}
\usepackage[a4paper]{geometry}
\usepackage{graphicx}
\usepackage[textwidth=8em,textsize=small]{todonotes}
\usepackage[utf8]{inputenc} 
\usepackage[round]{natbib}

\usepackage{amsmath,amssymb,amstext,amsthm,amsfonts}
\usepackage{appendix}
\usepackage{bm}  
\usepackage{bbm} 
\usepackage{color}

\usepackage{booktabs}
\usepackage{longtable}
\usepackage{array}
\usepackage{multirow}
\usepackage{wrapfig}
\usepackage{float}
\usepackage{colortbl}
\usepackage{pdflscape}
\usepackage{tabu}
\usepackage{threeparttable}
\usepackage{threeparttablex}
\usepackage[normalem]{ulem}
\usepackage[utf8]{inputenc}
\usepackage{makecell}
\usepackage{xcolor}
\usepackage{longtable}
\usepackage{algorithm}

\usepackage[caption = false]{subfig} 
\usepackage{enumerate}
\usepackage{float}

\newcommand{\bb}{{\boldsymbol{\beta}}}

\newcommand{\bu}{{\bf u}}
\newcommand{\bv}{{\bf v}}
\newcommand{\gb}{{\boldsymbol{\gamma}}}
\newcommand{\yb}{{\bm{y}}}
\newcommand{\lb}{{\bm{\lambda}}}
\newcommand{\ep}{{\boldsymbol{\varepsilon}}}
\newcommand{\xb}{{\bm{x}}}
\newcommand{\R}{\mathbb{R}}
\newcommand{\data}{\operatorname{data}}
\newcommand{\ds}{\displaystyle}
\newcommand{\simdot}{\stackrel{\cdot}{\sim} }

\newcommand{\ceil}[1]{\lceil {#1} \rceil}

\DeclareMathOperator{\n}{\mathcal{N}}
\DeclareMathOperator{\gig}{GIG}
\DeclareMathOperator{\invgamma}{Inv-Gamma}
\DeclareMathOperator{\ber}{Bernoulli}
\DeclareMathOperator{\bet}{Beta}
\DeclareMathOperator{\sumin}{\sum_{i=1}^n}
\DeclareMathOperator{\prodin}{\prod_{i=1}^n}
\DeclareMathOperator{\diag}{diag}
\DeclareMathOperator{\rest}{rest}
\DeclareMathOperator{\one}{\mathbbm{1}}
\DeclareMathOperator*{\median}{median}

\theoremstyle{remark}
\newtheorem*{remark}{Remark}
\newtheorem{theorem}{Theorem}
\newtheorem{proposition}{Proposition}

\usepackage{etoolbox}
\makeatletter
\patchcmd{\@makecaption}
{\parbox}
{\advance\@tempdima-\fontdimen2} 
{}{}
\makeatother

\allowdisplaybreaks

\title{A generalized Bayesian approach for high-dimensional robust regression with serially correlated errors and predictors}
\author{Saptarshi Chakraborty, Kshitij Khare and George Michailidis}
\date{}

\begin{document}
	
	\maketitle
	
	\begin{abstract}
		This paper introduces a loss-based generalized Bayesian methodology for high-dimensional robust regression with serially correlated errors and predictors. The proposed framework employs a novel scaled pseudo-Huber (SPH) loss function, which smooths the well-known Huber loss, effectively balancing quadratic ($\ell_2$) and absolute linear ($\ell_1$) loss behaviors. This flexibility enables the framework to accommodate both thin-tailed and heavy-tailed data efficiently. The generalized Bayesian approach constructs a working likelihood based on the SPH loss, facilitating, efficient and stable estimation while providing rigorous uncertainty quantification for all model parameters. Notably, this approach allows formal statistical inference without requiring ad hoc tuning parameter selection while adaptively addressing
		a wide range of tail behavior in the errors. By specifying appropriate prior distributions for the regression coefficients--such as ridge priors for small or moderate-dimensional settings and
		spike-and-slab priors for high-dimensional settings--the framework ensures principled inference. 
		We establish rigorous theoretical guarantees for accurate parameter estimation and correct predictor selection under sparsity assumptions for a wide range of data generating setups. Extensive simulation studies demonstrate the
		superior performance of our approach compared to traditional Bayesian regression methods based on $\ell_2$ and $\ell_1$-loss functions. The results highlight its flexibility and robustness, particularly in challenging high-dimensional settings characterized by data contamination.
	\end{abstract}

	\section{Introduction}\label{sec:intro}
	
	The presence of outliers and heavy-tailed data is common across a wide range of applications, where extreme values and anomalies are intrinsic to the system or phenomenon under study or arise from measurement errors. For example, in health sciences, patient data often contain outliers due to rare medical conditions \citep{rosenberg2002genetic,li2008worldwide}, errors in data collection \citep{lapinsky2006electromagnetic} or self-reported inaccuracies \citep{rosenman2011measuring,ezzati2006trends}. Various financial and economic indicators exhibit heavy-tailed behavior \citep{bradley2003financial}. In engineering, sensor networks and industrial processes can produce contaminated measurements due to faults or device malfunctions \citep{woodard2015survey,de2018improving}.
	
	Several concepts and techniques have been developed in the field of \textit{robust statistics} to assess and mitigate the impact of heavy-tailed data and outliers on the estimators of the parameters of the statistical model under consideration; see, e.g., 
	\cite{tukey1960survey,huber19721972,rousseeuw1991tutorial,hampel2001robust,Huber:1981, MMY:2006, Huber:Ronchetti:2009}. 
	In the context of linear regression, it has long been recognized that heavy-tailed observations and/or outliers can severely degrade the quality of regression estimators. To address this issue, robust loss function-based estimators have been developed and extensively analyzed in the literature, primarily in a low-dimensional setting. However, the literature for high-dimensional settings is rather sparse. A review of these methods is provided in the sequel.
	
	This paper aims to develop a robust estimation procedure for the regression coefficients in linear models under \textit{high-dimensional scaling}, \textit{extending beyond the assumption of independent and identically distributed} (iid) data. In  addition, it seeks to provide uncertainty quantification for the proposed estimator. Specifically, 
	consider the stochastic linear regression model for data $\{(y_i, \xb_i)\}_{i=1}^n$, wherein $y_i \in \mathbb{R}$ 
	and ${\bf x}_i \in \mathbb{R}^p$ denote the response and the predictor vector for the $i$-th observation, respectively, as given by 
	\begin{equation}\label{eq:stoch-regression}
		y_i = \xb_i^T \bb + \varepsilon_i \; \; 1 \leq i \leq n, 
	\end{equation}
	with $\bb \in \R^p$ denoting the vector of regression coefficients, and $\{\varepsilon_i\}_{i=1}^n$ the errors. Note that both the errors and the predictors are allowed to exhibit \textit{dependence}. Specifically, (a) the errors $\{\epsilon_i\}_{i=1}^n$ are identically distributed, but not necessarily independent, (b) the predictor vectors $\{\xb_i\}_{i=1}^n$ are identically distributed, but not necessarily independent, but (c) the error process is independent of the predictor process. 
	
	The primary objective of this paper is to develop a flexible Bayesian methodology and establish rigorous theoretical guarantees for the parameters of model \eqref{eq:stoch-regression} under the presence of corruption or heavy-tailed responses $y_i$, as detailed in the sequel. The methodology is tailored for the following two high-dimensional regimes: 
	(i) $p$ is comparable to $n$ (the ``large $p$, large $n$" setting), or (ii) $p$ is much larger than $n$ (the ``large $p$, small $n$" setting). In this broad and challenging setting, it is prudent to avoid specifying a data likelihood or making detailed assumptions about the error process, such as the existence of moments or other restrictive conditions. 
	
	Next, we provide a brief review of existing literature. In the frequentist domain, a popular approach to estimate the regression coefficient vector in model \eqref{eq:stoch-regression} in a robust manner, is to employ the Huber loss function \citep{Huber:1964}, given by 
	\begin{equation} \label{huber}
		\ell_{H,\alpha} (t) = \begin{cases}
			2 \alpha^{-1} |t| - \alpha^2 & |t| > \alpha^{-1}, \cr
			t^2 & |t| \leq \alpha^{-1}. 
		\end{cases}
	\end{equation}
	\noindent
	The loss $\ell_{H,\alpha}$ corresponds to the widely used $\ell_2$ loss function for smaller values of $t$, and to the $\ell_1$ loss for larger values, with the parameter $\alpha$ controlling the balance of the linear and quadratic components. 
	In high-dimensional settings, \cite{Lambert:Zwald:2011} and \cite{FLW:2017} consider $M$-estimation problems that combine the Huber loss with $\ell_1$-type penalty functions. 
	The high-dimensional asymptotic properties of the resulting estimators are established under the assumptions of iid errors and predictors, along with suitable moment conditions on the error distribution. For scenarios where data corruption is particularly severe, the $\ell_1$ loss function which completely omits the quadratic component of $\ell_{H,\alpha}$ is a widely used choice. Methodology and theory using the $\ell_1$ loss function (also known as the least absolute) is developed in \cite{Wang:2013}; see also, \citet{WLJ:2007}. 
	
	Our goal is to develop a Bayesian methodology that provides natural uncertainty quantification for regression parameters in the presence of heavy-tailed data or outliers, without requiring strong likelihood assumptions. Following \citet{BHW:2016}, we adopt a \textit{generalized Bayesian framework} that replaces the likelihood with a loss function, using its exponential negative value as a \textit{generalized likelihood}. This, combined with a prior distribution, yields a \textit{generalized posterior} for inference. For a regression model with Huber loss and prior density $\pi(\bb)$, the generalized posterior density is given by
	\begin{equation} \label{gpost}
		\pi_{H, \alpha} \left( \bb \mid \{(y_i, \xb_i)\}_{i=1}^n \right) = \frac{\exp \left( 
			- \sum_{i=1}^n \ell_{H,\alpha} ((\xb_i, y_i), \bb) \right) \pi (\bb)}{\int_{\mathbb{R}^p} \exp \left( - \sum_{i=1}^n \ell_{H,\alpha} ((\xb_i, y_i), \bb') \right) \pi (\bb') d \bb'} \; \; \ \ \forall \ \bb \in \mathbb{R}^p, 
	\end{equation}
	\noindent
	assuming the denominator is finite. This approach is both intuitive and theoretically justified, minimizing a relevant decision-theoretic loss over all distributions in the parameter space \cite{BHW:2016}.

	The non-smooth nature of the Huber loss $\ell_{H,\alpha}$ 
	complicates inference using the generalized posterior \eqref{gpost}. To address this, the pseudo-Huber loss \citep{hartley2003multiple} provides a smooth approximation:
	\begin{equation} \label{ph} \ell_{PH,\alpha} (t) = \alpha^2 \left( \sqrt{1 + \frac{t^2}{\alpha^2}} - 1 \right). \end{equation}
	This function behaves quadratically for small $t$ and approaches linearity for large values of $t$, with the parameter $\alpha$ controlling the transition between these two regimes. In regression settings, \cite{Park:Casella:2008} show that the exponential of the negative pseudo-Huber loss (summed over observations) corresponds, up to a multiplicative factor, to the likelihood of the data.  This is valid under the assumption that the errors are iid, with a distribution that corresponds to a specific Generalized Inverse Gaussian (GIG) scale mixture of Gaussian distributions. Assigning independent Laplace priors to $\bb$
	yields a generalized posterior analogous to \eqref{gpost}, replacing $\ell_{H, \alpha}$ replaced by $\ell_{PH, \alpha}$. This GIG representation enables an efficient Gibbs sampler, termed Bayesian Huberized Lasso (BHL) \citep{Park:Casella:2008}. Extensions in \cite{Kawakami:Hashimoto:2021} introduce hierarchical and empirical Bayes methods for estimating and leveraging $\alpha$.
	
	However, the pseudo-Huber loss function has a key drawback: while its limit is indeed $t^2$ as $\alpha \to \infty$, its other limit is \textit{not} $|t|$ as $\alpha \to 0$, as desired. This prevents it from seamlessly bridging $\ell_2$ and $\ell_1$ loss functions like the standard Huber loss, leading to potential performance degradation (see Section \ref{sec:scalingeffect}).

	The first key contribution of the paper is the introduction of the scaled pseudo-Huber (SPH) loss, a refined variant of the pseudo-Huber loss that correctly bridges $\ell_2$ and $\ell_1$ as $\alpha\rightarrow\infty$ and $\alpha\rightarrow 0$, respectively. The SPH loss is defined as
	\begin{equation} \label{sph}
		\ell_{SPH,\alpha} (t) = \alpha \sqrt{\alpha^2 + 1} \left( \sqrt{1 + \frac{t^2}{\alpha^2}} - 1 
		\right). 
	\end{equation}
	ensuring a smooth transition between quadratic and absolute loss behaviors. Moreover, its corresponding generalized likelihood retains an interpretation as a valid likelihood under iid errors according to a GIG scale mixture of Gaussian distributions, enabling scalable posterior sampling.
	Since Laplace prior distributions for the entries of $\bb$ have well-documented issues with posterior coverage \citep{castillo2015bayesian, bhadra2019lasso}, we consider two alternatives: (a) a standard multivariate Gaussian (``ridge") for $\bb$ for ``large $p$, large $n$" settings, and (b) a spike-and-slab prior to induce exact sparsity in $\bb$ for ``large $p$, small $n$" settings. For both alternatives, we develop efficient Gibbs sampling algorithms (see Section~\ref{sec:posterior-computation} and Supplementary Section~\ref{sec:details-posterior-comp}) which leverage the aforementioned scale mixture representation of the $\ell_{SPH, \alpha}$ loss function 
	
	The global contamination parameter $\alpha$ controls the robustness of the model, interpolating between $\ell_1$ and $\ell_2$-like behavior (corresponding respectively to high and low contamination in the responses). Instead of selecting $\alpha$ via computationally intensive model selection methods (e.g., AIC, BIC, or cross-validation), we adopt a fully Bayesian approach by assigning a (vague gamma) prior distribution on it and inferring it jointly with other parameters using an efficient stepping-out slice sampling \citep{neal2003slice} step. This enables full Bayesian inference while naturally accounting for uncertainty in $\alpha$. Additionally, the posterior distributions of observation-specific scale parameters, introduced in the model specification Section \ref{sec:robust-pseudo-huber}, provide a mechanism for identifying outliers (see Section \ref{sec:diagnostics}).

	The \textit{second key contribution} of the paper is establishing high-dimensional consistency results for the generalized posterior distributions under ridge and spike-and-slab prior distributions. Many optimization-based robust regression estimators can be viewed as posterior modes under an appropriate data model and prior distribution for $\bb$. While high-dimensional asymptotic properties of posterior mode estimators in robust regression have been established in, for example, \cite{Lambert:Zwald:2011, Nevo:Ritov:2016, FLW:2017, Loh:2017, SZF:2020, Loh:2021}, there are \textbf{no high-dimensional results regarding the consistency of the entire posterior distribution} in the existing literature for any of the relevant methods (remark following Theorem \ref{thm:high-dim-consistency}). Moreover, previous results assume (a) independent and identically distributed (i.i.d.) errors and predictors and (b) moment conditions on the error distribution. In contrast, we allow errors to follow a \textit{serially correlated second-order stationary process} with \textit{no moment assumptions} and model predictors as a \textit{mean-zero covariance stationary Gaussian process}, imposing only mild mixing conditions on both (see Assumptions A2-A3 or B2-B3 in Section \ref{sec:theory}).
	For the ridge prior setting, we prove that the generalized posterior concentrates around the true regression coefficients, without assuming sparsity, with the number of predictors growing at a rate $p\log(p)=o(n)$. In the spike-and-slab prior setting, assuming sparsity, we allow $p$ to grow sub-exponentially with $n$ 
	and establish that the posterior distribution asymptotically concentrates on the true sparsity pattern (Theorem \ref{thm:high-dim-consistency}).
	
	Sections \ref{sec:simulation-analysis} and \ref{sec:real-data-analysis} present extensive empirical analyses of our method. Simulations in Section~\ref{sec:simulation-analysis} assess estimation, prediction, variable selection, and uncertainty quantification across diverse data-generating settings, demonstrating that the SPH regression effectively adapts to both heavy- and thin-tailed errors, mimicking $\ell_1$- and $\ell_2$-based regressions while outperforming both in intermediate cases. An analysis of real US macroeconomic data in Section~\ref{sec:real-data-analysis} further shows the model's practical utility. Technical details, proofs, MCMC implementation, and simulation settings are provided in the Supplement.

	\noindent
	\textbf{Notation:} Our notation and probability mass/density functions for various probability distributions used in this paper are presented in Table~\ref{tab:notation}.

	\begin{table}[!h]
		\centering
		
		\begin{small}  
			\begin{tabular}{l|l} 
				\toprule
				Notation & Description \\
				\midrule
				$n$ & sample size \\
				$i$ & a typical observation; $i = 1, \dots, n$ \\
				$p$ & the number of predictors/covariates \\
				$j$ & a typical predictor/covariate; $j = 1, \dots, p$ \\
				$\mathbb{S}^{p \times p}$ & Space of all $p \times p$ symmetric positive definite matrices \\
				$\|\bm{x}\|$ & the $\ell_2$ norm for a vector $\bm{x}\in\mathbb{R}^p$ \\ 
				$\|A\|_q$ & the $\ell_q$ norm for a matrix $A$ \\
				\midrule
				Notation & Probability density/mass function\\ 
				\midrule
				$x \sim \operatorname{Inv-Gaussian}(\mu, \sigma)$& ${\sqrt {\frac {\sigma }{2\pi}}} \  x^{-3/2} \ \exp {\biggl (}-{\frac {\sigma (x-\mu )^{2}}{2\mu ^{2}x}}{\biggr )}; \quad x >0, \ \mu,  \sigma > 0$\\ 
				$x \sim \operatorname{GIG}(a, b, p)$& $\frac{(a/b)^{p/2}}{2 K_p(\sqrt{ab})} \ x^{p-1} \ \exp\left[-\frac{1}{2} \left(ax + \frac{b}{x}\right)\right],\  x>0, \ a, b > 0, \ -\infty < p < \infty$\\ 
				$x \sim \operatorname{Gamma}(a, b)$ & $f(x) = \frac{b^a}{\Gamma(a)} x^{a-1} \exp(-bx); \quad x >0;\ a, b > 0$ \\
				$x \sim \operatorname{Inv-Gamma}(a, b)$& $f(x) = {\frac {b ^{a }}{\Gamma (a )}}(1/x)^{a +1}\exp \left(-b /x\right); \quad x > 0;\ a, b > 0$\\ 
				$x \sim \operatorname{Beta}(a, b)$& $\frac{\Gamma(a+b)}{\Gamma(a) \Gamma(b)} x^{a-1} (1-x)^{b-1}; \ 0 < x < 1,\  a, b > 0$\\ 
				$x \sim \n(\mu, \sigma^2)$& $\frac{1}{\sqrt{2\pi} \sigma} \exp\left[-\frac{1}{2\sigma^2} \left(\frac{x-\mu}{\sigma} \right)^2\right]; \ -\infty < x < \infty; \ \sigma>0$\\
				$\bm x \sim \n_p(\bm \mu, \Sigma)$ & $\frac{1}{\sqrt{2\pi |\Sigma|}} \exp\left[-\frac{1}{2} \left(\bm x - \bm \mu\right)^T \Sigma^{-1} \left(\bm x - \bm \mu\right)\right]; \ \bm x \in \R^p; \ \bm \mu \in \R^p$, $\Sigma \in \mathbb{S}^{p \times p}$ \\
				$x \sim \operatorname{Bernoulli}(p)$ & $p^{x} (1-p)^{1-x} \quad x \in \{0,1\}; \ 0 \leq p \leq 1 $\\
				\bottomrule
			\end{tabular} 
			
		\end{small}
		
		\caption{\footnotesize Notation and density/mass functions for various probability distributions used in this paper. \label{tab:notation}}

	\end{table}

	\section{SPH based robust generalized Bayesian regression }\label{sec:robust-pseudo-huber}
	
	We begin the exposition with a key Gaussian scale-mixture representation for the pseudo-Huber loss function (proof given in Supplement~\ref{sec:proofs-pseudo-Huber}) that is used in the sequel. 
	
	\begin{proposition} \label{prop:pseudo-huber-mixture}
		Consider the hierarchical distribution for a real random variable: 
		$\ep \mid \lambda \sim \n(0, \lambda)$, with $\lambda \mid \alpha \sim \gig(a = 1+\alpha^2, b = \alpha^2, p = 1)$ for any fixed $\alpha > 0$. Then, the $\lambda$-marginalized density $f_\ep(\ep \mid \alpha)$ of $\ep$ at a fixed $\alpha > 0$ has the form:
		\[
		f_\ep(\ep \mid \alpha) \propto \exp\left[ - \alpha \sqrt{1 + \alpha^2} \left(\sqrt{1 + \left(\frac{\ep}{\alpha}\right)^2} - 1 \right) \right],
		\]
		which is the generalized density associated with the scaled pseudo-Huber loss function with tuning parameter $\alpha \in (0, \infty)$.
	\end{proposition}
	
	Within the framework of the generalized Bayes approach discussed in the introduction, the above $\lambda$-marginalized density $f_\ep(\ep \mid \alpha)$ can be thought of as the error distribution producing the \textit{generalized likelihood} associated with a SPH loss-based linear regression. Consequently, Proposition \ref{prop:pseudo-huber-mixture} enables the construction of the following hierarchical (generalized) likelihood  specification for the robust SPH regression:
	\begin{equation}\label{eq:pseudo-Huber-regression}
		y_i \mid \bb, \lambda_i \sim \n(\xb_i^T \bb, \lambda_i), \ \ 
		\lambda_i \mid \alpha \sim \gig(a = 1+\alpha^2, b = \alpha^2, p = 1) 
	\end{equation}
	wherein the parameters $\{\lambda_i: i=1, \dots, n\}$ are treated as latent/augmented data. 
	
	Remark \ref{rem:pseudo-Huber} in Supplement~\ref{sec:proofs-pseudo-Huber} shows that as $\alpha \to 0$, $f_\ep(\ep \mid \alpha)$ converges to a standard Laplace density, and as $\alpha \to \infty$, it approaches a standard normal density—aligning with the error distributions of $\ell_1$ (median) and $\ell_2$ regression, respectively. Thus, model \eqref{eq:pseudo-Huber-regression} seamlessly integrates these two frameworks. The $\ell_2$ regression is recovered by setting $\lambda_i \equiv \sigma^2$ for a common $\sigma^2 > 0$, while $\ell_1$ regression emerges by specifying $\lambda_i \sim \operatorname{Gamma}(1,1)$ (equivalent to $\gig(2,0,1)$).

	\begin{remark}
		Throughout, we assume the predictor variables to be centered and, therefore, ignore an additional intercept parameter $\mu$ in the model. If needed, a straightforward generalization of the model of the form:
		$y_i \mid \mu, \bb, \lambda_i \sim \n(\mu + \xb_i^T \bb, \lambda_i)$, with $\lambda_i \mid \alpha \sim \gig(a = 1+\alpha^2, b = \alpha^2, p = 1)$ enables incorporation of intercept terms. 
	\end{remark}

	\begin{remark}
		Following \cite{Kozumi:Kobayashi:2011}, one can introduce an additional \textit{global} scaling parameter $\sigma > 0$: 
		$y_i \mid \bb, \lambda_i, \sigma \sim \n(\xb_i^T \bb, \sigma^2 \lambda_i)$, 
		$\lambda_i \mid \alpha \sim \gig(a = 1+\alpha^2, b = \alpha^2, p = 1)$, 
		to potentially aid further flexibility to accommodate a richer set of data. 
	\end{remark}
	
	To complete the specification of the generalized Bayes posterior distribution of the model, prior distributions are assigned to the key regression parameter $\bb$ and the hyperparameter $\alpha$, as well as to the intercept parameter $\mu$ and/or the global scaling parameter $\sigma$, if included in the model. Next, we discuss specific choices for these prior distributions.

	\subsection{Specification of distributions for the parameters $\bb$ and $\alpha$} \label{sec:prior-specification}
	
	We consider independent prior distributions for the regression parameter $\bb$ and the pseudo-Huber balance hyperparameter $\alpha$. We consider two different specifications for the prior distribution of $\bb$: the first is better suited for a low-dimensional setting wherein the number of predictors is of the order of the sample size ($p=O(n)$), and the second is suited for high-dimensional data ($p>>n$).  These two prior distributions are listed next.
	
	\noindent
	\textit{(1)  A Gaussian, weakly informative prior distribution} of the form $\bb \sim \n(\bb_0, Q^{-1})$,
	where $\bb_0$ is a fixed prior mean and $Q$ a fixed prior precision matrix for the regression parameter $\bb$. Typically,  $\bb_0$  is set to $\bm 0$, and $Q$ to a diagonal matrix with moderately small diagonal entries, such as 0.01, yielding independent vague priors for the coordinates of $\bb$.
	
	\noindent
	\textit{(2) A hierarchical spike-and-slab prior distribution} of the form: 
	\begin{eqnarray}\label{eq:prior-spike-slab} 
		\beta_j \mid \gamma_j = 0 \sim \one_{\{0\}}, \ \
		\beta_j \mid \gamma_j = 1 \sim  \n(0, \tau^2); \  
		\gamma_j \sim \ber(q), \ \
		q \sim \bet(a_q, b_q), 
	\end{eqnarray}
	where $\gamma_j$ is a Bernoulli 0-1 random variable with $[\gamma_j = 1]$ implying that the $j$-th  predictor is ``active". Conditional on $\gamma_j = 1$, $\beta_j$ is endowed with a ``slab'' Gaussian distribution $\n(0, \tau^2)$ with some reasonably large $\tau$ such as $\tau = 100$. On the other hand, when $\gamma_j = 0$, $\beta_j$ has a degenerate distribution at zero. The a priori proportion $q$ of ``active'' predictors can be specified; we consider a Beta($a_q, b_q$) prior on $q$ for its data-driven estimation.
	
	The global contamination parameter $\alpha$ is assigned an induced prior through its square, namely:
	$\alpha^2 \sim \operatorname{Gamma}(a_\alpha,b_\alpha)$.
	In addition, if the model includes an intercept term $\mu$, a vague normal prior such as $\mu \sim N(0, \tau_\mu^2)$ is used for some reasonably large $\tau_\mu$ such as $\tau_\mu = 100$. Finally, if the model includes an additional global scaling parameter $\sigma$, we use an inverse gamma prior: $\sigma^2 \sim \operatorname{Inv-Gamma}(a_\sigma, b_\sigma)$ with some small $a_\sigma$ and $b_\sigma$ such as $a_\sigma = 0.01$, $b_\sigma = 0.01$.

	\subsection{Posterior Distribution Computation}
	\label{sec:posterior-computation}
	The complex structures of both the likelihood and the prior distribution---whether in low/moderate or high-dimensional settings---render the resulting posterior distributions intractable, precluding independent random sampling. Since principled uncertainty quantification is a central goal of this paper, we focus on MCMC sampling, which offers theoretically guaranteed posterior computation. Below, we outline an efficient Gibbs-type algorithm for MCMC sampling from the target posterior distribution. We first present MCMC sampling of the model parameters given a fixed value of the hyperparameter $\alpha$, followed by an approach for MCMC sampling of $\alpha$ to facilitate full Bayesian inference. 
	
	\medskip
	
	\noindent
	\textbf{MCMC sampling from the posterior distribution given $\alpha$.} Under setting (1) with the weakly informative prior $\bb \sim \n(\bb_0, Q^{-1})$, the conditional posterior distributions for the model parameters given a fixed $\alpha$ have closed-form expressions involving standard probability distributions (the joint posterior density given $\alpha$ is provided in Supplement \ref{sec:details-posterior-comp}.): Gaussian (for $\bb$ and the intercept $\mu$, if included in the model), generalized inverse Gaussian (for $\lambda_1, \dots, \lambda_n$), and inverse Gamma (for $\sigma^2$, if included in the model) that permit efficient random sampling.  Consequently, a standard Gibbs sampling algorithm can be constructed for efficient MCMC sampling from the $\alpha$-conditioned posterior. 
	
	In setting (2) with the hierarchical spike-and-slab prior \eqref{eq:prior-spike-slab} for $\bb$, the full conditional posterior distributions of the model parameters, given a fixed value of $\alpha$, maintain analogous closed-form expressions. These distributions are similar to those in the weakly informative Gaussian prior case for $\bb$, $\{\lambda_1, \dots, \lambda_n\}$, $\mu$, and $\sigma^2$ (if included in the model).  Additionally, the full conditional distribution for each spike and slab ``active'' predictor indicator $\gamma_j$ has a Bernoulli structure, while the prior proportion parameter $q$  has a full conditional beta distribution. Consequently, a similar Gibbs sampling algorithm can be derived for MCMC sampling from the resulting $\alpha$-conditioned posterior.  For computational efficiency, we update $\bb$ and $\bm \gamma$  coordinate-wise with $(\beta_j, \gamma_j)$ sampled jointly from their full conditional distribution. This coordinate-wise strategy avoids challenges related to numerical matrix inversion and enhances scalability, particularly in high-dimensional settings where the spike-and-slab prior is advantageous. Detailed steps of the Gibbs samplers for settings (1) and (2) are provided in Section \ref{sec:details-posterior-comp} of the Supplement.
	
	\medskip
	
	\noindent
	\textbf{MCMC sampling for $\alpha$.} The integral producing the marginal posterior density of $\alpha$ is not available in closed form, and the full conditional posterior density of $\alpha$, given the remaining model parameters, lacks a standard form for efficient random sampling. A rejection sampler can be derived using analytical upper bounds for the modified Bessel function-based terms to sample $\alpha$ from its conditional posterior distribution given $\lambda_1, \dots, \lambda_n$.  However, the general non-tightness of these bounds may cause substantial inefficiency. In \cite{Kawakami:Hashimoto:2021}, the authors approximate the full conditional posterior of $\alpha$ with an ``optimized" Gamma distribution, but the impact of this approximation on the Markov chain's distribution is not well understood. 
	
	Instead, we suggest a middle-of-the-road approach that entails sampling 
	$\alpha$ from its $\{\lambda_1, \dots, \lambda_n\}$-integrated conditional posterior given only $\bb$ (and $\mu$, and $\sigma^2$, if relevant). The corresponding density is available in closed form up to a normalizing constant (Eqn. in \eqref{eq:post-density-alpha2} in Supplement~\ref{sec:details-posterior-comp}), and we use stepping-out slice sampling \citep{neal2003slice} to generate draws from this univariate density. This approach (a) avoids the need for any approximations, (b) is computationally straightforward, and (c) reduces dependence between successive iterates in the Gibbs sampler through blocking, which can lead to improved mixing.

	\medskip
	
	\noindent
	\textbf{Computational complexity.} The computational complexity per iteration of the proposed slice-within-Gibbs sampler can be expressed in terms of $n$ and $p$. The stepping out slice sampling \citep{neal2003slice} for the SPH tuning parameter $\alpha$, and the ordinary Gibbs updates for the parameters $\mu$, $\sigma$, and $q$ (if included in the model) run in constant time $O(1)$ each and do not impact the overall cost complexity of the MCMC algorithm in terms of $n$ and $p$. For the normal (ridge) prior model, each MCMC iteration has complexity $O(n) + O(p^3)$ due to the $O(n)$ cost of sampling $\{\lambda_1, \dots, \lambda_n\}$ and the $O(p^3)$ cost of sampling $\bb$ which involves inverting a $p \times p$ matrix for multivariate normal generation. In the spike-and-slab model, the coordinate-wise updating of $(\bb, \bm \gamma)$ avoids any matrix inversion, reducing the cost to $O(p)$. Consequently, the total cost per MCMC iteration is $O(n) + O(p)$.

	\section{Theoretical guarantees for SPH regression}\label{sec:theory}
	
	\noindent
	Consider the linear regression model in \eqref{eq:stoch-regression}. 
	As discussed in Section \ref{sec:robust-pseudo-huber}, we consider two settings, the first wherein the number of predictors is of the order of the sample size and the second corresponding to high-dimensional scaling.
	For both settings, we consider the generalized likelihood function $\mathcal{L}(\bb):=\exp(-n H_\alpha (\bb))$, with 
	\begin{equation}\label{eq.gen-likelihood}
		H_\alpha (\bb) := \frac{1}{n} \sum_{i=1}^n \ell_{SPH,\alpha} (Y_i - {\bf x}_i^T \bb),
	\end{equation}
	and $\ell_{SPH,\alpha}$ corresponding to the scaled pseudo-Huber loss function defined in (\ref{sph}).

	\subsection{Consistency in the $p=\mathcal{O}(n)$ setting} \label{sec:theory-low-dim}
	
	As discussed in Section \ref{sec:prior-specification}, for this setting a Gaussian prior distribution on the regression coefficients is imposed, given by
	\begin{equation} \label{ridgeprior}
		\pi_{ridge} (\bb) \propto \exp(-\tau^2 \bb^T \bb) \; \; \; \forall \bb \in \mathbb{R}^p, 
	\end{equation}
	for some $\tau^2 > 0$. The posterior density for the posited working model is given by 
	\begin{equation} \label{ridge:posterior}
		\pi_{ridge} (\bb \mid Y) \propto \exp(-n H_\alpha (\bb) - \tau^2 \bb^T \bb) \; \; \; \forall \bb \ \in \mathbb{R}^p. 
	\end{equation}
	
	We consider an asymptotic setting wherein the number of regressors $p = p_n$ grows with the sample size $n$. For the purposes of asymptotic evaluation, we allow $\alpha = \alpha_n$ to vary with $n$ as well, but consider it to be fixed/known and do not place a prior distribution on $\alpha$ in the working model. The true data-generating model is given by 
	\begin{equation} \label{truemodel}
		Y_{i,n} = {\bf x}_{i,n}^T \bb_{0,n} + \epsilon_{i,n} \; \; \; i = 1,2, 
		\cdots, n. 
	\end{equation}
	
	\noindent
	for every $n \geq 1$, with $\bb_{0,n}$ denoting the vector of true regression coefficients. In particular, we make the following regularity assumptions regarding the data generating model and the prior precision parameter $\tau^2$. 
	
	\noindent
	$\bullet$ \textbf{Assumption A1} - $p_n \log p_n = o(n)$, $p_n \rightarrow \infty$, $\alpha_n \rightarrow \infty$ and $\alpha_n \sqrt{\frac{p_n}{n}} \rightarrow 0$. Here $\tilde{M}$ is an 
	appropriately chosen constant. 
	
	\smallskip
	
	\noindent
	The growth rate of $p_n$ in this setting is constrained by the lack of any low dimensional structure, such as sparsity on $\bb_{0,n}$. Note that $p_n$ is allowed to grow at a much faster rate (sub-exponentially) in the spike-and-slab based consistency analysis (Section \ref{sec:theory-high-dim}).
	
	\noindent
	$\bullet$ \textbf{Assumption A2} - For every $n \geq 1$, the predictor vectors $\{{\bf x}_{i,n}\}_{i=1}^n$ are independent of the errors $\{\epsilon_{i,n}\}_{i=1}^n$, and form a covariance stationary Gaussian sequence with $\Gamma_n (h) := Cov(\xb_{i,n}, \xb_{i+h,n})$ for every $-(n-1) \leq h \leq n-1$ and $1 \leq i, i+h \leq n$. There exists $\kappa_1 > 0$ (not depending on $n$) such that 
	$$
	0 < \kappa_1 < \lambda_{\min} (\Gamma_n (0)) \leq \lambda_{\max} (\Gamma_n (0)) < \kappa_1^{-1} < \infty, 
	\text{ and } 
	\kappa_2 := \sup_{n \geq 1} \sum_{h=0}^{n-1} \left\| \Gamma_n (h) 
	\right\|_2 < \infty. 
	$$
	
	\noindent
	$\bullet$ \textbf{Assumption A3} - For every $n \geq 1$, the errors $\{\epsilon_{i,n}\}_{i=1}^n$ form a second order stationary sequence. Also, for the uniformly bounded function $g(x) := E \left[ (1 + x^2 + (1/\kappa_1) Z^2)^{-3/2} Z^2 \right]$ (with $Z$ standard normal), we have 
	$$
	K_\epsilon := \sup_{n \geq 1} \left\{ Var(g(\epsilon_{1,n})) + 2 \sum_{i=2}^n |Cov(g(\epsilon_{1,n}), g(\epsilon_{i,n}))| \right\} < \infty. 
	$$
	
	\noindent
	Some common settings where Assumption A3 is satisfied are presented next.
	\begin{itemize}
		\item The error process forms an $m$-dependent second order stationary sequence (such as a moving average process); in this case $Cov(g(\epsilon_{1,n}), g(\epsilon_{i,n})) = 0$ for every $i > m$. 
		\item The errors form a second order stationary $\alpha$-mixing sequence (see for example \cite{Jones:2004}) with $\sum_{k=1}^\infty \alpha_\epsilon (k) < \infty$. Since $g$ is uniformly bounded by $\kappa_1$, it follows by \cite[Theorem A.5]{Ibragimov:1962} that $|Cov(g(\epsilon_{1,n}), g(\epsilon_{i,n}))| \leq 4 \kappa_1^2 \alpha_\epsilon(i-1)$ for every $i \geq 2$, and hence Assumption A3 is satisfied. 
		\item In particular, Assumption A3 is satisfied if the errors form a stationary and geometrically ergodic Markov chain (since such a Markov chain is exponentially fast $\alpha$-mixing and $g$ is uniformly bounded, see \cite{Chan:Geyer:1994}). 
	\end{itemize}
	
	\noindent       
	$\bullet$ \textbf{Assumption A4} - The prior distribution's precision parameter $\tau^2_n$ satisfies 
	$
	\tau^2_n = O(\alpha\sqrt{np_n}/\|\bb_{0,n}\|). 
	$
	
	\smallskip
	
	\noindent
	Note that under a Gaussian likelihood based working model, the posterior mode for $\bb$ (with the prior distribution specified in (\ref{ridgeprior})) is given by the ridge regression estimator $\hat{\bb}_{ridge} = (X^T X + \tau^2 I_p)^{-1} X^T {\bf y}$. It is clear that some upper bound on the parameter $\tau^2$ (depending also on $\bb_{0,n}$) is needed for consistency of $\hat{\bb}_{ridge}$. To see this, consider the special case when $X$ is semi-orthogonal, in particular, $X^T X = n I_p$. In this case 
	$$
	\hat{\bb}_{ridge} = \frac{n}{n + \tau^2} \bb_{0,n} + \frac{1}{n} X^T {\boldsymbol \epsilon}. 
	$$
	
	\noindent
	The $\|\ell_2\|$-norm of the second (error) term on the right-hand-side can be shown to converge to zero (in probability) by routine arguments assuming Gaussian errors, and it is clear that the condition $\frac{\tau^2}{n+\tau^2} \|\bb_{0,n}\|$ is necessary for consistency of $\hat{\bb}_{ridge}$. Assumption A4 can be thought as its counterpart in the current setting (with possibly non-Gaussian and correlated errors).

	\noindent
	Let $P_0$ denote the underlying probability measure corresponding to the true data generating model, and $E_0$ the expectation with respect to $P_0$. In the subsequent analysis, we will often 
	refer to $Y_{i,n}, \epsilon_{i,n}, \xb_{i,n}, Q_n, \bb_{0,n}$ by 
	$Y_i, \epsilon_i, \xb_i, Q, \bb_0$ for notational convenience. Since 
	$\ell_\alpha''(x) = \sqrt{1 + \alpha^{-2}} (1 + (x/\alpha))^{-3/2} > 0$ for every $x \in 
	\mathbb{R}$, it follows that the Hessian matrix of $H$ given by 
	$$
	\nabla^2 H_\alpha (\bb) = \frac{1}{n} \sum_{i=1}^n \ell_\alpha''(Y_i 
	- {\bf x}_i^T \bb) {\bf x}_i {\bf x}_i^T
	$$
	
	\noindent
	is positive definite for every $\bb \in \mathbb{R}^p$. It follows 
	that 
	$$
	Q_\alpha (\bb) := \alpha^{-1} H_\alpha (\bb) + \frac{\tau^2}{n 
		\alpha} \bb^T \bb
	$$
	
	\noindent
	is strictly convex and has a unique minimizer. This minimizer also corresponds to the posterior mode, and is denoted by 
	$\hat{\bb}_{pm, ridge}$. The first task is to study the asymptotic 
	properties of $\hat{\bb}_{pm, ridge}$ under the high-dimensional setting 
	described above. 
	\begin{theorem}[\em Posterior mode consistency with a ridge prior distribution] \label{thm:postmean}
		Under Assumptions A1-A4 
		$$
		P_0 \left( \|\hat{\bb}_{pm, ridge} - \bb_0\| > \tilde{M} \alpha_n \sqrt{\frac{p_n}{n}} \right) 
		\rightarrow 0
		$$
		
		\noindent
		as $n \rightarrow \infty$, for an appropriate constant $\tilde{M}$. 
	\end{theorem}
	
	With the consistency of the posterior mode in hand, we proceed to establish the consistency of the \textit{entire posterior distribution}. For this result, we need to slightly strengthen our set of assumptions by adding the following regularity conditions. 
	
	\noindent
	$\bullet$ \textbf{Assumption A5} - (a) The prior precision parameter $\tau^2$ satisfies $\tau^2 = O \left( \min \left( \frac{\alpha \sqrt{np}}{\|\bb_0\|}, \frac{n^2}{p} \right) \right)$, (b) the error process has a finite first moment, i.e., $E|\epsilon_1| < \infty$, and (c) there exists a constant $\kappa_3 > 0$ such that 
	$\lambda_{min} (\Theta_n) \geq \kappa_3$ for every $n \geq 1$. 
	Recall that $\Theta_n$ denote the $n \times n$ block partitioned matrix whose $(i,j)^{th}$ block is given by $\Gamma_n (i-j)$ for $1 \leq i,j \leq n$.
	
	\noindent
	The following result shows that the posterior distribution asymptotically 
	places all of its mass on a neighbourhood of radius $O(\alpha_n \sqrt{\frac{p_n}{n}})$ around the true parameter $\bb_0$. 
	\begin{theorem}[\em Posterior distribution consistency with a ridge prior dsitribution] \label{thm:postdist}
		Let $\Pi_{ridge} (\cdot \mid {\bf Y})$ denote the posterior distribution for the Bayesian working model based on the generalized likelihood (\ref{eq.gen-likelihood}) and prior distribution (\ref{ridgeprior}). Under Assumptions A1-A5, there exists a constant $\tilde{M}^*$ such that 
		$$
		E_0 \left[ \Pi \left( \|\bb - \bb_0\| > \tilde{M}^* \alpha_n \sqrt{\frac{p_n}{n}} \mid {\bf Y} \right) \right] \rightarrow 0 \quad \text{as} \ n \rightarrow \infty.
		$$
	\end{theorem}
	
	\noindent
	\begin{remark}
		With a Gaussian likelihood based working model, a ridge prior distribution on $\bb$, and serially correlated \textit{Gaussian errors and predictors} in the data generating model (with relevant regularity assumptions on their respective spectral densities), minor modifications to arguments in \cite{GKM:2021} lead to a posterior convergence rate of $\sqrt{\frac{p}{n}}$, when \textit{no low-dimensional structure} is imposed on $\bb_{0,n}$ and $p_n \log p_n = o(n)$. In the current setting, where minimal assumptions are placed on the error process (existence of first moment and weak dependence outlined in Assumption A3) in the data generating model, Theorem \ref{thm:postdist} establishes a convergence rate of $\alpha_n \sqrt{\frac{p}{n}}$. To summarize, the rate in Theorem \ref{thm:postdist} contains an extra factor of $\alpha_n$ as compared to the Gaussian error setting, but is obtained under \textit{significantly weaker assumptions} on the error process and also using a different, pseudo-Huber based, working model.
	\end{remark}
	
	\subsection{Sparsity selection consistency in a high-dimensional setting} \label{sec:theory-high-dim}
	
	Next, we focus on the high-dimensional setting where sparsity is induced in $\bb$ by the use of independent spike-and-slab prior distributions on the entries of $\bb$ as in (\ref{eq:prior-spike-slab}). The spike-and-slab posterior distribution can be obtained by combining this prior with the generalized likelihood in 
	(\ref{eq.gen-likelihood}). We begin by defining relevant sparsity-based notation.
	
	Note that every element of the set $\{0,1\}^p$ represents a possible sparsity pattern in the regression coefficient vector $\bb$. In particular, ${\bf s} \in \{0,1\}^p$ represents the 
	sparsity pattern where the coefficients with indices in $ind(\bf s) := \{j: s_j = 1\}$ are deemed 
	significant and other coefficients are deemed insignificant. Given a sparsity pattern 
	${\bf s}$, for any ${\bf a} \in \mathbb{R}^p$, define the sub-vector ${\bf a}_s$ as 
	${\bf a}_{s} = (a_j)_{j \in ind(\bf s)}$. Similarly, for any $p \times p$ matrix $A$, define 
	the submatrix $A_{\bf s}$ as $A_{\bf s} = ((a_{jk}))_{j,k \in ind(\bf s)}$. Finally, we 
	define $|s| := |\{j: s_j = 1\}|$, and for any ${\bf b} \in \mathbb{R}^{|s|}$, $Q_\alpha 
	({\bf b})$ will implicitly stand for the function $Q_\alpha ({\bf b}_{fill,s})$, where 
	the ${b}_{fill,s,s_j} = 1$ for $1 \leq j \leq |s|$ and all other entries of ${\bf b}_{fill,s}$ are zero. 
	
	Note that the spike-and-slab posterior distribution induces a probability distribution over the space of all possible sparsity patterns, or equivalently $\{0,1\}^p$. Let $\Pi_{SS}({\bf s} \mid {\bf Y})$ denote the probability mass assigned to the sparsity pattern ${\bf s}$ by the spike-and-slab posterior distribution. Routine calculations show that 
	\begin{equation} \label{postsparse}
		\Pi_{SS} \left( {\bf s} \mid {\bf Y} \right) \propto \left( \frac{q \tau}{(1-q) \sqrt{2 \pi}} \right)^{|{\bf s}|} \int \exp \left( -n\alpha Q_\alpha 
		(\bb_s) \right) d \bb_s, \ \ \text{for every } {\bf s} \in \{0, 1\}^p.
	\end{equation}
	
	Consider the true data generating model described in (\ref{truemodel}). Recall that $P_0$ denotes the underlying probability measure corresponding to the true data generating model, and $E_0$ the expectation with respect to $P_0$. Further, let  ${\bf s}_0 \in \{0,1\}^p$ represent the sparsity pattern corresponding to $\bb_0$ (the 
	``true" sparsity pattern). The first task will be to establish \textit{strong selection consistency}, i.e., 
	$\Pi_{SS} ({\bf s}_0 \mid {\bf Y}) \stackrel{P_0}{\rightarrow} 1$, 
	as $n \rightarrow \infty$. In other words, we want to show that with $P_0$-probability tending to $1$, the posterior distribution (on the sparsity patterns) 
	places almost all of its mass on the true sparsity pattern ${\bf s}_0$. This will be achieved by examining the ratio 
	\begin{equation} \label{post:ratio}
		\frac{\Pi_{SS} \left( {\bf s} \mid {\bf Y} \right)}{\Pi_{SS} \left( {\bf s}_0 \mid 
			{\bf Y} \right)} = \left( \frac{q \tau}{(1-q) \sqrt{2 \pi}} 
		\right)^{|{\bf s}|-|{\bf s}_0|} \frac{\int \exp \left( -n\alpha Q_\alpha 
			(\bb_s) \right) d \bb_{s}}{\int \exp \left( -n\alpha Q_\alpha 
			(\bb_{s_0}) \right) d \bb_{s_0}} 
	\end{equation}
	
	\noindent
	for different choices of the sparsity pattern ${\bf s}$.  
	\cite{Narisetty:He:2014} establish strong selection consistency for linear regression with a spike-and-slab prior distribution, assuming that the errors in both the true and the working model are independent and identically \textit{normally} distributed. Further, in their setting, the non-zero components of the true parameter $\bb_0$ remain unchanged as $n$ increases. Simiarly, we assume that {\em the set of indices corresponding to the non-zero entries in the true sparsity pattern ${\bf s}_0$ do not change with $n$}. We also  and impose the following regularity conditions, whcih closely resemble Assumptions A1-A4, with appropriate adaptations for the spike-and-slab setting. 
	
	\noindent
	$\bullet$ \textbf{Assumption B1} -  $p_n \rightarrow \infty$, $\alpha_n \rightarrow \infty$ and $\alpha_n^{2+\delta} \log p = o(n)$ for some $\delta > 0$. 
	
	\noindent
	$\bullet$ \textbf{Assumption B2} - For every $n \geq 1$, the predictor vectors $\{{\bf x}_{i,n}\}_{i=1}^n$ are independent of the errors $\{\epsilon_{i,n}\}_{i=1}^n$, and form a covariance stationary Gaussian sequence with $\Gamma_n (h) := Cov(\xb_{i,n}, \xb_{i+h,n})$ for every $-(n-1) \leq h \leq n-1$ and $1 \leq i, i+h \leq n$. There exists $\kappa_1 > 0$ (not depending on $n$) such that 
	$$
	0 < \kappa_1 < \lambda_{\min} (\Gamma_n (0)) \leq \lambda_{\max} (\Gamma_n (0)) < \kappa_1^{-1} < \infty, \text{ and }
	\kappa_2 := \sup_{n \geq 1} \sum_{h=0}^{n-1} \left\| \Gamma_n (h) 
	\right\|_2 < \infty. 
	$$
	\noindent
	$\bullet$ \textbf{Assumption B3} - For every $n \geq 1$, the errors $\{\epsilon_{i,n}\}_{i=1}^n$ form a second order stationary sequence which is either $m$-dependent or is $\alpha$-mixing with $\sum_{k=1}^\infty \alpha_\epsilon (k) < \infty$. 
	
	\noindent
	$\bullet$ \textbf{Assumption B4} - The prior mixture probability $q = q_n$ satisfies $q_n = p_n^{-\alpha^{2+\delta}}$. The prior slab precision parameter $\tau^2 > 0$  does not vary with $n$.

	\noindent
	\begin{remark}
		In \cite{GKM:2021}, the authors consider a linear regression with a spike-and-slab prior distribution and a Gaussian likelihood based working model. They extend the strong selection results of \cite{Narisetty:He:2014} to a setting where 
		the true error and predictor processes are stationary Gaussian processes with serial correlation. Apart from minor modifications concerning the boundedness of eigenvalues of spectral densities and fixing ${\bf s}_0$ with $n$, the key differences and tradeoffs in the assumptions required by \cite{GKM:2021} and Assumptions B1-B4 above are as follows: (a) Assumption B3 does not require Gaussianity and is significantly weaker than the corresponding assumption on the error process in \cite{GKM:2021}, while (b)in Assumption B1, $\log p = o(n/\alpha^{2+\delta})$, as opposed to $\log p = o(n)$ in 
		\cite{GKM:2021}, and $q_n = p_n^{-\alpha^{2+\delta}}$ as opposed to $q_n = p_n^{-C}$ (for an appropriate constant $C$) in \cite{GKM:2021}.It should also be noted that a pseudo-Huber loss based working model is used here, as compared to the Gaussian likelihood based working model in \cite{GKM:2021}.
	\end{remark}
	
	With Assumptions B1-B4 in hand, we proceed to analyze and bound the ratio $\frac{\Pi \left( {\bf s} \mid {\bf Y} \right)}{\Pi \left( {\bf s}_0 \mid {\bf Y} \right)}$ under different cases - the sparsity pattern $\mathbf{s}$ is a superset of the true one $\mathbf{s}_0$, $\mathbf{s}$ is a subset of $\mathbf{s}_0$, and finally none is a subset of the other one, but with some additional requirements on their size - to establish the following result. 
	\begin{theorem}[\em Strong selection consistency with spike-and-slab prior] \label{thm:high-dim-consistency}
		Consider the spike-and-slab prior distribution based working model in Section \ref{sec:prior-specification}, with the true data generating mechanism given by (\ref{truemodel}). Under Assumptions B1-B4, and restricting to {\it realistic} sparsity patterns, whose cardinality is less than or equal to $n/(\log(\max(n,p)))^{1+\delta}$, the working model posterior distribution on the space of sparsity patterns satisfies 
		$$
		\Pi_{SS} ({\bf s}_0 \mid {\bf Y}) \stackrel{P_0}{\rightarrow} 1, \quad \text{as} \ \  n \rightarrow \infty.$$
	\end{theorem}
	
	\begin{remark} \label{robust:literature:review}
		We carefully review relevant high-dimensional consistency results in the robust regression literature. To the best of our knowledge, existing high-dimensional analyses focus \textit{exclusively on the consistency of posterior modes} for various robust Bayesian models (note that most optimization-based estimators can be regarded as posterior modes under an appropriate Bayesian model), and do not establish consistency/convergence of the entire posterior distribution. In \cite{Lambert:Zwald:2011}, consistency and asymptotic normality of penalized estimators based on the Huber loss and the lasso/adaptive lasso penalty is established in the i.i.d. error and fixed $p$ setting. \cite{FLW:2017} extend the consistency results in the high-dimensional setting, where $p$ is allowed to grow sub-exponentially with $n$, but consider independent errors with bounded second moments (under the data-generating model). The predictor process is assumed to be i.i.d sub-Gaussian, and \cite{SZF:2020} explores truncation based adaptations and extensions to the setting when the predictors are heavy-tailed (with finite fourth moments) under the data generating model. \cite{Loh:2017} considers generalized $M$-estimators obtained by minimizing an objective which combines a ``robust" loss function (convex, bounded derivatives, etc.) and a separable penalty function (with suitable regularity), establishing consistency while allowing $p$ to grow sub-exponentially with $n$. The errors in the data-generating model are assumed to be independent. 
		In \cite{Nevo:Ritov:2016}, the authors establish consistency of the Bayes estimator under a bounded loss function with spike-and-slab prior distributions on the components of $\bb$. The working model and the data-generating models {\it both} assume i.i.d. errors with a common log-concave density.
	\end{remark}
	
	\section{Performance Evaluation based on Synthetic Data} \label{sec:simulation-analysis}
	
	This section presents extensive simulation results evaluating the frequentist properties of SPH against Bayesian $\ell_1$ and $\ell_2$ regressions under ridge and spike-and-slab prior distributions. We consider a range of data-generating scenarios, varying $n$, $p$, sparsity, error distributions, and correlations. Two main settings are explored: (i) low/moderate-dimensional ($p = o(n)$) and (ii) sparse high-dimensional. The true data generating model is specified as $y_i = \bm{x}_i^T \bb^{\text{true}} + \epsilon_i$, with randomly generated predictors $\bm{x}_i = (x_{i1}, \dots, x_{ij})^T \in \R^p$, response $y_i \in \R$, and errors $\epsilon_i \in \R$, and a prespecified (i.e., fixed) ``true'' regression parameter $\bb^{\text{true}}$. Errors follow an autoregressive lag-1 process with serial correlation $\rho_\epsilon \in \{0, 0.2, 0.4\}$, while predictors follow a vector autoregressive lag-1 (VAR(1)) process with serial (over $i$) correlation $\rho_x$ and the common predictor-coordinate (over $j$) correlation $\rho_x \in \{0, 0.4, 0.6\}$. 
	
	For the predictors $\{\bm x_i: i = 1, \dots, n\}$, a standard normal distribution is used as the underlying marginal distribution (across both $i$ and $j$) for the VAR(1) process across all simulation settings. For the marginal distribution of the errors $\epsilon_i$, a wide variety of distributions are used across simulation settings, including: (a) \textit{thin-tailed}, corresponding to the standard normal distribution; (b) \textit{moderate-tailed}, corresponding to the Student $t$ distribution with 4 and 8 degrees of freedom, as well as, 99\%-1\% and 95\%-5\% discrete mixtures of standard normal-standard Cauchy and standard normal-$\mathcal{N}(0, 10^2)$ distributions; (c) \textit{heavy-tailed}, corresponding to the Student $t$-distribution with 1 (i.e., the standard Cauchy) and 2 degrees of freedom, as well as 90\%-10\% discrete mixtures of standard normal-standard Cauchy and standard normal-$\mathcal{N}(0, 10^2)$ distributions; and (d) \textit{extremely heavy-tailed}, corresponding to 90\%-10\% and 50\%-50\% discrete mixtures of standard normal and Uniform($-10^{10}, 10^{10}$) distributions.  
	
	Collectively, a wide range of sample sizes $n$ (50 to 20,000), and predictor dimensions $p$ (10 to 250) are considered. In low/moderate-dimensional $p = o(n)$ settings, it was ensured that $n > p$, while for the sparse high-dimensional setups that $n \leq p$, respectively. The ``true'' regression coefficient $\bb^{\text{true}} = (\beta^{\text{true}}_j : j = 1, \dots, p)$ was generated according to: $
	\beta^{\text{true}}_j = 0.5 + (j-1)\ \frac{2}{p-1}$,
	for the low/moderate-dimensional $p = o(n)$ setups, and according to:  $\beta^{\text{true}}_j = 2$ for $j \leq \ceil{p/20}$ and 0, otherwise,
	for the sparse high-dimensional setups, where $\ceil{x}$ denotes the smallest positive integer greater than or equal to $x$. A detailed description of all individual data-generating settings considered in our simulation experiments is provided in Supplementary Tables \ref{tab:settings-extremely heavy-normal}-\ref{tab:settings-thin-spikeslab}.
	
	For each data-generating setting, defined by a specific choice of $n$, $p$, correlation (for $\epsilon$ and $x$), and error distribution, $R = 200$ independent replicates of datasets are generated. In each replicate, we fit the Bayesian SPH model, along with $\ell_1$ and $\ell_2$ regressions, to compare their performance. In the low/moderate-dimensional setups, the ridge prior is used for model fitting, while for the sparse high-dimensional setups,  the spike-and-slab prior distribution is employed. The models include an intercept term, but an additional common variance parameter $\sigma^2$ beyond $\{\lambda_i: i=1, \dots, n\}$ was not incorporated, except for $\ell_2$ regressions, which included a common variance $\lambda$ for all observations. 
	
	All model parameters, including the SPH tuning hyperparameter $\alpha$, are estimated in a fully Bayesian manner through posterior MCMC sampling. Specifically, the proposed MCMC algorithms in Supplement~\ref{sec:details-posterior-comp} are used to generate 10,000 approximate posterior draws for model inference, after discarding the initial 5,000 used for burn-in purposes. The evaluation metrics used to assess the quality of the regression coefficient estimates, their credible intervals, as well as the model's prediction and variable selection performance are presented in the next four subsections, along with the main findings.

	\subsection{Estimation performance} \label{sec:simulation-estimation}
	
	To comprehensively assess the Bayesian estimation of the regression parameters $\bb^{\text{true}}$, we consider the posterior mean squared error (posterior MSE), defined as  
	\[
	M_{j, \text{data}} = \text{posterior MSE}(j, \text{data}) = E \left[ (\beta_j - \beta_j^{\text{true}})^2 \mid \text{data} \right],
	\]
	which utilizes the \textit{entire posterior distribution} of $\beta_j$ given a dataset. These values are computed from posterior MCMC draws for $\bb$. For each of the three models, SPH, $\ell_1$, and $\ell_2$, we compute the posterior MSE coordinate-wise for $\bb$ in each dataset, yielding a separate MSE $M_{j, r}^{\text{model}}$ for each data-generating setting, where $r = 1, \cdots, R = 200$ indexes the data replicates and $j = 1, \dots, p_1$ indexes the $\bb$ coordinates (see Supplementary Tables \ref{tab:settings-extremely heavy-normal}-\ref{tab:settings-thin-spikeslab}). The parameter $p_1$ denotes the number of non-zero (``signal'') coefficients in $\bb^{\text{true}}$, with $p_1 = p$ in all settings with  $n > p$, while $p_1 = \ceil{p/20}$ in settings with $n \leq p$. As a summary measure for each model in each data-generating setting, we then focus on the median posterior MSE, defined as  
	\[
	\median_{j=1, \dots, p_1} \left( \median_{r=1, \dots, R} M_{j, r}^{\text{model}} \right),
	\]
	where $\median(\cdot)$ denotes the empirical median operator.
	\begin{figure}[htpb]
		\centering
		\includegraphics[width=1.0\linewidth]{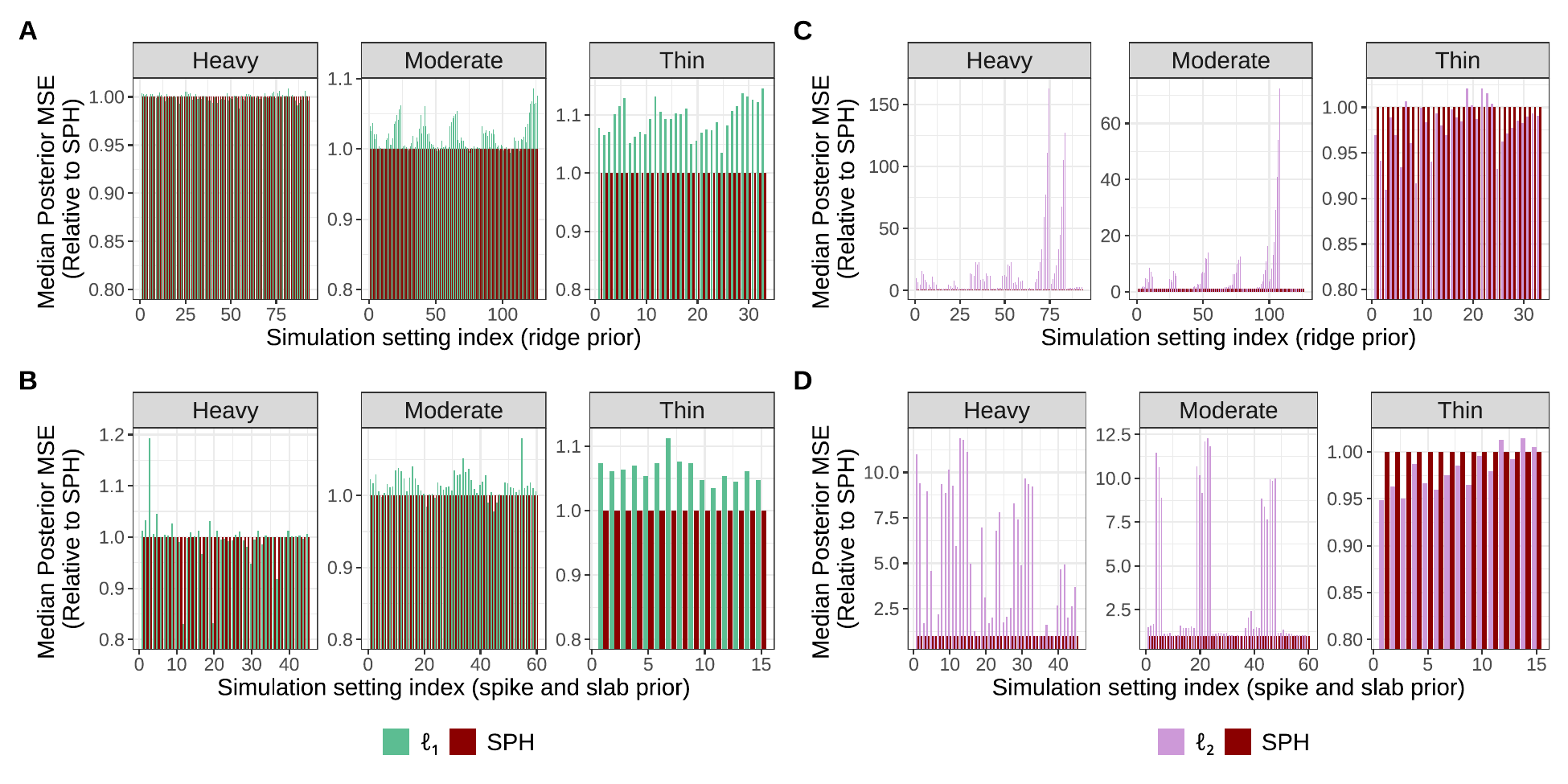}
		\caption{\footnotesize Median posterior MSEs (over replicates and $\bb$ coordinates) for Bayesian $\ell_1$, $\ell_2$, and SPH regression across simulation settings (detailed in Supplementary Tables \ref{tab:settings-extremely heavy-normal}-\ref{tab:settings-thin-spikeslab}). Panels A and C present low/moderate-dimensional settings with the ridge prior, while panels B and D depict sparse high-dimensional settings with the spike-and-slab prior. Each panel is grouped by error distributions--- heavy, moderate, and thin tails---displayed as subplots/facets. Median posterior MSE values are scaled relative to SPH in each setting, with results for SPH, $\ell_1$, and $\ell_2$ regressions shown in red, green, and purple.}
		\label{fig:simultion-post-rmse}
	\end{figure}
	
	Figure \ref{fig:simultion-post-rmse} depicts these median posterior MSEs obtained for the various simulation settings across the three models as vertical line/bar plots. The simulation settings are shown along the horizontal axis, with the median posterior MSE (relative to SPH for that setting) plotted along the vertical axis. Results are presented separately for low/moderate-dimensional setups (panels A, C) and sparse high-dimensional setups (panels B, D). Different error distribution groups---specifically, heavy, moderate, and thin---are considered within each setup and displayed as subplots within each panel.
	
	The figure illustrates that across all simulation settings---both low/moderate-dimensional using a ridge prior distribution for estimation (panels A, C) and high-dimensional involving a spike-and-slab prior (panels B, D)---the proposed SPH model achieves an impressive balance between $\ell_1$ and $\ell_2$ regressions in terms of parameter estimation accuracy. It closely approximates the better performing model in extreme situations, such as heavy-tailed distributions where $\ell_1$ regression is expected to be superior, and thin tailed distributions where $\ell_2$ regression is expected to perform better. Notably, in intermediate settings involving moderate-tailed, the SPH regression outperforms its $\ell_1$ and $\ell_2$ counterparts.

	\subsection{Prediction performance}
	\label{sec:prediction-perf}
	
	For prediction assessment, an independent test dataset is generated for each replicated training dataset used to fit the models. The test dataset preserves the same $n$, $p$, $\bb^{\text{true}}$, predictor and error correlation structure, and error distribution as the training data, but differed in the random elements of $\epsilon_i$, $\bm{x}_i$, and $y_i$. Subsequently, we focus on the expected posterior predictive distribution, averaged over the data distribution, to predict $y_i^{\text{test}}$ given $x_i^{\text{test}}$ and computed the prediction (posterior) MSE $\tilde{M}_{i, \text{data}}$ and its median over replicates and coordinates (i.e., observations), analogous to estimation MSE in Section~\ref{sec:simulation-estimation} (see Supplement~\ref{sec:detailed-prediction-perf} for detailed definitions).  
	
\begin{figure}[htpb]
		\centering
		\includegraphics[width=1.0\linewidth]{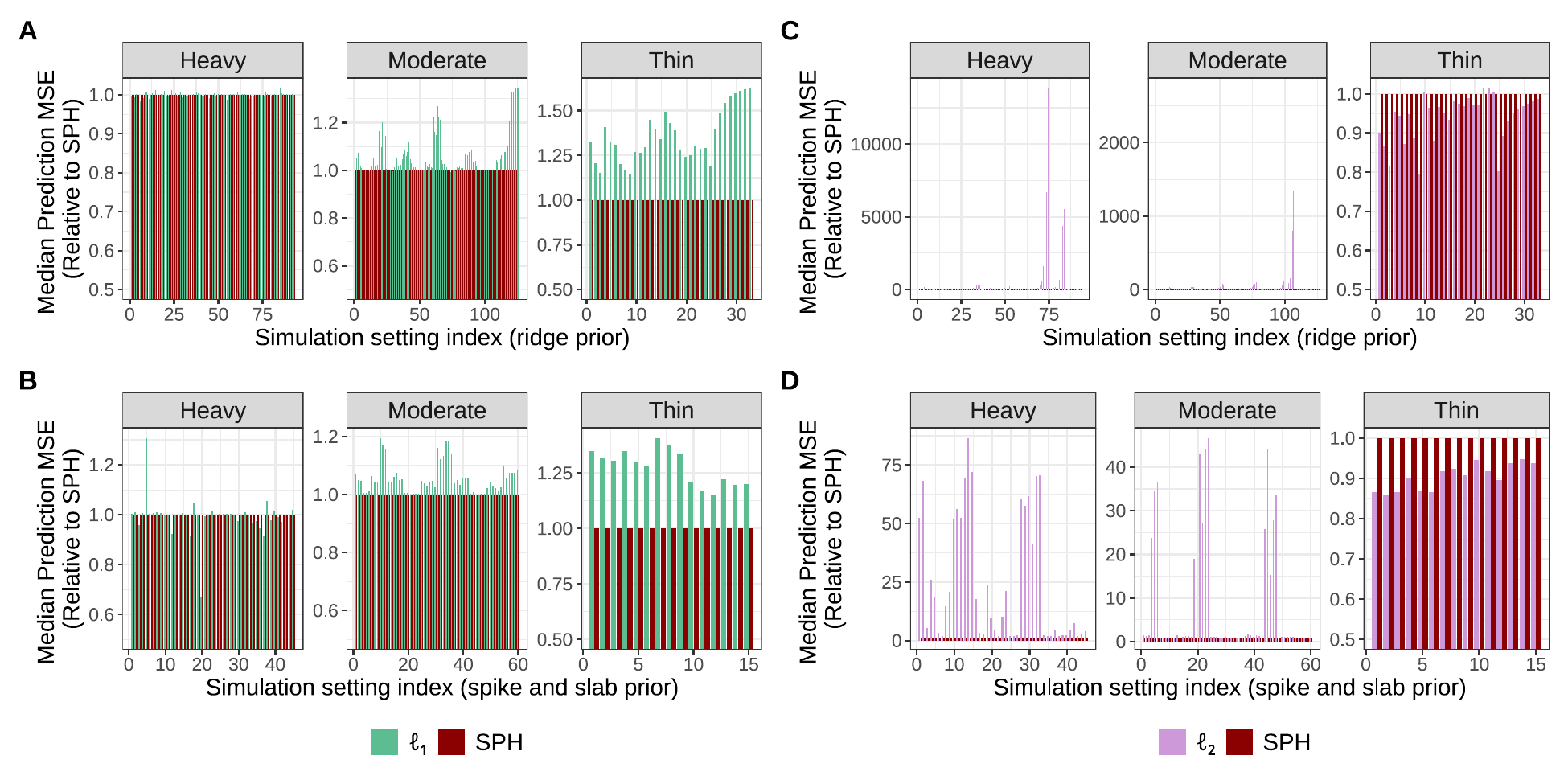}
		\caption{\footnotesize Median (over replicates and $\yb$ coordinates) prediction MSEs for Bayesian $\ell_1$, $\ell_2$, and SPH regression across simulation settings (detailed in Supplementary Tables \ref{tab:settings-extremely heavy-normal}-\ref{tab:settings-thin-spikeslab}). Panels A and C show low/moderate-dimensional setups with the ridge prior, while panels B and D depict sparse high-dimensional setups with the spike-and-slab prior. Each panel is grouped by error distributions--- heavy, moderate, and thin tails---displayed as subplots/facets. Median prediction MSE values are scaled relative to SPH in each setting, with results for SPH, $\ell_1$, and $\ell_2$ regressions shown in red, green, and purple.}
		\label{fig:simultion-post-predict-mse}
	\end{figure}

	Figure \ref{fig:simultion-post-predict-mse} depicts the median prediction MSEs across different simulation settings for the three models using vertical line/bar plots. Simulation settings are arranged along the horizontal axis, with the median posterior MSE (relative to SPH for that setting) displayed separately for low/moderate-dimensional setups (panels A, C) and sparse high-dimensional setups (panels B, D). Different error distribution groups--specifically, heavy, moderate, and thin--are considered within each setup and are labeled as facets in each panel.
	
	The figure conveys a similar message to the estimation performance assessment in Figure \ref{fig:simultion-post-rmse}. Across all simulation settings—low/moderate-dimensional setups with a ridge prior (panels A, C) and high-dimensional setups with a spike-and-slab prior (panels B, D)—the proposed SPH model strikes an impressive balance between $\ell_1$ and $\ell_2$ regressions in terms of prediction accuracy. In extreme scenarios, such as heavy-tailed distributions favoring $\ell_1$ regression, SPH closely aligns with $\ell_1$; whereas in thin-tailed settings suited for $\ell_2$ regression, SPH mirrors $\ell_2$'s performance. Notably, in moderate-tailed distributions, SPH surpasses both $\ell_1$ and $\ell_2$ regressions in predictive accuracy.

	\subsection{Interval Estimation Performance: Frequentist coverages of uncertainty (posterior credible) intervals}
	
	We evaluate the model's ability to quantify uncertainty through credible intervals for the regression parameters. To assess how well the different models capture uncertainty across various sample sizes and error distributions, we focus on settings with independent errors (zero serial correlation) and a low-dimensional $\bb$ ($p = 10$) to minimize confounding factors. We consider four error distribuiton categories: extremely heavy, heavy, moderate, and thin (see Supplementary Tables \ref{tab:settings-extremely heavy-normal}-\ref{tab:settings-thin-spikeslab}). Replicated datasets are generated across a wide sample size range from $n=50$ to $n=20,\!000$. For each replicate, we fit the $\ell_2$, $\ell_1$, and SPH models using the ridge prior, and obtain marginal credible intervals for each $\beta_j$ using equi-tailed posterior quantiles from MCMC draws. Frequentist coverage is computed as:  
	\[
	\text{coverage}(j, \text{model}) = \frac{1}{R} \sum_{r=1}^R \one\left(\hat \beta^{L, \text{model}}_{j, r} \leq \beta^{\text{true}}_{j} \leq \hat \beta^{U, \text{model}}_{j, r} \right),
	\]
	while the mean credible interval length is:  
	\[
	\text{mean length}(j, \text{model}) = \frac{1}{R} \sum_{r=1}^R \left(\hat \beta^{U, \text{model}}_{j, r} - \hat \beta^{L, \text{model}}_{j, r} \right),
	\]
	where $(\hat \beta^{L, \text{model}}_{j, r}$ and $\hat \beta^{U, \text{model}}_{j, r})$  denote the 90\% equi-tailed posterior credible interval for $\beta_j$ from MCMC draws for the model in each replicate $r$.
	
	Bayesian $\ell_1$ regression with fixed/low $p$ and vague priors on $\bb$ is known to exhibit poor frequentist coverage under high error contamination and model misspecification \citep{sriram2015sandwich, yang2016posterior}. This issue arises from the non-standard asymptotic behavior of the Bayesian $\ell_1$ posterior, causing a mismatch between the asymptotic (in $n$) sampling covariance of point estimates of $\bb$ (e.g., the posterior mean or mode) and the asymptotic posterior covariance of $\bb$, as typically used in Bernstein–von Mises-type asymptotic normal approximations of posterior distributions.  To address this, adjustments to the asymptotic posterior covariance of $\bb$ have been proposed. Specifically, for Bayesian $\ell_1$ regression with known $\sigma$, it has been established \citep{sriram2015sandwich, yang2016posterior} that an asymptotic normal approximation of the form $\mathcal{N}_p(E(\bb \mid \text{data}), V_n)$ with  
	\[
	V_n = \frac{1}{\sigma^2} \left( \text{var}(\bb \mid \text{data}) \ \bm{X}^T \bm{X} \ \text{var}(\bb \mid \text{data}) \right)
	\]
	for the posterior distribution of $\bb$ correctly aligns with the frequentist asymptotic sampling distribution of $E(\bb \mid \text{data})$. Here, $\text{var}(\bb \mid \text{data})$ is the posterior covariance matrix, computable via MCMC draws. This contrasts with Bayesian $\ell_2$ regression, where both the posterior distribution of $\bb$ and the sampling distribution of $E(\bb \mid \text{data})$ attain the same asymptotic normal distribuiton of the form  
	$\mathcal{N}_p(E(\bb \mid \text{data}), \text{var}(\bb \mid \text{data}))$. 
	
\begin{figure}[htpb]
		\centering
		\includegraphics[width=1.0\linewidth]{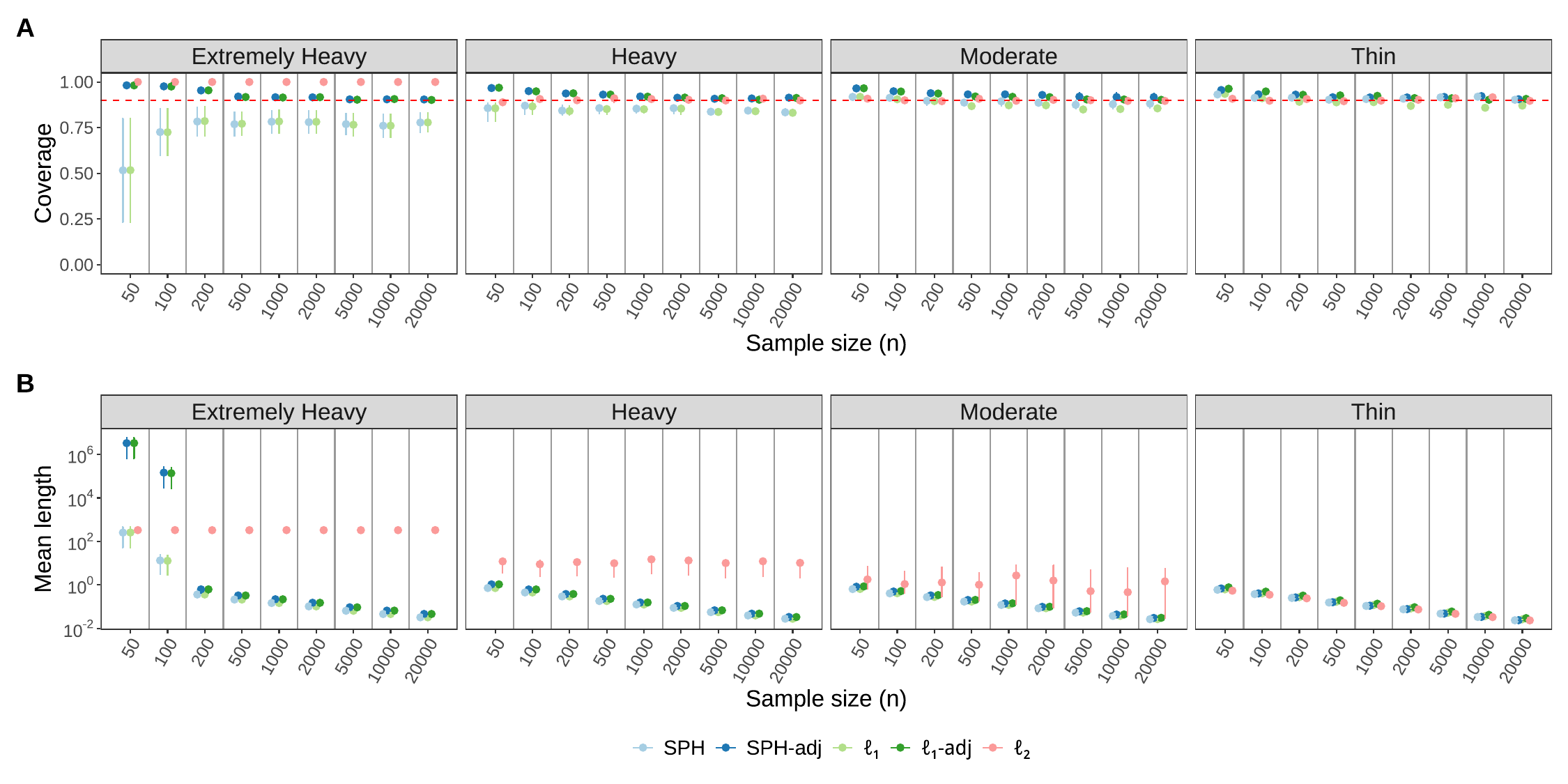}
		\caption{\footnotesize Replication-based coverages (Panel A) and mean lengths (Panel B; vertical axis plotted in a log-scale) of 90\% Bayesian credible (equi-tailed) intervals for Bayesian $\ell_1$, $\ell_1$-adj, $\ell_2$, SPH, and SPH-adj regression models across various error distribution categories (extremely heavy, heavy, moderate, and thin) and sample sizes ($n$). }
		\label{fig:simultion-coverage}
	\end{figure}

	In our simulation experiments we implement this adjustment to produce an adjusted $\ell_1$ posterior (``$\ell_1$-adj'') and evaluate the corresponding posterior credible intervals. Specifically, credible/confidence intervals for the coordinates of $\bb$ under $\ell_1$-adj are obtained through the corresponding approximate normal equi-tailed quantiles. Since the SPH loss converges to the $\ell_1$ loss under heavy-tailed settings, we also consider analogous adjustments to the SPH-based $\bb$ posteriors, resulting in the ``SPH-adj'' posterior and the corresponding adjusted credible intervals. The results are displayed in Figure~\ref{fig:simultion-coverage}. Frequentist coverages and average interval lengths are plotted on the vertical axis, with dots showing the median and error bars indicating the 10\% and 90\% empirical percentiles of the metrics across all simulation settings for each $n$ (horizontal axis). Colors differentiate between models.
	
	The figure illustrates the robustness of SPH and $\ell_1$, along with their posterior covariance-adjusted counterparts (SPH-adj and $\ell_1$-adj), in achieving adequate frequentist coverage, while maintaining narrow interval lengths across all sample sizes ($n$) and error distributions. Notably, SPH-adj and $\ell_1$-adj attain near-nominal 90\% coverage for all $n \geq 200$ and data-generating settings, while keeping mean interval lengths small. However, for $n \leq 100$ under extremely heavy-tailed errors, the adjustments produce overly wide intervals, which become reasonable again for $n \geq 200$. The adjustment is essential for $\ell_1$ to prevent undercoverage under moderate and thin-tailed errors, whereas SPH naturally achieves sufficient coverage in these cases---by mimicking $\ell_2$---while benefiting from the adjustment under heavy-tailed errors. Across all settings, SPH and $\ell_1$ consistently yield shorter intervals than  $\ell_2$, except for thin-tailed errors. Although $\ell_2$ regression attains near nominal coverage in most cases (except for extremely heavy-tailed errors, where it severely overcovers), it does so at the cost of substantially wider intervals---often over an order of magnitude wider than SPH and $\ell_1$---except in thin-tailed error settings.  
	
	\subsection{Variable Selection Performance under Spike and Slab prior}
	
	To assess variable selection under the spike-and-slab prior distribution in the $n>p$ regime, we focus on corresponding simulation settings. For each replicated dataset,  we obtain a Bayesian point estimate of the variable selection indicator vector $\hat{\bm{\gamma}}^{\text{model}} = (\hat{\gamma}^{\text{model}}_j: j=1, \dots, p)$, computed separately for each model and each coordinate $j$ of $\bb$ as  
	$\hat{\gamma}_j^{\text{model}} = \one\left[\Pr\left(\gamma_j^{\text{model}} = 1 \mid \text{data}\right) > 0.5\right]$,
	where the posterior probabilities are calculated using posterior MCMC draws. Variable selection performance is then evaluated using the Matthews Correlation Coefficient (MCC), which quantifies agreement between the $\hat{\bm{\gamma}}^{\text{model}}$ and the ``true active'' variable indicators 
	$\bm{\gamma}^{\text{true}} = \left( \one \left[\beta_j^{\text{true}} \neq 0\right]: j = 1, \dots, p \right)$.
	To summarize performance across replicates, the empirical median of the MCC values is computed for each combination of $n$, $p$, predictor and error correlation, and error tail category (heavy, moderate, and thin) across all simulation settings. The results are presented in Table~\ref{tab:varselect-MCC}.
	\begin{table}[t]
		\centering
		\begin{small}
			\renewcommand{\arraystretch}{0.5}
			\begin{tabular}[t]{cc>{}c|cc>{}c|cc>{}c|ccc}
				\toprule
				\multicolumn{3}{c}{ } & \multicolumn{3}{c}{Error Tail: Heavy} & \multicolumn{3}{c}{Error Tail: Moderate} & \multicolumn{3}{c}{Error Tail: Thin} \\
				\cmidrule(l{3pt}r{3pt}){4-6} \cmidrule(l{3pt}r{3pt}){7-9} \cmidrule(l{3pt}r{3pt}){10-12}
				Correlation & $n$ & $p$ & $\ell_1$ & $\ell_2$ & SPH & $\ell_1$ & $\ell_2$ & SPH & $\ell_1$ & $\ell_2$ & SPH\\
				\midrule
				& 75 & 100 & \textbf{1.00} & 0.08 & \textbf{1.00} & \textbf{1.00} & 0.88 & \textbf{1.00} & \textbf{1.00} & \textbf{1.00} & \textbf{1.00}\\
				\cmidrule{2-12}
				& 75 & 200 & 0.86 & 0.03 & \textbf{0.88} & \textbf{1.00} & 0.81 & \textbf{1.00} & \textbf{1.00} & \textbf{1.00} & \textbf{1.00}\\
				\cmidrule{2-12}
				\multirow{-3}{*}{\centering\arraybackslash None} & 75 & 250 & 0.51 & 0.02 & \textbf{0.52} & 0.98 & 0.72 & \textbf{0.99} & \textbf{1.00} & 0.98 & \textbf{1.00}\\
				\cmidrule{1-12}
				& 75 & 100 & 0.98 & 0.16 & \textbf{0.99} & \textbf{1.00} & 0.90 & \textbf{1.00} & \textbf{1.00} & \textbf{1.00} & \textbf{1.00}\\
				\cmidrule{2-12}
				& 75 & 200 & \textbf{0.96} & 0.11 & 0.94 & \textbf{1.00} & 0.87 & \textbf{1.00} & \textbf{1.00} & \textbf{1.00} & \textbf{1.00}\\
				\cmidrule{2-12}
				& 75 & 250 & \textbf{0.76} & 0.05 & 0.71 & \textbf{1.00} & 0.82 & \textbf{1.00} & \textbf{1.00} & \textbf{1.00} & \textbf{1.00}\\
				\cmidrule{2-12}
				& 100 & 100 & \textbf{1.00} & 0.16 & \textbf{1.00} & 0.99 & 0.94 & \textbf{1.00} & \textbf{1.00} & \textbf{1.00} & \textbf{1.00}\\
				\cmidrule{2-12}
				& 100 & 200 & 0.99 & 0.12 & \textbf{1.00} & \textbf{1.00} & 0.92 & \textbf{1.00} & \textbf{1.00} & \textbf{1.00} & \textbf{1.00}\\
				\cmidrule{2-12}
				\multirow{-6}{*}{\centering\arraybackslash Low} & 100 & 250 & 0.97 & 0.10 & \textbf{0.98} & \textbf{1.00} & 0.89 & \textbf{1.00} & \textbf{1.00} & \textbf{1.00} & \textbf{1.00}\\
				\cmidrule{1-12}
				& 75 & 100 & \textbf{0.92} & 0.21 & 0.91 & \textbf{0.99} & 0.90 & \textbf{0.99} & 0.99 & \textbf{1.00} & 0.99\\
				\cmidrule{2-12}
				& 75 & 200 & \textbf{0.89} & 0.25 & 0.88 & \textbf{0.99} & 0.88 & \textbf{0.99} & \textbf{1.00} & \textbf{1.00} & \textbf{1.00}\\
				\cmidrule{2-12}
				& 75 & 250 & \textbf{0.75} & 0.18 & \textbf{0.75} & \textbf{0.99} & 0.81 & \textbf{0.99} & \textbf{1.00} & \textbf{1.00} & \textbf{1.00}\\
				\cmidrule{2-12}
				& 100 & 100 & \textbf{0.92} & 0.34 & \textbf{0.92} & \textbf{0.99} & 0.94 & \textbf{0.99} & 0.99 & \textbf{1.00} & 0.99\\
				\cmidrule{2-12}
				& 100 & 200 & \textbf{0.93} & 0.25 & \textbf{0.93} & \textbf{0.99} & 0.92 & \textbf{0.99} & \textbf{1.00} & \textbf{1.00} & \textbf{1.00}\\
				\cmidrule{2-12}
				\multirow{-6}{*}{\centering\arraybackslash Moderate} & 100 & 250 & \textbf{0.93} & 0.24 & \textbf{0.93} & \textbf{1.00} & 0.93 & 0.99 & \textbf{1.00} & \textbf{1.00} & \textbf{1.00}\\
				\bottomrule
			\end{tabular}
		\end{small}
		
		\caption{\footnotesize \label{tab:varselect-MCC}Overall variable selection performances as measured by MCC (summarized via medians across replicates and simulation settings) under simulation settings with models fitted using spike and slab priors. In each simulation setting with a specific combination of correlation, $n$, $p$, and error tail, the highest MCC obtained from the three models $\ell_1$, $\ell_2$, and SPH are highlighted via \textbf{bold} text.}
	\end{table}

	The table demonstrates SPH's impressive variable selection performance across all simulation settings. It consistently achieves the highest or nearly highest MCC values among the three models, regardless of the correlation structure, $n$, or $p$. SPH achieves nearly perfect variable selection under moderate and thin-tailed errors and maintains high performance for most heavy-tailed settings, except when $n \ll p$ (e.g., $n = 75$ and $p = 250$), where the MCC values are moderate. Bayesian $\ell_1$ regression performs comparably to SPH in most cases, whereas $\ell_2$ regression consistently lags, except under thin-tailed errors, where it achieves higher MCC values. All models exhibit a decline in variable selection performance under heavy-tailed error distribution when predictor and error correlations increase, which reduces the effective sample sizes.

	\section{An Application to Forecasting the US GDP}
	\label{sec:real-data-analysis}
	
	Accurate GDP forecasts are vital for a diverse set of stakeholders, including policymakers, businesses, and investors as they guide decisions on monetary policy, resource allocation, and market strategies. Various statistical models have been employed for the forecasting task, including autoregressive distributed lag regression models \citep{ghosh2023bayesian}, vector autoregressive models \citep{koop2013forecasting} and factor models \citep{higgins2014gdpnow}. A common characteristic of these modeling strategies is the inclusion of a large number of macroeconomic and financial indicators, which significantly enhances their forecasting performance \citep{cimadomo2022nowcasting}.
	
	We apply the SPH regression model to forecast GDP based on data from 1960Q1 to 2023Q4, using the quarterly GDP growth rate as the outcome and a broad set of macroeconomic indicators as predictors. These include GDP components (production, consumption, investment, trade), monetary and fiscal policy measures, employment metrics, price levels, and financial indices. The model incorporates both current and lagged predictors, with transformations ensuring stationarity as recommended by \citet{mccracken2016fred,mccracken2020fred}.
	
	Two major economic disruptions in the dataset stand out: the 2008 Great Recession (2008Q1–2009Q2), which significantly reduced GDP, and the COVID-19 pandemic (2020Q1–2022Q4), during which fiscal measures temporarily boosted GDP in certain quarters.
	To account for these disruptions, we conduct two separate analyses: (1) the pre-post-recession analysis, forecasting GDP from 2007Q2–2011Q1, and (2) the pre-post-COVID analysis, forecasting GDP from 2019Q4–2023Q4. Both use a rolling window framework, where regression coefficients are estimated using a fixed window (1960Q2–2006Q4 for the first and 1960Q2–2019Q2 for the second), shifting forward one quarter at a time to incorporate recent data and assess predictive accuracy.
	
	For comparison purposes, six models are fitted: generalized Bayesian $\ell_1$ (``L1''), $\ell_2$ (``L2''), and SPH regression, each with either a normal/ridge (``N") or spike-and-slab (``SS") prior distribution, across all training datasets. Models are fitted using 10,000 post-burn-in draws (after discarding the initial 5,000 burn-in draws) of the proposed MCMC samplers. To assess the impact of outliers (extreme GDP growth fluctuations), outlier-filtered versions of the $\ell_1$ and SPH models are also considered. Outliers are identified using posterior draws of $\lambda_i$, where a data point $i'$ is flagged if the upper 95\% posterior credible limit of $\lambda_{i'}$ is significantly distant from the rest, based on Tukey’s boxplot method (see Supplement~\ref{sec:diagnostics} for details on outlier filtering). These points are removed, and models are refitted using 10,000 posterior MCMC draws (after 5,000 burn-in iterations). Outlier filtering is not applicable to the $\ell_2$ model, as it lacks the micro-level contamination parameters $\lambda_i$.

	For each model fit, posterior draws for the next-quarter GDP forecast are obtained and compared to observed values to compute the prediction (posterior) MSE (Eq. \eqref{defn:pred-post-mse}). These calculations are conducted across all time points under the rolling window scheme, separately for each analysis (pre-post COVID and pre-post recession) and each model/prior combination, including the original and outlier-filtered refits for the $\ell_1$ and SPH models. Results are then scaled relative to the prediction MSEs of the SPH-N outlier-filtered refit.
	
	\begin{figure}[htpb]
		\centering
		\includegraphics[width=0.8\linewidth]{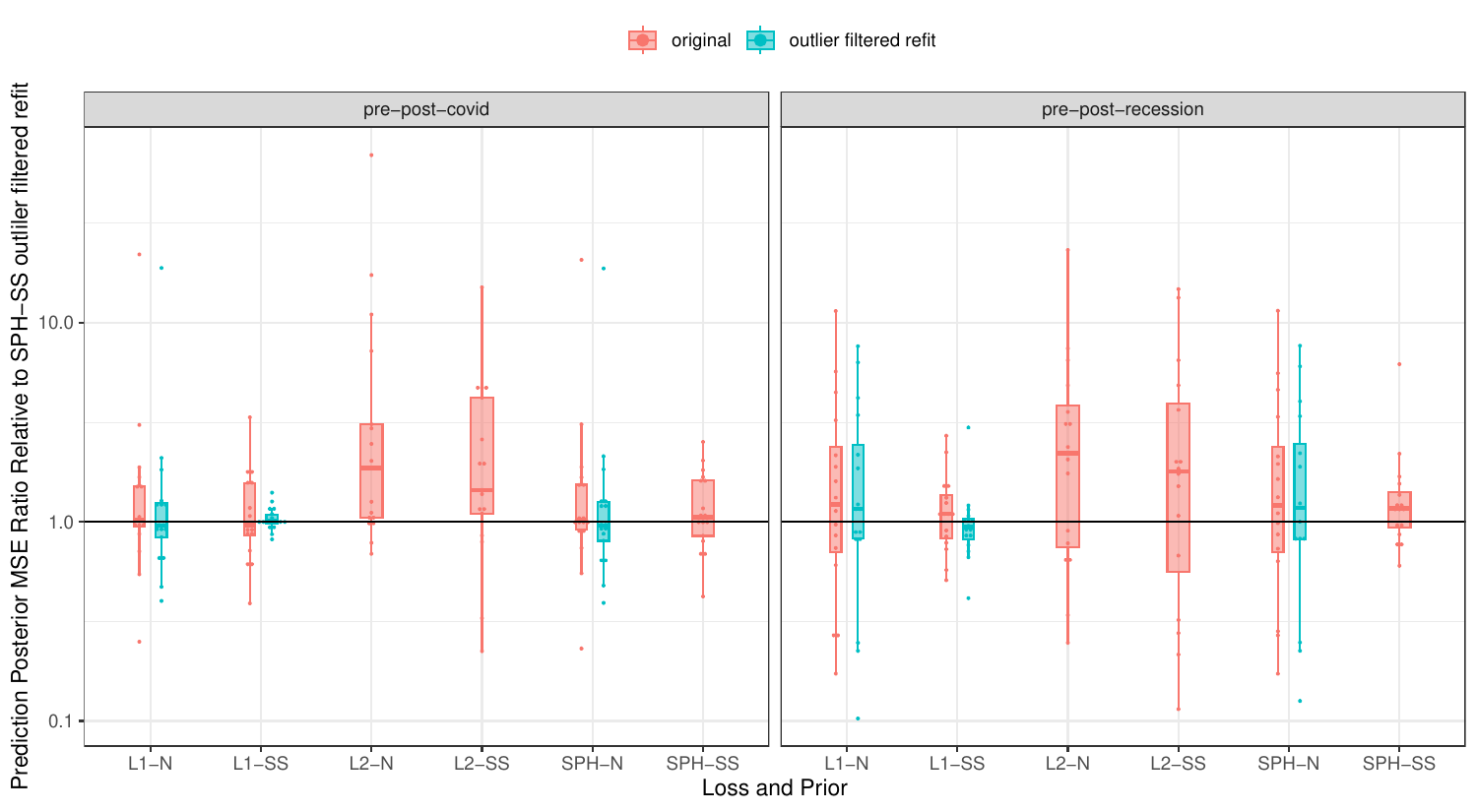}
		\caption{\footnotesize Comparing the prediction (posterior) MSEs for the different models, priors, and fit combinations relative to the SPH-SS outlier-filtered refit model for pre-post-COVID (left panel) and pre-post-recession (right panel) analyses. The boxplots display the prediction MSE ratios for various models (L1-N, L1-SS, L2-N, L2-SS, SPH-N, and SPH-SS) for the original fit (red) and outlier-filtered refit (blue). The L2 model lacks an outlier-filtered refit version as it does not include the $\lambda_i$ parameters.  The horizontal black line at 1 represents the baseline performance of the SPH-SS outlier-filtered refit model. Lower MSE ratios indicate better predictive performance compared to the baseline.}
		\label{fig:real-data-prediction-mse}
	\end{figure}
	
	Figure~\ref{fig:real-data-prediction-mse} visualizes the scaled prediction MSEs using a boxplot with embedded dots, illustrating the distribution of posterior MSEs across time points, grouped by model/prior combinations (horizontal axis) and fitting instances (before/after outlier removal; color-coded; no outlier removal for $\ell_2$ due to the absence of $\lambda_i$ parameters). The figure shows that, except for L1-N, most boxplots lie above 1, indicating worse prediction MSEs than the SPH-N refit. The performance of SPH-N and L1-N refits is largely comparable, as expected in the presence of outliers, where the SPH loss closely mimics $\ell_1$.
	
	We note that many studies on GDP forecasting post-pandemic exclude data from 2020Q1–2020Q3, following the strategy and recommendations in \cite{schorfheide2021real}, to improve forecast accuracy. This exclusion addresses the large outliers caused by the Covid-19 disruption, which otherwise severely affect the models' forecasting ability, albeit at the cost of misspecifying the true dynamics of the data-generating mechanism. In contrast, the SPH regression model, even when not enhanced with outlier filtering, handles such disruptions well without requiring the exclusion of these critical data points.

	\section{Concluding Remarks}
	
	\noindent
	This paper develops a generalized Bayesian framework for high-dimensional regression under contamination of the response, leveraging a novel scaled pseudo-Huber loss function. The SPH loss adaptively balances $\ell_1$ and $\ell_2$ regression, ensuring robustness while maintaining efficiency. We establish posterior consistency under high-dimensional scaling, accommodating both i.i.d. and temporally correlated settings, significantly extending existing methods. Extensive simulations and a real-world forecasting application with time-dependent data validate the framework’s effectiveness.
	
	Additionally, a filtering strategy for heavily contaminated data, adopted in the application, further improves forecasting performance, aligning with recent work on robust regression under contamination in both errors and covariates \cite{PJL:2024}. This suggests a promising direction for future research: integrating filtering techniques within a high-dimensional Bayesian paradigm with sparsity-inducing priors and establishing corresponding posterior contraction rates.

	\bibliographystyle{rss}
	\bibliography{refereces}

	\newpage
	\appendix
	\renewcommand{\thesection}{S.\arabic{section}}
	\renewcommand{\theequation}{S.\arabic{equation}}
	\renewcommand{\thefigure}{S.\arabic{figure}}
	\renewcommand{\thetable}{S.\arabic{table}}
	
	\counterwithin{equation}{section}
	\counterwithin{figure}{section}
	\counterwithin{table}{section}

	\begin{center}
		\Large {Supplement for ``A generalized Bayesian approach for high-dimensional robust regression with serially correlated errors and predictors''}
	\end{center}
	
	\medskip
	
	\medskip

	\noindent This supplement provides detailed information on various technical developments, including the MCMC samplers for the proposed model under ridge and spike-and-slab prior distributions, detailed proofs of the theorems presented in the main text, and tabulated descriptions of the simulation settings used in our experiments. Equations, sections, tables, and figures in this document are labeled with an ``S." prefix (e.g., Equations (S.1), (S.2); Section S.1; Table S.2; Figure S.3, etc.).

	This supplement provides detailed information on various technical developments, including the MCMC samplers for the proposed model under ridge and spike-and-slab prior distributions, detailed proofs of the theorems presented in the main text, and tabulated descriptions of the simulation settings used in our experiments. Equations, sections, tables, and figures in this document are labeled with an ``S." prefix (e.g., Equations (S.1), (S.2); Section S.1; Table S.2; Figure S.3, etc.).

	\section{Technical Developments for Section \ref{sec:robust-pseudo-huber}}
	\label{sec:proofs-pseudo-Huber} 
	
	\noindent
	\textbf{Proof of Proposition \ref{prop:pseudo-huber-mixture}}:
	
	The joint density of $(\ep, \lambda)$ is given by:
	\begin{align*}
		f_{\ep, \lambda}(\ep, \lambda) &= \frac{1}{\sqrt{2\pi}} \lambda^{-1/2} \exp \left[ -\frac12 
		\frac{\ep^2}{\lambda} \right] \ \times \ \frac{\sqrt{1 + \alpha^2}/\alpha}{2 K_1(\alpha \sqrt{1 + \alpha^2})} \exp\left[ -\frac12 \left\{ (1+\alpha^2) \lambda + \frac{\alpha^2}{\lambda} \right\} \right] \\
		&= C_1(\alpha) \ \lambda^{-1/2} \ \exp\left[ -\frac12 \left\{ (1 + \alpha^2) \lambda + \frac{\alpha^2 + \ep^2}{\lambda} \right\} \right]; \quad \lambda > 0, -\infty < \ep < \infty,  
	\end{align*}
	where $C_1(\alpha) = \frac{\sqrt{1 + \alpha^2}}{2 \sqrt{2\pi} \alpha K_1(\alpha \sqrt{1 + \alpha^2})}$ and $K_1(\cdot)$ denotes the Bessel function of the second kind. We consider the transformation $(\ep, \lambda) \mapsto (\ep, \kappa)$ where $\kappa = 1/\lambda$. The absolute value of the Jacobian of the transformation is simply $1/\kappa^2$. Therefore, in the transformed scale, the joint density of $(\ep, \kappa)$ is:
	\[
	f_{\ep, \kappa}(\ep, \kappa) = C_1(\alpha) \  \kappa^{-3/2} \ \exp\left[ -\frac12 \left\{  (\alpha^2 + \ep^2)\kappa  + \frac{1 + \alpha^2}{\kappa} \right\} \right].
	\]
	Note that for any fixed $\ep \in (-\infty, \infty)$, the right hand side above without the proportionality constant $C_1(\alpha)$ is the kernel of an Inverse-Gaussian$\ds \left(\mu = \frac{\sqrt{1 + \alpha^2}}{\sqrt{\alpha^2 + \ep^2}}, \sigma = 1+\alpha^2 \right)$ density for $\kappa$. Thus, 
	\begin{align*}    
		&\quad \int_{0}^\infty \kappa^{-3/2} \ \exp\left[ -\frac12 \left\{  (\alpha^2 + \ep^2)\kappa  + \frac{1+\alpha^2}{\kappa} \right\} \right] \ d\kappa \\
		&= \frac{\sqrt{2\pi}}{\sqrt{1 + \alpha^2}} \ \exp \left[-\sqrt{1 + \alpha^2} \sqrt{\alpha^2 + \ep^2 }\right] \\
		&= \frac{\sqrt{2\pi}}{\sqrt{1 + \alpha^2}} \  \exp\left[ - \alpha \sqrt{1 + \alpha^2} \left(\sqrt{1 + \left(\frac{\ep}{\alpha}\right)^2} \right) \right] \\
		&= \frac{\sqrt{2\pi}}{\sqrt{1 + \alpha^2}} \ \exp\left(-\alpha\sqrt{1 + \alpha^2}\right) \ \exp\left[ - \alpha \sqrt{1 + \alpha^2} \left(\sqrt{1 + \left(\frac{\ep}{\alpha}\right)^2} - 1 \right) \right].
	\end{align*}
	Therefore, the marginal density of $\epsilon$ is obtained as
	\[
	f_\ep(\ep \mid \alpha) = \int_0^\infty f_{\ep, \kappa}(\ep, \kappa) \ d\kappa = C_2(\alpha) \ \exp\left[ -  \alpha \sqrt{1 + \alpha^2} \left(\sqrt{1 + \left(\frac{\ep}{\alpha}\right)^2} - 1 \right) \right], 
	\]
	where $C_2(\alpha) = C_1(\alpha) \ \frac{\sqrt{2\pi}}{\sqrt{1 + \alpha^2}} \ \exp\left(- \alpha \sqrt{1 + \alpha^2}\right) 
	= \frac{1}{2 \alpha K_1\left(\alpha \sqrt{1 + \alpha^2}\right)} \ \exp\left(-\alpha \sqrt{1 + \alpha^2}\right)
	$ is free of $\ep$. This completes the proof. \hfill$\Box$
	
	\bigskip
	
	\begin{remark}\label{rem:pseudo-Huber}
		Since $\lim_{\alpha \to 0} \alpha K_1\left(\alpha \sqrt{1 + \alpha^2}\right) = 1$, it follows that as $\alpha \to 0$, $f_\ep(\ep \mid \alpha) \to \frac{1}{2} \exp\left( - |\ep | \right)$ which is the density of a standard Laplace distribution. On the other hand, as $\alpha \to \infty$, $\sqrt{1 + \alpha^2} \simdot \alpha$ and $C_2(\alpha) \simdot \tilde C_2(\alpha)$ where $\tilde C_2(\alpha) = \frac{1}{2 \alpha  K_1 (\alpha^2) \ \exp(\alpha^2)}$. The notation ``$\simdot$'' represents asymptotic equivalence between two functions $f_1(\alpha)$ and $f_2(\alpha)$, defined as $f_1(\alpha) \simdot f_2(\alpha)$ as $\alpha \to \infty$, if and only if $\lim_{\alpha \to \infty} \frac{f_1(\alpha)}{f_2(\alpha)} = 1$. For positive real $\alpha \to \infty$, $K_1(\alpha) = \sqrt{\frac{\pi }{2\alpha}} \ \exp(-\alpha) \ \left(1 + o\left(\frac{1}{\alpha}\right)\right)$ \citep[p. 378, 9.7.2]{abramowitz:stegun:1988}; hence as $\alpha \to \infty$
		\[
		\frac{1}{\tilde C_2(\alpha)} = 2 \alpha  \sqrt{\frac{\pi }{2\alpha^2}} \ \exp\left(-\alpha^2\right) \ \left(1 + o\left(\frac{1}{\alpha^2}\right)\right) \ \exp\left(\alpha^2\right) = \sqrt{2\pi} \left(1 + o\left(\frac{1}{\alpha^2}\right)\right) \to \sqrt{2 \pi}.
		\] 
		Further, $\exp\left[ - \alpha \sqrt{1 + \alpha^2} \left(\sqrt{1 + \left(\frac{\ep}{\alpha}\right)^2} - 1 \right) \right] \simdot \exp\left[ - \alpha^2 \left(\sqrt{1 + \left(\frac{\ep}{\alpha}\right)^2} - 1 \right) \right]$ as $\alpha \to \infty$ and
		\[
		\lim_{\alpha \to \infty} \alpha^2 \left(\sqrt{1 + \left(\frac{\ep}{\alpha}\right)^2} - 1  \right) =  \lim_{t \to 0} \frac{\sqrt{1 + t \ep^2} - 1}{t} = \lim_{t \to 0} \frac{\ep^2}{2 \sqrt{1 + t \ep^2}} = \frac{\ep^2}{2}
		\]
		where the second last last equality is a consequence of the L'Hospital rule. Together, this implies $f_\ep(\ep \mid \alpha) \to \frac{1}{\sqrt{2\pi}} \exp(- \frac{\ep^2}{2})$, the standard normal density, as $\alpha \to \infty$.
	\end{remark}
	
	\section{Additional Details on Posterior Distribution Computations}
	\label{sec:details-posterior-comp} 
	
	\subsection{Posterior MCMC sampling for the Gaussian prior distribution}
	For the Gaussian, weakly informative prior distribution, some standard calculations lead to the following simplified form of the posterior distribution of the model parameters. Let $\lb = (\lambda_1, \dots, \lambda_n)^T$ and $\Lambda = \diag(\lb)$, we get:
	\begin{align*}
		\quad & \pi(\bb, \lb, \sigma^2, \alpha^2 \mid \data) \\
		\propto \ & \left\{\prodin  \lambda_i^{-1/2} \right\} (\sigma^2)^{-n/2} \exp\left[ -\frac{1}{2\sigma^2} (\yb - X\bb)^T \Lambda^{-1} (\yb - X\bb) \right] \\
		& \times  2^{-n} (1 + \alpha^2)^{n/2} (\alpha^2)^{-n/2} \left[K_{1}\left(\sqrt{\alpha^2(1 + \alpha^2)}\right)\right]^{-n}   \exp \left[ -\frac12 \left(\sumin \frac{\alpha^2}{\lambda_i} + (1+\alpha^2) \sumin{\lambda_i} \right) \right] \\
		& \times (\sigma^2)^{-p/2} |Q|^{1/2} \exp\left[ -\frac{1}{2\sigma^2} (\bb - \bb_0)^T Q (\bb - \bb_0) \right] \\
		& \times (\sigma^2)^{-a_\sigma - 1} \exp\left(-\frac{1}{\sigma^2} b_\sigma\right) \times  \\
		& \times (\alpha^2)^{a_\alpha - 1} \exp(- b_\alpha \alpha^2) \\
		\propto \ & \exp\left[ -\frac{1}{2\sigma^2} (\yb - X\bb)^T \Lambda^{-1} (\yb - X\bb) \right] \\
		& \times  \prodin \left\{ \lambda_i^{\frac{1}{2} - 1} \exp \left[ -\frac12 \left(\frac{\alpha^2}{\lambda_i} + \alpha^2 {\lambda_i} \right) \right] \right\}  \\
		& \times \exp\left[ -\frac{1}{2\sigma^2} (\bb - \bb_0)^T Q (\bb - \bb_0) \right]  \\
		& \times (\sigma^2)^{-\left( \frac{n+p}{2} + a_\sigma\right) - 1} \exp\left(-\frac{1}{\sigma^2} b_\sigma\right) \\
		& \times (\alpha^2)^{a_\alpha - 1} \exp(- b_\alpha \alpha^2) 
	\end{align*}
	
	\noindent While direct independent sampling from this density is infeasible, we propose an efficient slice-within-Gibbs sampler for MCMC sampling from this posterior density. Starting from some initial values (we used the frequentist estimates of $\mu$, $\sigma$, and $\bb$ in our computations), the algorithm iteratively generates posterior draws for the model parameters. Steps involved in one iteration of the sampler is presented in Algorithm~\ref{algo:mcmc-sampler-ridge}.

	
	

	\begin{algorithm}[htpb]
		\caption{One iteration of a slice-within-Gibbs sampler for posterior sampling for the SPH regression model under the ridge prior}
		
		\label{algo:mcmc-sampler-ridge}
		
		\begin{enumerate}
			\item Generate the intercept $\mu$ (if included in the model) from  
			\[
			\mu \mid \bb, \lambda_1, \dots, \lambda_n, \sigma^2, \bm y, X \sim \n\left(v_{\lambda;\mu} m_{\lambda; \mu}, \sigma^2 v_{\lambda;\mu} \right)
			\]
			where $v_{\lambda;\mu} = 1\left/\left(\frac{1}{\tau^2_\mu} + \sum_{i=1}^n \frac{1}{\lambda_i} \right) \right.$ and $m_{\lambda; \mu} = \sum_{i=1}^n \frac{1}{\lambda_i} (y_i - \xb_i^T \bb )$, while setting $\sigma \equiv 1$ if not included in the model.
			
			\item Generate $\bb$ from
			\[
			\bb \mid \mu, \lambda_1, \dots, \lambda_n, \sigma, \bm y, X \sim \n\left( V_\lb m_\lb, \sigma^2 V_\lb \right)
			\]
			where $\Lambda = \diag(\lambda_1, \dots, \lambda_n)$, 
			$\ds V_{\lb} = \left(X^T \Lambda^{-1} X + Q \right)$, and 
			$\ds m_{\lb} = X^T \Lambda^{-1} (\yb - \mu \bm{1}_n) + Q \bb_0$, $\bm{1}_n$ being the $n$-component vector of all ones, while setting $\sigma \equiv 1$ and $\mu = 0$ if these parameters are not included in the model.
			
			\item Generate $\lambda_1, \dots \lambda_n$ independently from 
			\[
			\lambda_i \mid \mu, \bb, \sigma, \alpha, \bm y, X \sim \gig \left( a = \alpha^2, b = \alpha^2  + \frac{1}{\sigma^2} (y_i - \mu - \xb_i^T \bb)^2, p = \frac{1}{2} \right)
			\]
			for $i = 1, \dots, n$, while setting $\sigma \equiv 1$ and $\mu = 0$ if not included in the model. Efficient sampling from this $\gig$ distribution can be made by noticing $\gig(a, b, p) = \frac{1}{\gig(b, a, -p)}$ and a GIG distribution with $p = -1/2$  collapses into an ordinary inverse Gaussian distribution which permits computationally efficient random variate generation. 
			
			\item Generate the common scale parameter $\sigma$, if included in the model, from 
			\[
			\sigma^2 \mid \mu, \bb, \lambda_1, \dots, \lambda_n, \bm y, X \sim \invgamma\left(a = \tilde a, b = \tilde b\right)
			\]
			where $\ds \tilde a =  \frac{n + p}{2} + a
			_\sigma$ and  
			\[
			\ds \tilde b = \frac{1}{2}\left\{ (\yb - \mu \bm{1}_n - X\bb)^T \Lambda^{-1} (\yb - \mu \bm{1}_n - X\bb) + (\bb - \bb_0)^T Q (\bb - \bb_0)\right\} + b_\sigma.
			\]

			\item Generate MCMC samples for $\alpha^2$ from the conditional density
			\begin{equation} \label{eq:post-density-alpha2}
				p(\alpha^2 \mid \mu, \bb, \sigma, \yb, X) \propto \left(\frac{1}{\sqrt{\alpha^2} K_1( \sqrt{\alpha^2 (1 + \alpha^2)} }\right)^n  \exp \left( -\sqrt{1 + \alpha^2} \ \sum_{i=1}^n \sqrt{\alpha^2 + \tilde \varepsilon_i^2} \right)
			\end{equation}
			where $\ds \tilde \varepsilon_i = \frac{y_i - \mu - \xb_i^T \bb}{\sigma}$; $i = 1, \dots, n$ are scaled residuals, while setting $\mu = 0$ and $\sigma = 1$ if these parameters are not included in the model. The above density can be computed up to arbitrary precision leveraging numerical expansions for the Bessel functions but cannot be sampled efficiently. Instead, we suggest using a stepping-out slice sampler \citep{neal2003slice} for MCMC sampling from this univariate density.  
		\end{enumerate}
	\end{algorithm}

	\begin{remark}
		Algorithm~\ref{algo:mcmc-sampler-ridge} is designed for posterior sampling from the SPH regression model (generalized) posterior.  Straightforward modifications can be made to the algorithm, particularly in Steps 3, 4 and 5, for sampling $\lambda_i^2$, $\sigma$ and $\alpha^2$, respectively, to cater to the $\ell_2$ and $\ell_1$ regression problems.  Specifically, step 5 is skipped altogether in these cases. For $\ell_1$ regression, $\lambda_i$; $i = 1, \dots, n$ are generated independently in step 3 from
		\begin{equation*} 
			\lambda_i \mid \mu, \bb, \sigma, \alpha, \bm y, X \sim \gig \left( a = 2, b = \frac{1}{\sigma^2} (y_i - \xb_i^T \bb)^2, p = \frac{1}{2} \right).
		\end{equation*} 
		For $\ell_2$ regression, because there is only one common error variance parameter, steps 3 and 4 are merged. One generates a common $\lambda$ from 
		\[
		\lambda \mid \mu, \bb, \bm y, X \sim \invgamma\left(a = \tilde a, b = \tilde b^*\right)
		\]
		where $\ds \tilde a =  \frac{n + p}{2} + a
		_\sigma$ as in Algorithm~\ref{algo:mcmc-sampler-ridge}, and
		\[
		\ds \tilde b^* = \frac{1}{2}\left\{ (\yb - \mu \bm{1}_n - X\bb)^T (\yb - \mu \bm{1}_n - X\bb) + (\bb - \bb_0)^T Q (\bb - \bb_0)\right\} + b_\sigma.
		\]
		and afterward sets $\lambda_i = \lambda$ for all $i = 1, \dots, n$ and $\sigma = 1$ (i.e., $\sigma$ is not included in the model). 
		The remaining Steps 1 and 2 for generating $\mu$ (if included in the model) and $\bb$ remain unaltered. 
	\end{remark}

	\subsection{Posterior MCMC sampling for the spike-and-slab prior distribution }
	
	For the hierarchical spike-and-slab prior distribution, we note that conditional on $\bb$, the full conditional posterior densities for $\mu$, $\lb$, $\sigma$ and $\alpha$  remain the same as those provided in Algorithm\ref{algo:mcmc-sampler-ridge}. However, the full conditional distributions of $\bb$ and $\sigma^2$ have a different form, and in addition, there is a need to sample the predictor activation variables $\gb = (\gamma_1, \dots, \gamma_p)^T$ and $q$. For computational efficiency particularly in high dimension we propose generating $\bb$ and $\gb$ coordinate wise, with $(\beta_j, \gamma_j)$ sampled jointly from their full conditional density. Below we first derive these full conditional distributions.

	\medskip
		
		\textbf{Full conditional posterior distribution of $(\beta_j, \gamma_j)$ for each $j=1, \dots, p$.}  Due to the degenerate nature of the spike distribution, the joint full conditional distributions of the entire $(\bb, \bm \gamma)$ vector becomes intractable. Instead, we focus on the full conditional posterior distribution of each coordinate $(\beta_j, \gamma_j)$ for posterior Gibbs sampling.  Straightforward algebra shows that
		\[
		p(\bm y \mid \beta_j, \gamma_j = 0, \beta_{-j}, \gamma_{-j}, \mu, \sigma^2, \lambda_1, \dots, \lambda_n) \propto \mathcal{N}(y \mid X_{-j} \beta_{-j}, \sigma^2 \Lambda)
		\]
		and 
		\[
		p(\bm y \mid \beta_j, \gamma_j = 1, \bm\beta_{-j}, \gamma_{-j}, , \mu, \sigma^2, \lambda_1, \dots, \lambda_n) \propto \mathcal{N}\left( y \mid X_{-j} \bm\beta_{-j} + X_j \beta_j, \sigma^2 \Lambda \right)
		\]
		Therefore, the $\beta_j$ integrated (marginal) likelihood is:
		\[
		p(\bm y \mid \gamma_j = 0, \beta_{-j}, \gamma_{-j}, \mu, \sigma^2, \lambda_1, \dots, \lambda_n) = \mathcal{N}(y \mid X_{-j} \beta_{-j}, \sigma^2 \Lambda)
		\]
		and
		\[
		p(\bm y \mid \gamma_j = 1, \gamma_{-j}, \beta_{-j}, \sigma^2, \lambda_1, \dots, \lambda_n) = \mathcal{N}\left( y \mid X_{-j} \beta_{-j}, \sigma^2 \Lambda + \tau^2 X_j X_j^\top \right).
		\]
		Combining we get
		\[
		\mathrm{LR}_j = \frac{P(y \mid \gamma_j=1, \beta_{-j}, \gamma_{-j}, \rest)}{P(y \mid \gamma_j=0, \beta_{-j}, \gamma_{-j}, \rest)}
		= \frac{\mathcal{N}\Bigl(r_j \mid 0,\,\Sigma_1\Bigr)}
		{\mathcal{N}\Bigl(r_j \mid 0,\,\Sigma_0\Bigr)},
		\]
		where $r_j = y - X_{-j}\beta_{-j}$ denotes the partial residual and 
		\[
		\Sigma_0 = \sigma^2\Lambda,\quad \Sigma_1 = \sigma^2\Lambda + \tau^2 X_jX_j^\top.
		\]
		To simplify $LR_j$, we first employ the matrix determinant lemma, to get
		\[
		\frac{\det(\Sigma_0)}{\det(\Sigma_1)}
		= \Bigl(1 + \frac{\tau^2}{\sigma^2}t_j\Bigr)^{-1},
		\]
		with
		\[
		t_j = X_j^\top\Lambda^{-1}X_j.
		\]
		Next, using the Sherman--Morrison formula, we get
		\[
		r_j^\top\Bigl(\Sigma_1^{-1}-\Sigma_0^{-1}\Bigr)r_j
		= -\frac{\tau^2}{\sigma^2+\tau^2t_j}\,(s_j)^2,
		\]
		where
		\[
		s_j = X_j^\top\Lambda^{-1}r_j.
		\]
		Combining these two pieces and defining,
		\[
		\log \mathrm{LR}_j = -\tfrac{1}{2}\,\log\!\Bigl(1+\frac{\tau^2t_j}{\sigma^2}\Bigr)
		+ \tfrac{1}{2}\,\frac{(s_j\,\tau^2)^2}{\sigma^2(\sigma^2+\tau^2t_j)},
		\]
		the full conditional distribution of $(\beta_j, \gamma_j)$ for $j=1, \dots, p$ is obtained as:
		\begin{gather*}
			\gamma_j \mid \text{rest}\sim  \mathrm{Bernoulli}(p_j), \text{ with }
			p_j = \displaystyle \frac{q\,\exp\bigl(\log \mathrm{LR}_j\bigr)}{\,q\,\exp\bigl(\log \mathrm{LR}_j\bigr)+(1-q)}, \\
			\beta_j \mid \gamma_j, \text{rest} \sim
			\begin{cases}
				0, & \gamma_j = 0,\\[1mm]
				\mathcal{N}(\mu_j,\,V_j), & \gamma_j = 1,
			\end{cases}
		\end{gather*}
		where 
		\[
		V_j = \frac{\sigma^2}{\,t_j + \frac{\sigma^2}{\tau^2}\,},\quad
		\mu_j = \frac{V_j}{\sigma^2}\,s_j.
		\]
		
		\noindent
		Algorithm~\ref{algo:mcmc-sampler-ss} displays the steps involved in one iteration of a slice-within-Gibbs sampler for posterior sampling from the joint posterior of an SPH regression with a spike and slab prior.

		\begin{algorithm}[htpb]
			\caption{One iteration of a slice-within-Gibbs sampler for posterior sampling for the SPH regression model under the ridge prior}
			
			\label{algo:mcmc-sampler-ss}
			
			\begin{enumerate}
				\item Generate the intercept $\mu$ from its full conditional posterior distribution as provided in Step 1 of Algorithm~\ref{algo:mcmc-sampler-ridge}.
				
				\item Generate $\bb$ and $\gb$ coordinate wise,  with $(\beta_j, \gamma_j)$ sampled jointly from their full conditional density: \begin{gather*}
					\gamma_j \mid \mu, \bb_{-j}, \sigma, q, \bm y, X \sim  \mathrm{Bernoulli}(p_j), \text{ with }
					p_j = \displaystyle \frac{q\,\exp\bigl(\log \mathrm{LR}_j\bigr)}{\,q\,\exp\bigl(\log \mathrm{LR}_j\bigr)+(1-q)}, \\
					\beta_j \mid \gamma_j, \mu, \bb_{-j}, \sigma, q, \bm y, X \sim
					\begin{cases}
						0, & \gamma_j = 0,\\[1mm]
						\mathcal{N}(\mu_j,\,V_j), & \gamma_j = 1,
					\end{cases}
				\end{gather*}
				where 
				\begin{gather*}
					\log \mathrm{LR}_j = -\tfrac{1}{2}\,\log\!\Bigl(1+\frac{\tau^2t_j}{\sigma^2}\Bigr)
					+ \tfrac{1}{2}\,\frac{(s_j\,\tau^2)^2}{\sigma^2(\sigma^2+\tau^2t_j)} , \text{ and} \\
					V_j = \frac{\sigma^2}{\,t_j + \frac{\sigma^2}{\tau^2}\,},\quad
					\mu_j = \frac{V_j}{\sigma^2}\,s_j, \text{ with} \\
					r_j = y - \mu - X_{-j}\beta_{-j},  t_j = X_j^\top\Lambda^{-1}X_j, 
					s_j = X_j^\top\Lambda^{-1}r_j
				\end{gather*}
				
				\item Generate the common scale parameter $\sigma$ if included in the model from the conditional gamma posterior density
				\[
				\sigma^2 \mid \bb, \mu, \lambda_1, \dots, \lambda_n, \yb, X  \sim \invgamma(a = \tilde{a}, , \ b = \tilde{b})
				\]
				with $\ds \tilde a = a_\sigma + \frac{n + p_1}{2}$, 
				\[
				\tilde b =  b_\sigma + \frac{1}{2}\left\{ \left(\yb -\mu \bm{1}_n - X \bb\right)^T \Lambda  \left(\yb -\mu \bm{1}_n - X \bb\right) + \frac{1}{\tau^2} \sum_{j \in \Gamma_1} \beta_j^2 \right\}
				\]
				and $\Gamma_j = \{j:  \beta_j \neq 0\}$.

				\item Perform a stepping-out slice sampling to generate $\alpha^2$ from its conditional posterior density as provided in Step 5 of Algorithm~\ref{algo:mcmc-sampler-ss}.   
				
				\item   Generate $q$ from the following conditional distribution: 
				\[
				q \mid \gamma_1, \dots, \gamma_p \sim \bet(a_q + p_1, b_q + p - p_1)
				\]
				where $p_1 = \#\{j : \gamma_j = 1\}$.
				
			\end{enumerate}
		\end{algorithm}

		\subsection{Empirical assessment of the effect of `scaling' the pseudo-Huber loss} \label{sec:scalingeffect}
		
		As noted in the Introduction, a key novelty of the developed methodology is the proposed scaling of the pseudo-Huber loss, which ensures that the loss asymptotically becomes the \textit{exact} $\ell_1$ and $\ell_2$ losses as $\alpha \to 0$ and $\alpha \to \infty$, respectively. The unscaled pseudo-Huber loss, by contrast, does not converge to  $\ell_1$  when $\alpha \to 0$ and thus is not guaranteed to provide robust Bayesian inference in the presence of heavy contamination—precisely where the  $\ell_1$ loss is preferred over the  $\ell_2$  loss.
		
		To evaluate the impact of the lack of convergence of the unscaled pseudo-Huber loss on inference, we considered the first simulation experiment described in Section \ref{sec:diagnostics} with a 90\%-10\% mix of contaminated and non-contaminated observations in each of the three simulated datasets of sizes  $n=20$ (small), $n=50$ (medium), and $n=500$ (large). On each data set, we fitted two generalized Bayesian pseudo-Huber models: one with a scaled pseudo-Huber loss and one with an unscaled loss, using the proposed MCMC algorithm and its modification (analogous to \cite{}) to handle the unscaled loss, respectively. For comparison, we also fitted Bayesian $\ell_1$ and $\ell_2$ regression models using MCMC sampling. Each MCMC was run for 10,000 iterations after discarding the initial 10,000 iterations as burn-in.
		
		To further assess the contamination diagnostic method proposed in Section \ref{sec:diagnostics}, we obtained the scaled posterior standard deviations $\{\tilde{s}_i\}$ of $\{\lambda_i\}$ from each scaled pseudo-Huber fit and applied an empirical quantile-based outlier detection approach on these $\{\tilde{s}_i\}$ values using the default boxplot function in R. The observations $\{i\}$ corresponding to the identified outliers in $\{\tilde{s}_i\}$ were deemed contaminated and were subsequently discarded from the original training datasets. We then reran the Bayesian scaled pseudo-Huber model on these \textit{filtered} datasets using MCMC sampling.
		
\begin{figure}[!h]
			\centering
			\includegraphics[width=1.0\linewidth]{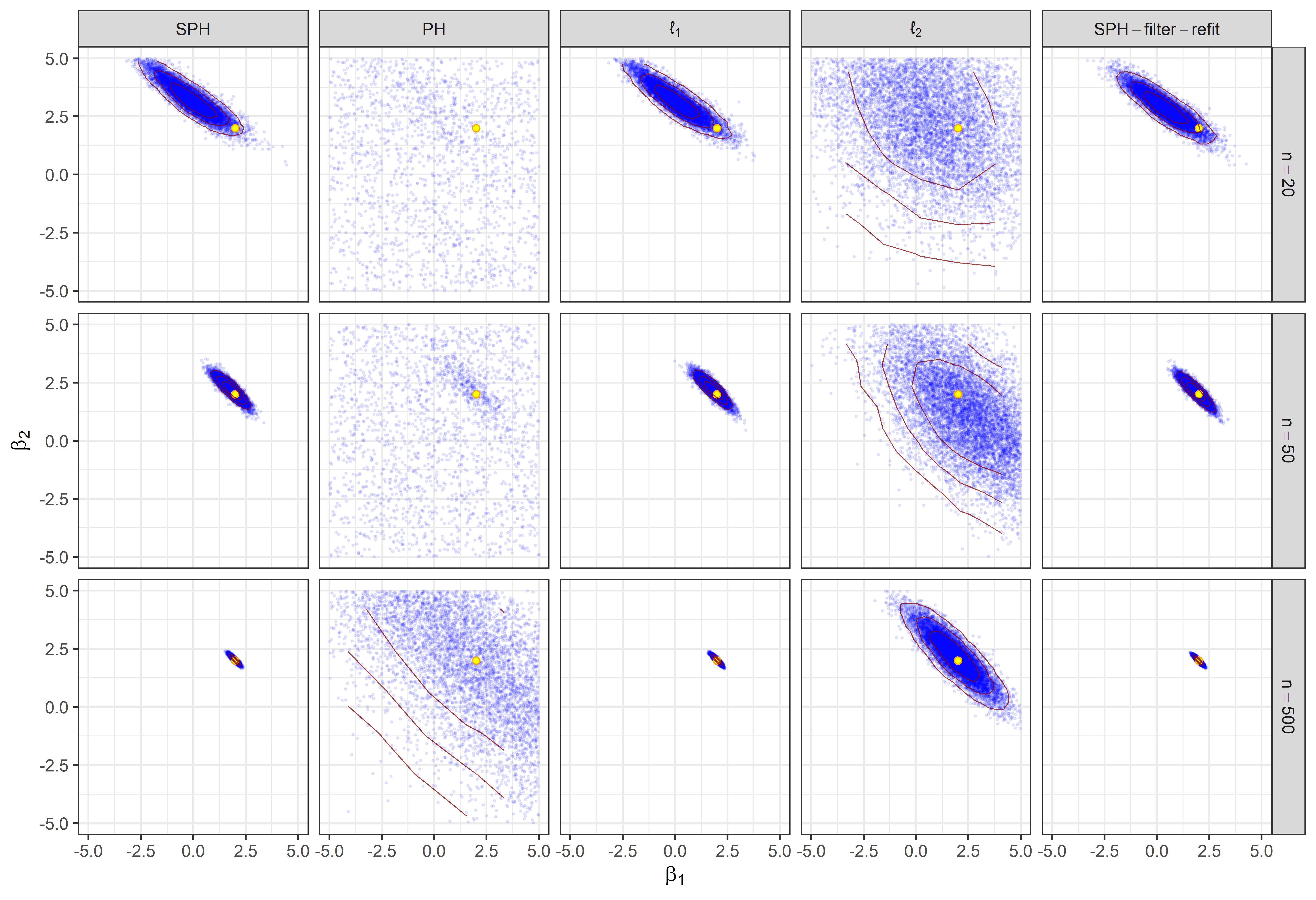}
			\caption{\small Visualizing the joint generalized posterior distributions of the first two coordinates  $(\beta_1, \beta_2)$  of $\bb$,   under different losses and a weakly informative Gaussian  prior belief distribution for  $\bb$, through point clouds and density contours. The contour lines represent the joint highest posterior density  sets for $(\beta_1, \beta_2)$  at  50\%,  80\%, 90\%, and 95\% probability levels.}
			\label{fig:contour-SPH-UPH-L1-L2}
		\end{figure}

		Posterior draws for $\bb = (\beta_1, \dots, \beta_5)^T$ are collected from each model fit on each dataset, and the first two coordinates of these draws were visualized as scatterplots. These scatterplots are displayed in Figure~\ref{fig:contour-SPH-UPH-L1-L2} with overlaid contour lines (red curves) showing the 50\%, 80\%, and 95\% highest posterior density regions for $(\beta_1, \beta_2)$ computed from the posterior MCMC draws. The figure also visualizes the corresponding true value $(2, 2)$ (yellow dot) of $(\beta_1, \beta_2)$.

		The following observations can be drawn from Figure~\ref{fig:contour-SPH-UPH-L1-L2}. First, the scaled pseudo-Huber Bayesian model exhibits strong estimation performance for $(\beta_1, \beta_2)$, its posterior closely aligning with the true values across all data set sizes (small, medium, and large). The posteriors concentrate well around the true values, with this concentration increasing as the sample size grows. Second, the posterior distributions for the scaled pseudo-Huber model closely resemble those of the $\ell_1$ regression model. This similarity is expected, given the substantial contamination in the generated data, which causes the scaled pseudo-Huber fit to behave similarly to the $\ell_1$ model fit. 
		Third, the unscaled pseudo-Huber fits exhibit a highly erratic pattern, with no clear posterior concentration around the true values. This stands in stark contrast to the well-behaved posteriors of the scaled pseudo-Huber model, highlighting the limitations of the unscaled pseudo-Huber model for inference in high-contamination settings. The $\ell_2$ model fits also display some instability, though to a lesser extent than the unscaled pseudo-Huber fits. Finally, applying contamination filtering (based on the diagnostic proposed in Section~\ref{sec:diagnostics}) followed by refitting leads to a modest but positive improvement in posterior accuracy, aligning the estimates more closely with the true values.

		\subsection{A brief note on outlier filtering} \label{sec:diagnostics}
		
		Note that outlier filtering proves particularly useful in the real data application. Next, we briefly elaborate on the filtering strategy, which utilizes information from the \textit{micro-level} parameters $\lambda_i$. These parameters, based on the hierarchical model structure, serve as the individualized Gaussian scale parameter for the response $y_i$, with independent Generalized Inverse Gaussian (GIG) priors assigned to $\lambda_i$. The posterior distribution is influenced by both the working model and the data. For contaminated observations with heavy-tailed errors, the marginal posterior distribution of the corresponding $\lambda_i$ tends to exhibit greater variability.  To identify contaminated observations. we propose examining the marginal upper posterior 95\% percentile points $s_i$ with $P(\lambda_i \leq s_i \mid \text{data}) = 0.95$ for all observations $i = 1, \dots, n$. These percentile points are computed using posterior MCMC draws for $\lambda_i$'s. Observations with notably large $s_i$ can be flagged using standard outlier detection methods, such as Tukeys' boxplot method (or its variants), which uses 3rd quartile + 1.5 interquartile range (computed on $s_1, \dots, s_n$) as a threshold  \citep{chambers2018graphical, mcgill1978variations}. These methods are implemented in common boxplot computation routines, including the default boxplot method in \texttt{R}, which we use in our computations. This heuristic is expected to perform reasonably well in applications with low-to-moderate contamination levels. 
		
		To visualize the heuristic's performance as the contamination proportion varies, we conduct two simulation experiments. In each experiment, data are generated from a linear regression model with $p=5$ predictors and regression coefficients $\bb = (2, 2, 0, 0, 0)^T$. The predictors are generated from autoregressive AR(1) processes with a standard normal base distribution and a serial correlation coefficient of 0.4. The first two predictors, corresponding to the non-zero regression coefficients, are highly correlated with a coefficient of 0.9, while the remaining three predictors are independent of each other and the first two.
		
		In the first simulation, errors are generated from a 90\%-10\% mixture of (a) an AR(1) process with a standard normal base distribution and a serial correlation of 0.2 (the uncontaminated distribution) and (b) an independent Cauchy(0, 5) distribution (the contaminant distribution). Contaminated observations are labeled. In the second experiment, a 50\%-50\% mixture of the same distributions is used to generate the errors. In each setting, data sets with $n=20$ (small), $n=50$ (medium), and $n=500$ (large) are generated and fitted the proposed SPH regression with a ridge prior on the regression coefficients using the proposed MCMC sampler (10,000 final draws after discarding the initial 10,000 draws as burn-in).
		
\begin{figure}[!h]
			\centering
			\includegraphics[width=1\linewidth]{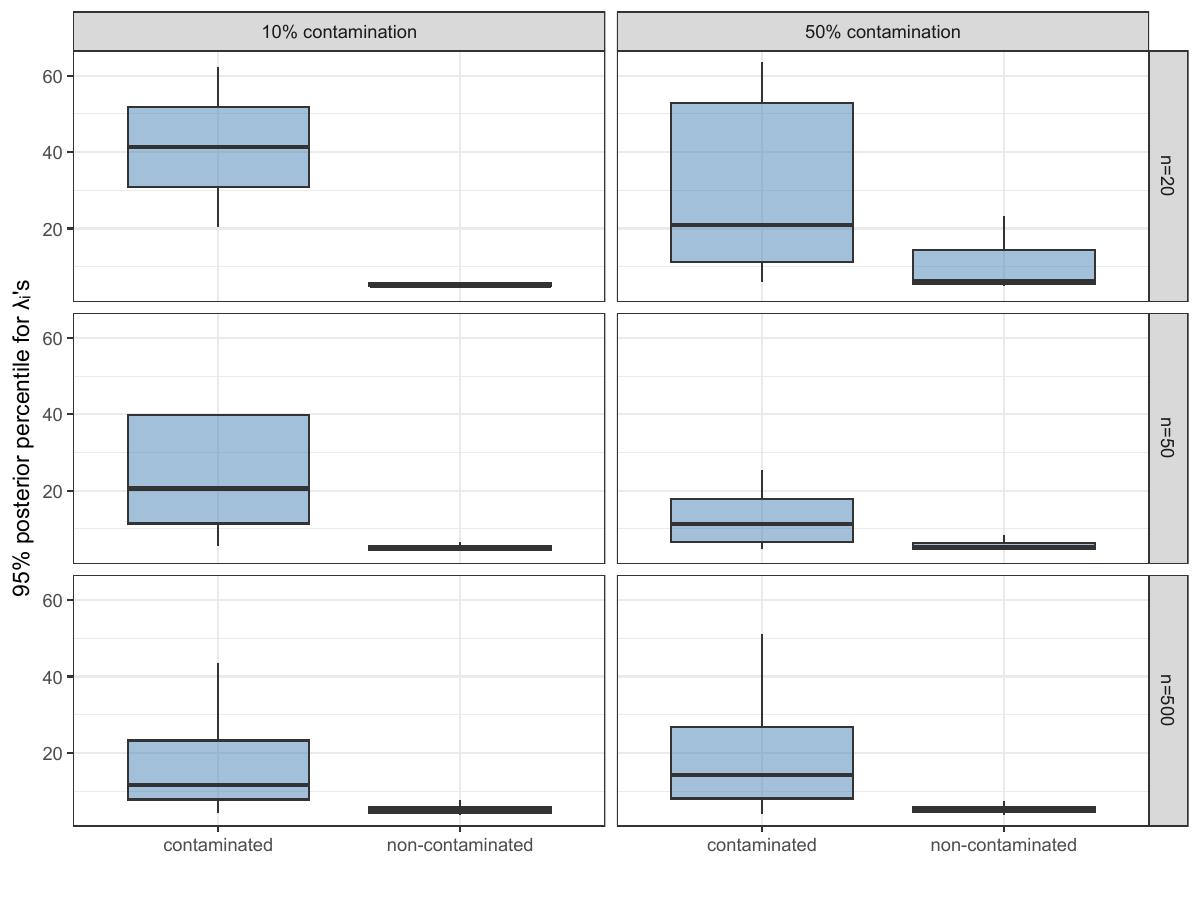}
			\caption{\footnotesize Boxplots for upper 95\% percentile points for $\{\lambda_i: i=1, \dots, n\}$ (vertical axis) plotted separately for contaminated and non-contaminated observations (horizontal axis). Panels reflect sample sizes (rows) and contamination proportions (columns) underlying the data-generating setups. Extremely large contaminated values deemed as outliers in their respective boxplots are removed.}
			\label{fig:contour-SPH-UPH-L1-L2-sd}
		\end{figure}

		We then obtain the marginal posterior upper 95\% percentiles $\{s_1, \dots, s_n\}$ of the micro-level contamination parameter $\{\lambda_i: i=1, \dots, n\}$. These values are depicted as boxplots in Figure \ref{fig:contour-SPH-UPH-L1-L2-sd}, with separate boxplots for the contaminated and non-contaminated observations (true labels). As illustrated, there is a clear distinction between $s_i$ values for contaminated and non-contaminated observations across all data sizes and contamination proportions. The $\{{s}_i\}$ values for contaminated observations are notably higher. Thus, a standard empirical threshold-based outlier detection method (e.g., Tukey's boxplot method) applied to these $\{{s}_i\}$ values is expected to identify the ``true'' contaminated observations with reasonable precision.
		
		\section{Proofs of Posterior Consistency Results in Section \ref{sec:theory}} \label{sec:proofs-consistency}

		\subsection{Proof of Theorem \ref{thm:postmean}} 
		
		Suppose we have $\inf_{{\bf u}: \|{\bf u}\| = 1} Q_\alpha (\bb_0 + \delta_n \bu) > Q_\alpha (\bb_0)$. It would then imply that $Q_{\alpha}$ has a local minimum in the set $\{\bb: \|\bb - \bb_0\| \leq \delta_n\}$. Since $Q_\alpha$ is a strictly convex function, this would imply that $\|\hat{\bb}_{pm} - \bb_0\| \leq \delta_n$. Hence, to establish the result, it is enough to show that 
		$$
		P_0 \left( \inf_{{\bf u}: \|{\bf u}\| = 1} Q_\alpha (\bb_0 + \delta_n \bu) > Q_\alpha (\bb_0) \right) \rightarrow 1 
		$$
		
		\noindent
		as $n \rightarrow \infty$. 
		
		With this goal in mind, we arbitrarily fix $\bu$ such that $\|\bu\| = 1$. 
		Using the 
		second order Taylor expansion of $f_\bu (t) = Q_\alpha (\bb_0 + t\delta_n 
		\bu)$ around $t=0$, we get 
		\begin{eqnarray}
			Q_\alpha (\bb_0 + \delta_n \bu) - Q_\alpha (\bb_0) 
			&=& f_\bu (1) - f_\bu (0) \nonumber\\
			&=& \frac{\delta_n}{n \alpha} \sum_{i=1}^n \ell_{SPH,\alpha}' (\epsilon_i) 
			\xb_i^T \bu + \frac{\delta_n^2}{2 n \alpha} \sum_{i=1}^n \ell_{SPH,\alpha}'' (\epsilon_i - t^* \delta_n^2 \xb_i^T \bu) (\xb_i^T \bu)^2 + 
			\nonumber\\
			& & \frac{\tau^2 \delta_n^2}{n \alpha} \bu^T \bu + 
			\frac{2 \tau^2 \delta_n}{n \alpha} \bu^T \bb_0. \label{taylor}
		\end{eqnarray}
		
		\noindent
		where $t^* \in (0,1)$. Since $(\epsilon_i - t^* \delta_n \xb_i^T \bu)^2 
		\leq 2 \epsilon_i^2 + 2 \delta_n (\xb_i^T \bu)^2$, and $\ell_{SPH,\alpha}''(y) = \sqrt{1+\alpha^{-2}}(1 + \alpha^{-2} y^2)^{-3/2}$, it follows that 
		\begin{eqnarray}
			Q_\alpha (\bb_0 + \delta_n \bu) - Q_\alpha (\bb_0) 
			&\geq& \frac{\delta_n}{n \alpha} \sum_{i=1}^n \ell_{SPH,\alpha}' (\epsilon_i) 
			\xb_i^T \bu - \frac{2 \tau^2 \delta_n}{n \alpha} \|\bu\| \|\bb_0\| + \nonumber\\
			& & \frac{\delta_n^2\sqrt{1+\alpha^{-2}}}{2 n \alpha} \sum_{i=1}^n \left( 1 + 2\alpha^{-2} \epsilon_i^2 + 2 \delta_n^2 \alpha^{-2} (\xb_i^T \bu)^2 
			\right)^{-3/2} (\xb_i^T \bu)^2. \nonumber\\
			& & \label{taylor1}
		\end{eqnarray}
		
		\noindent
		Since $2 \alpha^{-2} < 1$ and $2 \alpha^{-2} \delta_n^2 < 1$ for large 
		enough $n$ (by Assumption A1), we have 
		\begin{eqnarray}
			& & \inf_{{\bf u}: \|{\bf u}\| = 1} (Q_\alpha (\bb_0 + \delta_n \bu) - 
			Q_\alpha (\bb_0)) \nonumber\\
			&\geq& \inf_{{\bf u}: \|{\bf u}\| = 1} \frac{\delta_n^2 \sqrt{1+\alpha^{-2}}}{2 n \alpha} 
			\sum_{i=1}^n \left( 1 + \epsilon_i^2 + (\xb_i^T \bu)^2 \right)^{-3/2} (\xb_i^T \bu)^2 - \sup_{{\bf u}: \|{\bf u}\| = 1} 
			\left| \frac{\delta_n}{n \alpha} \sum_{i=1}^n \ell_{SPH,\alpha}' (\epsilon_i) 
			\xb_i^T \bu \right| - \nonumber\\
			& & \frac{2 \tau^2 \delta_n}{n \alpha} \|\bb_0\|. \label{diff}
		\end{eqnarray}
		
		\noindent
		Next, we focus on the second term on the RHS in (\ref{diff}). Let $K_1 > 
		0$ be arbitrarily fixed. Since $\{\xb_i\}_{i=1}^n$ and ${\boldsymbol 
			\epsilon} = \{\epsilon_i\}_{i=1}^n$ are independent, it follows that 
		for any $\bu$ with $\|\bu\| \leq 1$ 
		\begin{eqnarray}
			P_0 \left( \frac{1}{n \alpha} \sum_{i=1}^n \ell_{SPH,\alpha}' (\epsilon_i) 
			\xb_i^T \bu > K_1 \sqrt{\frac{p}{n}} \right) 
			&=& E_0 \left[ P_0 \left( \frac{s}{n \alpha} \sum_{i=1}^n \ell_{SPH,\alpha}' 
			(\epsilon_i) \xb_i^T \bu > K_1 \sqrt{\frac{p}{n}} \mid {\boldsymbol 
				\epsilon} \right) \right] \nonumber\\
			&\leq& E_0 \left[ E_0 \left[ \exp \left( \frac{s}{n \alpha} \sum_{i=1}^n 
			\ell_{SPH,\alpha}' (\epsilon_i) \xb_i^T \bu - s K_1 \sqrt{\frac{p}{n}} \right) 
			\mid {\boldsymbol \epsilon} \right] \right]. \label{exponent}
		\end{eqnarray}
		
		\noindent
		By Assumption A2 and $|\ell_{SPH,\alpha}' (\epsilon_i)| \leq \alpha \sqrt{1+\alpha^{-2}}$, it 
		follows that conditional on ${\boldsymbol \epsilon}$, the random variable 
		$\alpha^{-1} \sum_{i=1}^n \ell_{SPH,\alpha}' (\epsilon_i) \xb_i^T \bu$ has 
		a Gaussian distribution with mean zero and variance $v_n$, where 
		\begin{eqnarray*}
			v_n 
			&=& \sum_{i=1}^n \frac{\ell_{SPH,\alpha}'^2 (\epsilon_i)}{\alpha^2} 
			{\bf u}^T \Gamma_n (0) {\bf u} + \sum_{1 \leq i \neq j \leq n} \frac{\ell_{SPH,\alpha}' (\epsilon_i) \ell_{SPH,\alpha}' (\epsilon_j)}{\alpha^2} 
			{\bf u}^T \Gamma_n (j-i) {\bf u}\\
			&\leq& (1+\alpha^{-2}) \left( n {\bf u}^T \Gamma_n (0) {\bf u} + \sum_{k=1}^{n-1} (n-k) \left| {\bf u}^T \left( \Gamma_n (k) + \Gamma_n (-k) \right) {\bf u} \right| \right)\\
			&\leq& (1+\alpha^{-2}) \left( n \|\Gamma_n (0)\|_2 + \sum_{k=1}^{n-1} (n-k) \| \Gamma_n (k) + \Gamma_n (-k) \|_2 \right)\\
			&\leq& (1+\alpha^{-2}) (2n \sum_{k=0}^{n-1} \| \Gamma_n (k) \|_2)\\
			&\leq& 2 (1+\alpha^{-2}) \kappa_2 n. 
		\end{eqnarray*}
		
		\noindent
		The last two inequalities follow from Assumption A2 and the fact that $\Gamma_n (-k) = \Gamma_n (k)^T$. It follows by (\ref{exponent}) 
		that 
		\begin{eqnarray*}
			P_0 \left( \frac{1}{n \alpha \sqrt{1+\alpha^{-2}}} \sum_{i=1}^n \ell_{SPH,\alpha}' (\epsilon_i) 
			\xb_i^T \bu > K_1 \sqrt{\frac{p}{n}} \right) 
			&=& E_0 \left[ \exp \left( \frac{2 \kappa_2 n s^2}{2 n^2} 
			- s K_1 \sqrt{\frac{p}{n}} \right) \right]\\
			&=& E_0 \left[ \exp \left( \frac{\kappa_2 s^2}{n} 
			- s K_1 \sqrt{\frac{p}{n}} \right) \right]
		\end{eqnarray*}
		
		\noindent
		for every $s > 0$. Choosing $s = K_1 \sqrt{np}/(2 \kappa_2)$, we get 
		$$
		P_0 \left( \frac{1}{n \alpha \sqrt{1+\alpha^{-2}}} \sum_{i=1}^n \ell_{SPH,\alpha}' (\epsilon_i) 
		\xb_i^T \bu > K_1 \sqrt{\frac{p}{n}} \right) \leq \exp \left( - \frac{K_1^2 
			p}{4 \kappa_2} \right) 
		$$
		
		\noindent
		for every $\bu$ such that $\|\bu\| \leq 1$. Since $\xb_i^T \bu$ has a symmetric distribution around $0$, it follows that 
		$$
		P_0 \left( -\frac{1}{n \alpha \sqrt{1+\alpha^{-2}}} \sum_{i=1}^n \ell_{SPH,\alpha}' (\epsilon_i) 
		\xb_i^T \bu > K_1 \sqrt{\frac{p}{n}} \right) \leq \exp \left( - \frac{K_1^2 
			p}{4 \kappa_2} \right) 
		$$
		
		\noindent
		for every $\bu$ such that $\|\bu\| \leq 1$, which implies 
		\begin{equation} \label{mibd}
			P_0 \left( \left| \frac{1}{n \alpha \sqrt{1+\alpha^{-2}}} \sum_{i=1}^n \ell_{SPH,\alpha}' (\epsilon_i) 
			\xb_i^T \bu \right| > K_1 \sqrt{\frac{p}{n}} \right) \leq \exp \left( - 
			\frac{K_1^2 p}{4 \kappa_2} \right) 
		\end{equation}
		
		\noindent
		for every $\bu$ such that $\|\bu\| \leq 1$. To get a bound on the supremum over all appropriate $\bu$, we employ a technique similar to \cite{Vershynin:2011}. By \cite[Lemma 5.2]{Vershynin:2011}, there exists a set $S_{10}$ with the property that $S_{10} \subseteq \{\bu: \; \|\bu\| \leq 1\}$, $|S_{10}| \leq 21^p$, and for any $\bu$ with $\|\bu\| \leq 1$, there exists ${\bf w}_{(\bu)} \in S_{10}$ such that $\|\bu - {\bf w}_{(\bu)}\| \leq 0.1$. Now, for any $\bu$ with $\|\bu\| \leq 1$, we have 
		\begin{eqnarray*}
			& & \left| \frac{1}{n \alpha} \sum_{i=1}^n \ell_{SPH,\alpha}' (\epsilon_i) 
			\xb_i^T \bu \right|\\
			&\leq& \left| \frac{1}{n \alpha} \sum_{i=1}^n \ell_{SPH,\alpha}' (\epsilon_i) 
			\xb_i^T (\bu - {\bf w}_{(\bu)}) \right| + \left| \frac{1}{n \alpha} 
			\sum_{i=1}^n \ell_{SPH,\alpha}' (\epsilon_i) \xb_i^T {\bf w}_{(\bu)} \right|\\
			&\leq& 0.1 \left| \frac{1}{n \alpha} \sum_{i=1}^n \ell_{SPH,\alpha}' (\epsilon_i) 
			\xb_i^T (10(\bu - {\bf w}_{(\bu)})) \right| + \max_{{\bf w} \in S_{10}} \left| \frac{1}{n \alpha} 
			\sum_{i=1}^n \ell_{SPH,\alpha}' (\epsilon_i) \xb_i^T {\bf w} \right|\\
			&\leq& 0.1 \sup_{{\bf u}: \|{\bf u}\| \leq 1} \left| \frac{1}{n \alpha} \sum_{i=1}^n \ell_{SPH,\alpha}' (\epsilon_i) 
			\xb_i^T \bu \right| + \max_{{\bf w} \in S_{10}} \left| \frac{1}{n \alpha} 
			\sum_{i=1}^n \ell_{SPH,\alpha}' (\epsilon_i) \xb_i^T {\bf w} \right|. 
		\end{eqnarray*}
		
		\noindent
		It follows that 
		$$
		\sup_{{\bf u}: \|{\bf u}\| \leq 1} \left| \frac{1}{n \alpha} \sum_{i=1}^n 
		\ell_{SPH,\alpha}' (\epsilon_i) \xb_i^T \bu \right| \leq \frac{10}{9} 
		\max_{{\bf w} \in S_{10}} \left| \frac{1}{n \alpha} \sum_{i=1}^n 
		\ell_{SPH,\alpha}' (\epsilon_i) \xb_i^T {\bf w} \right|. 
		$$
		
		\noindent
		Using this inequality along with the union-sum inequality, and noting that 
		(\ref{mibd}) holds for an arbitrary $K_1 > 0$, we obtain 
		\begin{eqnarray}
			& & P_0 \left( \sup_{{\bf u}: \|{\bf u}\| = 1} \left| \frac{1}{n \alpha \sqrt{1+\alpha^{-2}}} \sum_{i=1}^n \ell_{SPH,\alpha}' (\epsilon_i) \xb_i^T \bu \right| > K_1 \sqrt{\frac{p}{n}} \right) \nonumber\\
			&\leq& P_0 \left( \max_{{\bf w} \in S_{10}} \left| \frac{1}{n \alpha \sqrt{1+\alpha^{-2}}} \sum_{i=1}^n \ell_{SPH,\alpha}' (\epsilon_i) \xb_i^T {\bf w} \right| > \frac{9 K_1}{10} \sqrt{\frac{p}{n}} \right) \nonumber\\
			&\leq& 21^p \exp \left( -\frac{81K_1^2 p}{400 \kappa_2} \right) \nonumber\\
			&=& \exp \left( - \left\{ \frac{81K_1^2}{400 \kappa_2} - \log 21 \right\} p \right) \rightarrow 0 \mbox{ as } n \rightarrow \infty 
			\label{lterm}
		\end{eqnarray}
		
		\noindent
		if $K_1$ is chosen to be $\frac{40 \sqrt{\kappa_2 \log 21}}{9}$. Next, we focus our attention on the first term in (\ref{diff}). Again, fix $\bu$ with $\|\bu\| = 1$ arbitrarily. Define the random variables 
		$$
		Z_i (\bu) := \left( 1 + \epsilon_i^2 + \frac{(\xb_i^T \bu)^2}{\kappa_1 \bu^T \Gamma_n (0) \bu} \right)^{-3/2} 
		\frac{(\xb_i^T \bu)^2}{\bu^T \Gamma_n (0) \bu} \; \; \forall \ 1 \leq i \leq n. 
		$$
		
		\noindent
		It follows by Assumptions A2 and A3 that $\{Z_{i} (\bu)\}_{i=1}^n$ 
		are i.i.d. random variables and are uniformly bounded by $\kappa_1$. 
		Note that $G (\bu) := \xb_1^T \bu/\sqrt{\bu^T \Gamma_n (0) \bu}$ has 
		a standard normal distribution and is independent of $\epsilon_1$. 
		Hence 
		\begin{eqnarray*}
			E_0 [Z_1 (\bu)] 
			&=& E_0 \left[ \left( 1 + \epsilon_1^2 + (1/\kappa_1) 
			G(\bu)^2 \right)^{-3/2} G(\bu)^2\right]\\
			&:=& M_1. 
		\end{eqnarray*}
		
		\noindent
		Based on the arguments above, it follows that $M_1$ is a strictly 
		positive constant which does not depend on $\bu$ and $n$. Also, by 
		the definition of the function $g$ in Assumption A3, it follows that 
		$g(\epsilon_i) = E[Z_i (\bu) \mid {\boldsymbol \epsilon}]$ (and 
		$E_0 [Z_i (\bu)] = E[g(\epsilon_i)]$ by tower property). Note that 
		\begin{eqnarray}
			& & P_0 \left( \left| \frac{1}{n} \sum_{i=1}^n Z_i (\bu) - E_0 [Z_1 (\bu)] \right| > \frac{M_1}{2} \right) \nonumber\\
			&\leq& P_0 \left( \left| \sum_{i=1}^n Z_i (\bu) - \sum_{i=1}^n g(\epsilon_i) \right| > \frac{nM_1}{4} \right) + P_0 \left( \left| \sum_{i=1}^n g(\epsilon_i) - n E_0 [Z_1 (\bu)] \right| > \frac{nM_1}{4} \right) \nonumber\\
			&=& P_0 \left( \left| \sum_{i=1}^n Z_i (\bu) - \sum_{i=1}^n E[Z_i (\bu) \mid {\boldsymbol \epsilon}] \right| > \frac{nM_1}{4} \right) + P_0 \left( \left| \sum_{i=1}^n g(\epsilon_i) - n E_0 [g(\epsilon_1)] \right| > \frac{nM_1}{4} \right) \nonumber\\
			&=& E_0 \left[ P_0 \left( \left| \sum_{i=1}^n Z_i (\bu) - \sum_{i=1}^n E[Z_i (\bu) \mid {\boldsymbol \epsilon}] \right| > \frac{nM_1}{4} \mid {\boldsymbol \epsilon} \right) \right] + P_0 \left( \left| \sum_{i=1}^n g(\epsilon_i) - n E_0 [g(\epsilon_1)] \right| > \frac{nM_1}{4} \right). \label{union}
		\end{eqnarray}
		
		\noindent
		We first derive an upper bound for $V(\sum_{i=1}^n Z_i (\bu) \mid 
		{\boldsymbol \epsilon})$. Note that $Z_i (\bu)$ is a uniformly bounded function of $\xb_i^T \bu$ (which has a normal distribution, even if we condition on ${\boldsymbol \epsilon}$). Using the fact that the maximal correlation between two normal random variables $Z_1$ and $Z_2$ is given by $|Corr(Z_1, Z_2)|$ (see for example 
		\cite{Lancaster:1957}), based on the stationarity of the predictor process, and Assumption A2, we obtain 
		\begin{eqnarray*}
			Cov(Z_i (\bu), Z_j (\bu) \mid {\boldsymbol \epsilon}) 
			&\leq& 4 \kappa_1^2 |Corr(\xb_i^T \bu, \xb_j^T \bu \mid {\boldsymbol \epsilon})|\\
			&\leq& \frac{4 \kappa_1^2 |\bu^T \Gamma_n (i-j) \bu|}{\bu^T \Gamma_n (0) \bu}\\
			&\leq& 4 \kappa_1 |\bu^T \Gamma_n (i-j) \bu|\\
			&\leq& 2 \kappa_1 \|\Gamma_n (i-j) + \Gamma_n (i-j)\|. 
		\end{eqnarray*}
		
		\noindent
		Let ${\bf 1}_n$ denote the vector of all ones in $\mathbb{R}^n$. It 
		follows by Assumption A2 that 
		\begin{eqnarray*}
			V(\sum_{i=1}^n Z_i (\bu) \mid {\boldsymbol \epsilon}) 
			&=& \sum_{i=1}^n V(Z_i (\bu) \mid {\boldsymbol \epsilon}) + 
			\sum_{1 \leq i \neq j \leq n} Cov(Z_i (\bu), Z_j (\bu) \mid 
			{\boldsymbol \epsilon})\\
			&\leq& 4 n \kappa_1 \sum_{h=0}^{n-1} \left\| \Gamma_n (h) \right\|_2\\ 
			&=& 4n \kappa_1 \kappa_2. 
		\end{eqnarray*}
		
		\noindent
		Using the independence of the predictors and the errors, along with 
		Bernstein's concentration inequality for bounded random variables, we 
		obtain 
		\begin{eqnarray}
			& & E_0 \left[ P_0 \left( \left| \sum_{i=1}^n Z_i (\bu) - \sum_{i=1}^n E[Z_i (\bu) \mid {\boldsymbol \epsilon}] \right| > \frac{nM_1}{4} \mid {\boldsymbol \epsilon} \right) \right] \nonumber\\
			&\leq& E_0 \left[ \exp \left( -\frac{\frac{1}{32} n^2 M_1^2}{V(\sum_{i=1}^n Z_{i} (\bu) \mid {\boldsymbol \epsilon})  + \frac{n \kappa_1}{6} M_1} \right) \right] \nonumber\\
			&\leq& \exp \left( -\frac{\frac{1}{32} n^2 M_1^2}{4 n \kappa_1^2 + 16 n \kappa_1 \kappa_2 + \frac{n \kappa_1 M_1}{6}} \right) \nonumber\\
			&=:& \exp \left(-n M_2 \right), \nonumber\\
			& & \label{cond}
		\end{eqnarray}
		
		\noindent
		where $M_2 = \frac{3 M_1^2}{384 \kappa_1^2 + 1536 \kappa_1 \kappa_2 + 16 \kappa_1 M_1}$. We now focus on the second term in (\ref{union}). Note that by second order stationarity of the error sequence 
		$$
		\frac{1}{n} V(\sum_{i=1}^n g(\epsilon_i)) = V(g(\epsilon_1))+ 
		\sum_{i=2}^n \left( 1 - \frac{i}{n} \right) Cov(g(\epsilon_1), 
		g(\epsilon_i)) \leq V(g(\epsilon_1))+ \sum_{i=2}^n 
		|Cov(g(\epsilon_1), g(\epsilon_i))| \leq K_\epsilon. 
		$$
		
		\noindent
		Leveraging the uniform boundedness of $g$, Bernstein's concentration inequality and Assumption A3, we obtain
		\begin{eqnarray}
			P_0 \left( \left| \sum_{i=1}^n g(\epsilon_i) - n E_0 [g(\epsilon_1)] \right| > \frac{nM_1}{4} \right) 
			&\leq& \exp \left( -\frac{\frac{1}{32} n^2 M_1^2}{V(\sum_{i=1}^n g(\epsilon_i)) + \frac{n \kappa_1}{6} M_1} \right) \nonumber\\
			&\leq& \exp \left( -\frac{\frac{1}{32} n^2 M_1^2}{n K_\epsilon + \frac{n \kappa_1}{6} M_1} \right) =: \exp \left(-n M_3 \right), \label{vavg}
		\end{eqnarray}
		
		\noindent
		where $M_3 = \frac{3 M_1^2}{96 K_\epsilon + 16 \kappa_1 M_1}$. Since 
		$$
		Z_i (u) \leq \left( 1 + \epsilon_i^2 + (\xb_i^T \bu)^2 \right)^{-3/2} \frac{(\xb_i^T 
			\bu)^2}{\kappa_1}, 
		$$
		
		\noindent
		it follows by (\ref{union}), (\ref{cond}) and (\ref{vavg}) that  
		\begin{eqnarray}
			P_0 \left( \frac{1}{n} \sum_{i=1}^n \left( 1 + \epsilon_i^2 + (\xb_i^T 
			\bu)^2 \right)^{-3/2} (\xb_i^T \bu)^2 < \frac{\kappa_1 M_1}{2} \right)
			&\leq& P_0 \left( \frac{1}{n} \sum_{i=1}^n Z_i (\bu) < \frac{M_1}{2} \right) 
			\nonumber\\
			&=& P_0 \left( \frac{1}{n} \sum_{i=1}^n Z_i (\bu) < 
			E_0 [Z_1 (\bu)] - \frac{M_1}{2} \right) \nonumber\\
			&\leq& 2 \exp \left( -\min(M_2, M_3) n \right). \label{bound}
		\end{eqnarray}
		
		\noindent
		We now use another covering argument to get a bound on the infimum over all 
		appropriate $\bu$. By \cite[Lemma 5.2]{Vershynin:2011}, there exists a set $S_{1/p}$ with the property that $S_{1/p} \subseteq \{\bu: \; \|\bu\| \leq 1\}$, $|S_{1/p}| \leq (2p+1)^p$, and for any $\bu$ with $\|\bu\| = 1$, there exists ${\bf w}_{(\bu)} \in S_{1/p}$ such that $\|\bu - {\bf w}_{(\bu)}\| \leq p^{-1}$. We define $\tilde{\bf w}_{(\bu)} = (1/\|{\bf w}_{(\bu)}\|) {\bf w}_{(\bu)}$ so that $\|\tilde{\bf w}_{(\bu)}\| = 1$. 
		Since 
		$$
		\left| 1 - \|{\bf w}_{(\bu)}\| \right| = \left| \|\bu\| - \|{\bf w}_{(\bu)}\| \right| \leq \|\bu - {\bf w}_{(\bu)}\| \leq \frac{1}{p}, 
		$$
		
		\noindent
		we get 
		$$
		\|\bu - \tilde{\bf w}_{(\bu)}\| \leq \|\bu - {\bf w}_{(\bu)}\| + \left| 
		\frac{1}{\|{\bf w}_{(\bu)}\|} - 1 \right| \|{\bf w}_{(\bu)}\| \leq 
		\frac{2}{p}. 
		$$
		
		\noindent
		We denote the collection of all possible $\tilde{\bf w}_{(\bu)}$ (as $\bu$ 
		varies over $\{\bu: \; \|\bu\| = 1\}$) by $\tilde{S}_{1/p}$. It follows 
		that $|\tilde{S}_{1/p}| \leq (2p+1)^p$. Now, for any $a > 0$, consider 
		the function 
		$$
		g_a (x) = \frac{x^2}{(1 + a + x^2)^{-3/2}}. 
		$$
		
		\noindent
		Simple calculations show that 
		$$
		|g_a '(x)| \leq \left| \frac{2x}{(1 + a + x^2)^{-3/2}} \right| + \left| 
		\frac{3x^3}{(1 + a + x^2)^{-5/2}} \right| \leq 5. 
		$$
		
		\noindent
		Hence for every $\bu$, we have 
		\begin{eqnarray}
			\left\| \frac{\partial}{\partial \bu} \left( \frac{1}{n} \sum_{i=1}^n 
			\left( 1 + \epsilon_i^2 + (\xb_i^T \bu)^2 \right)^{-3/2} (\xb_i^T \bu)^2 
			\right) \right\| 
			&=& \left\| \frac{\partial}{\partial \bu} \left( \frac{1}{n} 
			\sum_{i=1}^n g_{\epsilon_i^2} (\xb_i^T \bu) \right) \right\| \nonumber\\
			&=& \left\| \frac{1}{n} \sum_{i=1}^n g_{\epsilon_i^2}' (\xb_i^T \bu) \xb_i 
			\right\| \nonumber\\
			&\leq& \frac{1}{n} \sum_{i=1}^n \left| g_{\epsilon_i^2} (\xb_i^T \bu) 
			\right| \|\xb_i\| \nonumber\\
			&\leq& \frac{5}{n} \sum_{i=1}^n \|\xb_i\| \nonumber\\
			&\leq& 5 \sqrt{\frac{1}{n} \sum_{i=1}^n \xb_i^T \xb_i} 
			\label{derivative}
		\end{eqnarray}
		
		\noindent
		The last inequality follows by Jensen's inequality, using the 
		concavity of the square-root function. Let $\xb \in \mathbb{R}^{np}$ 
		be the vector obtained by stacking $\xb_1, \xb_2, \cdots, \xb_n$ on 
		top of each other. Let $\Theta_n$ denote the $n \times n$ block partitioned matrix whose $(i,j)^{th}$ block is given by $\Gamma_n (i-j)$ for $1 \leq i,j \leq n$. Then, by Assumption A2, $\xb$ has a multivariate 
		normal distribution with mean ${\bf 0}$ and covariance matrix 
		$\Theta_n$. 
		
		Next, we bound the largest 
		eigenvalue of $\Theta_n$ in terms of $\kappa_2$, which leverages the proof of Theorem 2.3 in 
		\cite{Basu:Michailidis:2015}, but is presented here for completeness. 
		Consider the function 
		$$
		f_n (\theta) = \frac{1}{2 \pi} \sum_{k=-(n-1)}^{n-1} \Gamma_n (k) 
		e^{-ik\theta}, \; \; \; \theta \in [-\pi,\pi]. 
		$$
		
		\noindent
		The existence, boundedness and continuity of $f_n$ follows from 
		Assumption A2. For any $\tilde{\bu} \in \mathbb{R}^{np}$ with $\|\tilde{\bu}\|_2 = 1$, partition $\tilde{\bu}$ as $\left( (\tilde{\bu}^1)^T, (\tilde{\bu}^2)^T, \cdots, (\tilde{\bu}^2)^T\right)^T$. Define $G(\theta) = \sum_{k=1}^n \bu^k e^{-ik\theta}$, and note that 
		$$
		\int_{-\pi}^\pi G^* (\theta) G(\theta) d \theta = \sum_{k=1}^n \sum_{k'=1}^n \int_{-\pi}^\pi (\tilde{\bu}^k)^T \tilde{\bu}^{k'} e^{i(k-k')\theta} d\theta = 2 \pi \sum_{k=1}^n (\tilde{\bu}^k)^T \tilde{\bu}^{k'} = 2 \pi. 
		$$
		
		\noindent
		By Assumption A2, and the triangle inequality for the $\| \cdot \|_2$-norm (for matrices with complex valued entries), it follows 
		that $\|f_n (\theta)\|_2 \leq \kappa_2/\pi$ for every $\theta \in [-\pi, \pi]$. Note also that $f_n (\theta)$ is Hermitian and all its eigenvalues are real for every $\theta \in [-\pi, \pi]$. Using the block partitioned form of $\Theta_n$, and the definition of $f_n$, we obtain 
		\begin{eqnarray*}
			\tilde{\bu}^T \Theta_n \tilde{\bu} 
			&=& \sum_{k=1}^n \sum_{k'=1}^n (\tilde{\bu}^k)^T \Gamma_n (k-k') \tilde{\bu}^{k'}\\
			&=& \sum_{k=1}^n \sum_{k'=1}^n \int_{-\pi}^\pi (\tilde{\bu}^k)^T f_n (\theta) e^{i(k-k')\theta} \tilde{\bu}^{k'} d \theta\\
			&=& \int_{-\pi}^\pi G(\theta)^* f_n (\theta) G(\theta) d \theta\\
			&\leq& \frac{\kappa_2}{\pi} \int_{-\pi}^\pi G(\theta)^* G(\theta) d \theta\\
			&=& 2 \kappa_2. 
		\end{eqnarray*}
		
		\noindent
		We conclude that $\|\Theta_n\|_2 \leq 2 \kappa_2$. 
		
		Let $C_n := \{\frac{1}{n} \sum_{i=1}^n \xb_i^T \xb_i \leq 4p \kappa_2\}$. 
		Since $\xb^T \xb = \sum_{i=1}^n \xb_i^T \xb_i$ and $\|\Theta_n\|_2 \leq 2 
		\kappa_2$, the event $\{\xb^T \Theta_n^{-1} \xb \leq 2np\}$ is a 
		subset of $C_n$. Note that $\xb^T \Theta_n^{-1} \xb$ has a 
		$\chi^2_{np}$ distribution under $P_0$. Using standard tail 
		concentation bounds for $\chi^2$ random variables (see for example 
		\cite[Lemma 4.1]{CKG:2020}), it follows that 
		\begin{eqnarray*}
			P_0 (C_n^c) 
			&\leq& P_0 \left( \xb^T \Theta_n^{-1} \xb \geq 2np \right)\\
			&\leq& 2 \exp \left( -\frac{4n^2 p^2}{4np + 2np} \right)\\
			&=& 2 \exp \left( -\frac{2np}{3} \right) \rightarrow 0 
		\end{eqnarray*}
		
		\noindent
		as $n \rightarrow \infty$. It follows by the mean value theorem and 
		(\ref{derivative}) that for every $\bu$ with $\|\bu\| = 1$ 
		\begin{eqnarray}
			& & \left| \frac{1}{n} \sum_{i=1}^n 
			\left( 1 + \epsilon_i^2 + (\xb_i^T \bu)^2 \right)^{-3/2} (\xb_i^T \bu)^2 - 
			\frac{1}{n} \sum_{i=1}^n 
			\left( 1 + \epsilon_i^2 + (\xb_i^T \tilde{\bf w}_{(\bu)})^2 \right)^{-3/2} (\xb_i^T \tilde{\bf w}_{(\bu)})^2\right| \nonumber\\
			&\leq& 10 \sqrt{p \kappa_2} \|\bu - \tilde{\bf w}_{(\bu)}\| 
			\nonumber\\
			&\leq& \frac{20 \kappa_2}{\sqrt{p}} \label{qdiff}
		\end{eqnarray}
		
		\noindent
		on $C_n$. Hence, 
		$$
		\inf_{{\bf u}: \|{\bf u}\| = 1} \frac{1}{n} \sum_{i=1}^n \left( 1 + \epsilon_i^2 + (\xb_i^T \bu)^2 \right)^{-3/2} (\xb_i^T \bu)^2 \geq 
		\min_{{\bf w} \in S_{1/p}} \frac{1}{n} \sum_{i=1}^n \left( 1 + \epsilon_i^2 + (\xb_i^T {\bf w})^2 \right)^{-3/2} (\xb_i^T {\bf w})^2 - \frac{20 \kappa_2}{\sqrt{p}}. 
		$$
		
		\noindent
		on $C_n$. It follows by (\ref{bound}) and Assumption A1 that 
		\begin{eqnarray}
			& & P_0 \left( \inf_{{\bf u}: \|{\bf u}\| = 1} \frac{1}{n} \sum_{i=1}^n \left( 1 + \epsilon_i^2 + (\xb_i^T \bu)^2 \right)^{-3/2} (\xb_i^T \bu)^2 < \frac{\kappa_1 M_1}{2} - \frac{20 \kappa_2}{\sqrt{p}} \right) 
			\nonumber\\
			&\leq& P_0 \left( \min_{{\bf w} \in \tilde{S}_{1/p}} \frac{1}{n} \sum_{i=1}^n \left( 1 + \epsilon_i^2 + (\xb_i^T {\bf w})^2 \right)^{-3/2} (\xb_i^T {\bf w})^2 < \frac{\kappa_1 M_1}{2} \right) + P_0 (C_n^c) \nonumber\\
			&\leq& P_0 (C_n^c) + \sum_{{\bf w} \in \tilde{S}_{1/p}} P_0 \left( \frac{1}{n} \sum_{i=1}^n \left( 1 + \epsilon_i^2 + (\xb_i^T {\bf w})^2 \right)^{-3/2} (\xb_i^T {\bf w})^2 < \frac{\kappa_1 M_1}{2} \right) \nonumber\\
			&\leq& P_0 (C_n^c) + 2 (2p+1)^p \exp \left( -\min(M_2, M_3) n \right) 
			\nonumber\\
			&=& P_0 (C_n^c) + 2 \times \exp \left( -\min(M_2, M_3) n + p \log(2p+1) \right) \rightarrow 0 \label{qterm}
		\end{eqnarray}
		
		\noindent
		as $n \rightarrow \infty$. For large enough $n$, $\frac{\kappa_1 M_1}{2} - \frac{20 \kappa_2}{\sqrt{p}} > \frac{\kappa_1 M_1}{4}$. Using (\ref{diff}), (\ref{lterm}), \ref{qterm}), and Assumption A4, we get 
		\begin{eqnarray*}
			& & P_0 \left( \inf_{{\bf u}: \|{\bf u}\| = 1} (Q_\alpha (\bb_0 + \delta_n \bu) - Q_\alpha (\bb_0)) > \frac{\sqrt{1+\alpha^{-2}} \delta_n^2 \kappa_1 M_1}{8 \alpha} - (K_1\sqrt{1+\alpha^{-2}}+2) \delta_n \sqrt{\frac{p}{n}} \right)\\
			&\geq& 1 - \left( 2 \exp \left( -\frac{2np}{3} \right) + \exp \left( - \left\{ \frac{81K_1^2}{400 \kappa_2} - \log 21 \right\} p \right) +  2 \times \exp \left( -\min(M_2, M_3) n + p \log(2p+1) \right) \right)
		\end{eqnarray*}
		
		\noindent
		for large enough $n$. Since $\delta_n = \tilde{M} \alpha_n \sqrt{\frac{p}{n}}$, it follows that 
		\begin{eqnarray*}
			P_0 \left( \inf_{{\bf u}: \|{\bf u}\| = 1} (Q_\alpha (\bb_0 + \delta_n \bu) - Q_\alpha (\bb_0)) > \sqrt{1+\alpha^{-2}} (K_1+2) \delta_n \sqrt{\frac{p}{n}} \right) \rightarrow 1 
		\end{eqnarray*}
		
		\noindent
		for a large enough choice of $\tilde{M}$.  \hfill$\Box$
		
		\subsection{Proof of Theorem \ref{thm:postdist}}
		
		We first establish that the posterior distribution asymptotically places 
		all of its mass in a neighborhood of radius $K'' \alpha$ around $\bb_0$, 
		for an appropriate $K''$. Note that 
		\begin{eqnarray}
			\Pi \left( \|\bb - \bb_0\| > K'' \alpha \mid {\bf Y} \right) 
			&=& \frac{\int_{\|\bu\|> K'' \alpha } \exp(-n\alpha Q_\alpha (\bb_0 + \bu)) d \bu}{\int_{\mathbb{R}^p} \exp(-n\alpha Q_\alpha (\hat{\bb}_{pm} + {\bf v})) d {\bf v}} \nonumber\\
			&=& \exp(n \alpha Q_\alpha (\hat{\bb}_{pm})) \frac{\int_{\|\bu\|> K'' \alpha } \exp(-n\alpha Q_\alpha (\bb_0 + \bu)) d \bu}{\int_{\mathbb{R}^p} \exp(-n\alpha \left\{ Q_\alpha (\hat{\bb}_{pm} + {\bf v}) - Q_\alpha (\hat{\bb}_{pm}) \right\}) d {\bf v}} \nonumber\\
			& & \label{posterior1}
		\end{eqnarray}
		
		\noindent
		for any $K'' > 0$. A specific choice of $K''$ will be made later. Using the second order Taylor expansion of $\tilde{f}_\bv (t) = Q_\alpha (\hat{\bb}_{pm} + t \bv)$ around $t=0$, we get 
		\begin{eqnarray*}
			& & n \alpha \left( Q_\alpha (\hat{\bb}_{pm} + \bv) - Q_\alpha 
			(\hat{\bb}_{pm}) \right)\\
			&=& n \alpha \left( \tilde{f}_\bv (1) - \tilde{f}_\bv (0) \right) 
			\nonumber\\
			&=& \left\{ 2 \tau^2 \hat{\bb}_{pm}^T + \sum_{i=1}^n \ell_{SPH,\alpha}' (Y_i - 
			\xb_i^T \hat{\bb}_{pm}) \xb_i^T \right\} \bv + \frac{1}{2} \sum_{i=1}^n \ell_{SPH,\alpha}'' (Y_i - \xb_i^T \hat{\bb}_{pm} - t^* (\bu) \xb_i^T \bv) (\xb_i^T \bv)^2 + \tau^2 \bv^T \bv 
		\end{eqnarray*}
		
		\noindent
		where $t^* (\bv) \in (0,1)$. Since $\hat{\bb}_{pm}$ is the unique minimizer of $Q_\alpha$, it 
		follows that 
		\begin{equation} \label{ts}
			n \alpha \left( Q_\alpha (\hat{\bb}_{pm} + \bv) - Q_\alpha 
			(\hat{\bb}_{pm}) \right) = \frac{1}{2} \sum_{i=1}^n \ell_{SPH,\alpha}'' (Y_i - \xb_i^T \hat{\bb}_{pm} - t^* (\bv) \xb_i^T \bv) (\xb_i^T \bv)^2 + \tau^2 \bv^T \bv. 
		\end{equation}
		
		\noindent
		Since $0 \leq \ell_{SPH,\alpha}'' (y) \leq 1$ for every $y \in \mathbb{R}$, it 
		follows that 
		\begin{equation} \label{ts1}
			\frac{n \alpha}{\sqrt{1+\alpha^{-2}}} \left( Q_\alpha (\hat{\bb}_{pm} + \bv) - 
			Q_\alpha (\hat{\bb}_{pm}) \right) \leq \bv^T \frac{\sum_{i=1}^n \xb_i \xb_i^T}{2} \bv + \tau^2 \bv^T \bv 
		\end{equation}
		
		\noindent
		for every $\bu \in \mathbb{R}^p$. It follows from (\ref{posterior1}) 
		that 
		\begin{eqnarray}
			& & \Pi \left( \|\bb - {\bb}_{0}\| > K'' \alpha \mid {\bf Y} \right) \nonumber\\
			&\leq& \exp(n \alpha Q_\alpha (\hat{\bb}_{pm})) \frac{\int_{\|\bu\|> K'' \alpha } \exp(-n\alpha Q_\alpha (\bb_0 + \bu)) d \bu}{\int_{\mathbb{R}^p} \exp \left( - \sqrt{1+\alpha^{-2}} \bv^T \frac{\sum_{i=1}^n 
					\xb_i \xb_i^T}{2} \bu - \sqrt{1+\alpha^{-2}} \tau^2 \bv^T \bv \right) d \bu}. \label{gbound}
		\end{eqnarray}
		
		\noindent
		Next, let ${\bf v} \in \mathbb{R}^p$ with $\|v\| = 1$. Then, 
		$$
		{\bf v}^T \left( \frac{1}{n} \sum_{i=1}^n \xb_i \xb_i^T \right) {\bf v} = \frac{1}{n} {\bf Z}^T Q {\bf Z}, 
		$$
		
		\noindent
		where ${\bf Z} \sim \mathcal{N}_n ({\bf 0}, I_n)$ under $P_0$, and 
		the $(r,s)^{th}$ element of $Q$ is given by ${\bf v}^T \Gamma_n (r-s) 
		{\bf v}$. Using $\|Q\|_F^2 \leq n \|Q\|$, $E_0 \left[ \frac{1}{n} {\bf Z}^T Q {\bf Z} \right] = {\bf v}^T \Gamma_n (0) {\bf v}$, along with the Hanson-Wright inequality of \cite{Rudelson:Vershynin:2013}, we obtain 
		$$
		P_0 \left( \left| {\bf v}^T \left( \frac{1}{n} \sum_{i=1}^n \xb_i \xb_i^T \right) {\bf v} - {\bf v}^T \Gamma_n (0) {\bf v} \right| > 
		\|Q\| \eta \right) \leq 2 \exp \left( - cn \min(\eta^2, \eta) \right)
		$$
		
		\noindent
		for every $\eta > 0$. By a very similar argument to the one at the end of Page 1547 in \cite{Basu:Michailidis:2015}, it follows that 
		$\|Q\| \leq \|\Theta_n\| \leq 2 \kappa_2$. Hence, 
		$$
		P_0 \left( \left| {\bf v}^T \left( \frac{1}{n} \sum_{i=1}^n \xb_i \xb_i^T \right) {\bf v} - {\bf v}^T \Gamma_n (0) {\bf v} \right| > 
		\frac{10 \kappa_2}{\sqrt{c}} \sqrt{\frac{p}{n}} \right) \leq 2 \exp 
		\left( -25p \right). 
		$$
		
		\noindent
		Using Lemma B.2 in \cite{GKS:2019}, it follows that 
		\begin{equation} \label{covbound}
			P_0 \left( \left\| \frac{1}{n} \sum_{i=1}^n \xb_i \xb_i^T - \Gamma_n (0) 
			\right\| > \frac{10 \kappa_2}{\sqrt{c}} \sqrt{\frac{p}{n}} \right) 
			\leq 2 \exp \left( -p (25 - 2 \log(21)) \right) \rightarrow 0
		\end{equation}
		
		\noindent
		as $n \rightarrow \infty$. It follows by Assumption A1, Assumption A2 and 
		(\ref{gbound}) that on an event with $P_0$-probability converging to 
		one, we have 
		\begin{equation} \label{nonquad}
			\Pi \left( \|\bb - {\bb}_{0}\| > K'' \alpha \mid {\bf Y} \right) 
			\leq \left( \frac{3 \tau^2 + 3 {\kappa_1^{-1} n}}{2 \pi} \right)^{p/2} \exp(n \alpha Q_\alpha (\hat{\bb}_{pm})) 
			{\int_{\|\bu\|> K'' \alpha } \exp(-n\alpha Q_\alpha (\bb_0 + \bu)) d \bu}. 
		\end{equation}
		
		\noindent
		Note that for any $t \in \mathbb{R}$ we have 
		$$
		\sqrt{1 + \alpha^2} (|t| - \alpha) \leq \ell_{SPH,\alpha} (t) \leq \sqrt{1 + \alpha^2} |t|. 
		$$
		
		\noindent
		Since $y_i - \xb_i^T (\bb_0 + \bu) = \epsilon_i - \xb_i^T \bu$, $y_i - 
		\xb_i^T \hat{\bb}_{pm} = \epsilon_i - \xb_i^T (\hat{\bb}_{pm} - \bb_0)$ 
		and $|\epsilon_i - \xb_i^T \bu| \geq |\xb_i^T \bu| - |\epsilon_i|$, 
		it follows after straightforward calculations that 
		\begin{eqnarray}
			& & \exp(n \alpha Q_\alpha (\hat{\bb}_{pm})) {\int_{\|\bu\|> K'' \alpha } \exp(-n\alpha Q_\alpha (\bb_0 + \bu)) d \bu} \nonumber\\
			&\leq& \exp \left( n \alpha \sqrt{1+\alpha^2} + 2 \sqrt{1 + \alpha^2} \sum_{i=1}^n |\epsilon_i| + \sqrt{1 + \alpha^2} \sum_{i=1}^n |\xb_i^T (\hat{\bb}_{pm} - \bb_0)| + \tau^2 \|\hat{\bb}_{pm}\|^2 - \tau^2 \|\bb_0\|^2 \right) \times 
			\nonumber\\
			& & \int_{\|\bu\|> K'' \alpha } \exp \left( - \alpha \sum_{i=1}^n |\xb_i^T \bu| - 2 \tau^2 \bu^T \bb_0 \right) d \bu \nonumber\\
			&\leq& \exp \left( n \alpha \sqrt{1+\alpha^2} + 2 \sqrt{1 + \alpha^2} \sum_{i=1}^n |\epsilon_i| + \sqrt{1 + \alpha^2} \sum_{i=1}^n |\xb_i^T (\hat{\bb}_{pm} - \bb_0)| + \tau^2 \|\hat{\bb}_{pm}\|^2 - \tau^2 \|\bb_0\|^2 \right) \times 
			\nonumber\\
			& & \int_{\|\bu\|> K'' \alpha } \exp \left( - n \alpha \|\bu\| \left( \frac{1}{n} \sum_{i=1}^n |\xb_i^T \tilde{\bu}| - \frac{2 \tau^2 \|\bb_0\|}{n \alpha} \right) \right) d \bu, \label{absbound}
		\end{eqnarray}
		
		\noindent
		where $\tilde{\bu} = \bu/\|\bu\|$. By the Cauchy-Schwarz inequality, Assumption 2, (\ref{covbound}) and Theorem \ref{thm:postmean} it follows that 
		$$
		\sum_{i=1}^n |\xb_i^T (\hat{\bb}_{pm} - \bb_0)| \leq \sqrt{n} \sqrt{\sum_{i=1}^n (\hat{\bb}_{pm} - \bb_0)^T \xb_i \xb_i^T (\hat{\bb}_{pm} - \bb_0)} \leq \sqrt{2\kappa_1} \alpha \tilde{M} \sqrt{np}
		$$
		
		\noindent
		on a set with $P_0$ probability converging to $1$. Also, by the strong law of large numbers, Assumption A4' and Theorem \ref{thm:postmean}, we get 
		$$
		2 \sqrt{1 + \alpha^2} \sum_{i=1}^n |\epsilon_i| + \tau^2 
		\|\hat{\bb}_{pm}\|^2 - \tau^2 \|\bb_0\|^2 \leq 4n E_0 |\epsilon_1| + 2 
		\tau^2 \|\bb_0\| \|\hat{\bb}_{pm} - \bb_0\| + \tau^2 \|\hat{\bb}_{pm} - 
		\bb_0\|^2 \leq K_2 n \alpha^2 
		$$
		
		\noindent
		for an approrpriate constant $K_2$ on an event with $P_0$ probability 
		converging to $1$. It follows by (\ref{absbound}) that 
		\begin{eqnarray}
			& & \exp(n \alpha Q_\alpha (\hat{\bb}_{pm})) {\int_{\|\bu\|> K'' \alpha } \exp(-n\alpha Q_\alpha (\bb_0 + \bu)) d \bu} \nonumber\\
			&\leq& \exp( 2 (K_2 + 1) n \alpha^2) \int_{\|\bu\|> K'' \alpha } \exp \left( - n \|\bu\| \left( \frac{\alpha}{n} \sum_{i=1}^n |\xb_i^T \tilde{\bu}| - \frac{2 \tau^2 \|\bb_0\|}{n} \right) \right) d \bu \label{lapbound}
		\end{eqnarray}
		
		\noindent
		on an event with $P_0$ probability converging to $1$. Let $c := \frac{\log 
			2 \pi}{8 \sqrt{\kappa_2}}$. Fix $\bv$ with $\|\bv\| = 1$ arbitrarily. 
		Then, by Markov's inequality 
		\begin{eqnarray}
			P_0 (\sum_{i=1}^n |\xb_i^T \bv| < nc) 
			&\leq& P_0 \left( \exp \left(-2 \sqrt{\kappa_2} \sum_{i=1}^n |\xb_i^T \bv| 
			\right) > \exp(-nc) \right) \nonumber\\
			&\leq& \exp(2 \sqrt{\kappa_2} nc) E_0 \left[ \exp \left( -2 \sqrt{\kappa_2} \sum_{i=1}^n |\xb_i^T \bv| \right) \right]. \label{eqmarkov}
		\end{eqnarray}
		
		\noindent
		Recall that ${\bf x} \in \mathbb{R}^{np}$ is the vector obtained by 
		stacking $\xb_1, \xb_2, \cdots, \xb_n$ on top of each other, and $\xb$ 
		has a multivariate distribution with mean ${\bf 0}$ and covariance matrix 
		$\Theta_n$. It follows that $X \bv = (I_n \otimes \bv^T) \xb$ has a 
		multivariate normal distribution with mean ${\bf 0}$ and covariance 
		matrix $(I_n \otimes \bv^T) \Theta_n (I_n \otimes \bv)$. It follows by 
		Assumptions A2 and A5 that 
		$$
		\kappa_3 \leq \lambda_{min} \left( (I_n \otimes \bv^T) \Theta_n (I_n 
		\otimes \bv) \right) \leq \lambda_{max} \left( (I_n \otimes \bv^T) 
		\Theta_n (I_n \otimes \bv) \right) \leq \kappa_2. 
		$$
		
		\noindent
		Combining this fact with (\ref{eqmarkov}), we get 
		$$
		P_0 (\sum_{i=1}^n |\xb_i^T \bv| < nc) \leq \exp(2 \sqrt{\kappa_2} nc) 
		\left( \frac{\kappa_2}{\kappa_3} \right)^{p/2} E_0 \left[ \exp 
		\left( -2 \sum_{i=1}^n |Z_i| \right) \right], 
		$$
		
		\noindent
		where $\{Z_i\}_{i=1}^n$ have an i.i.d. standard normal distribution under 
		$P_0$. Using the Mills ratio identity, it follows that 
		\begin{eqnarray}
			P_0 (\sum_{i=1}^n |\xb_i^T \bv| < nc) 
			&\leq& \exp(2 \sqrt{\kappa_2} nc) \left( \frac{\kappa_2}{\kappa_3} \right)^{p/2} \left( E_0 \left[ \exp \left( -2 |Z_1| \right) \right] \right)^n \nonumber\\
			&\leq& \exp(2 \sqrt{\kappa_2} nc) \left( \frac{\kappa_2}{\kappa_3} \right)^{p/2} \left( 
			2 \exp \left( 2 \right) P_0 (Z_1 > 2) \right)^n \nonumber\\
			&\leq& \exp(2 \sqrt{\kappa_2} nc) \left( \frac{\kappa_2}{\kappa_3} \right)^{p/2} \left( \sqrt{\frac{1}{2 \pi}} \right)^n \nonumber\\
			&=& \exp \left( -\frac{n \log 2\pi}{4} \right) \left( \frac{\kappa_2}{\kappa_3} \right)^{p/2}. \label{millsbound}
		\end{eqnarray}
		
		\noindent
		Recall the construction of the set $S_{1/p}$ (in the proof of Theorem \ref{thm:postmean}) with the property that $S_{1/p} \subseteq \{\bv: \; \|\bv\| \leq 1\}$, $|S_{1/p}| \leq (2p+1)^p$, and for any $\bv$ with $\|\bv\| \leq 1$, there exists ${\bf w}_{(\bv)} \in S_{1/p}$ such that $\|\bv - {\bf w}_{(\bv)}\| \leq p^{-1}$. Recall also, the construction $\tilde{\bf w}_{(\bv)} = (1/\|{\bf w}_{(\bv)}\|) {\bf w}_{(\bv)}$ (so that $\|\tilde{\bf w}_{(\bv)}\| = 1$) with the property 
		$$
		\|\bv - \tilde{\bf w}_{(\bv)}\| \leq \frac{2}{p}, 
		$$
		
		\noindent
		and that $\tilde{S}_{1/p}$ denotes the collection of all $\tilde{\bf w}_{(\bv)}$ (as $\bv$ varies over $\{\bv: \; \|\bv\| \leq 1\}$). Now, for any $\bv$ with $\|\bv\| \leq 1$, we have 
		\begin{eqnarray*}
			\sum_{i=1}^n |\xb_i^T \bv|  
			&\geq& \sum_{i=1}^n |\xb_i^T \tilde{\bf w}_{(\bv)}| - \sum_{i=1}^n |\xb_i^T (\bv - \tilde{\bf w}_{(\bv)})|. 
		\end{eqnarray*}
		
		\noindent
		It follows that 
		\begin{eqnarray}
			\inf_{\bv: \; \|\bv\| \leq 1} \sum_{i=1}^n |\xb_i^T \bv| 
			&\geq& \inf_{{\bf w} \in \tilde{S}_{1/p}} \sum_{i=1}^n |\xb_i^T \tilde{\bf w}| - \frac{2}{p} \sum_{i=1}^n \|\xb_i\| \nonumber\\
			&\geq& \inf_{\bv: \; \|\bv\| \leq 1} \sum_{i=1}^n |\xb_i^T \bv| - \frac{4 n \sqrt{\kappa_2}}{\sqrt{p}} \label{infbound}
		\end{eqnarray}
		
		\noindent
		on an event with $P_0$-probability converging to $1$ (see the definition 
		of the set $C_n$ in the proof of Theorem \ref{thm:postmean}). Using Assumption A1 and 
		(\ref{infbound}), for large enough $n$, we get 
		\begin{eqnarray}
			& & P_0 \left( \inf_{\bv: \; \|\bv\| \leq 1} \sum_{i=1}^n |\xb_i^T \bv| < \frac{nc}{2} \right) \nonumber\\
			&\leq& P_0 \left( \inf_{{\bf w} \in \tilde{S}_{1/p}} \sum_{i=1}^n |\xb_i^T \tilde{\bf w}| < \frac{nc}{2} + \frac{4 n \sqrt{\kappa_2}}{\sqrt{p}} \right) + P_0 (C_n^c) \nonumber\\
			&\leq& P_0 \left( \inf_{{\bf w} \in \tilde{S}_{1/p}} \sum_{i=1}^n |\xb_i^T \tilde{\bf w}| < nc \right) + P_0 (C_n^c) \nonumber\\
			&\leq& \sum_{{\bf w} \in \tilde{S}_{1/p}} P_0 \left( \sum_{i=1}^n |\xb_i^T \tilde{\bf w}| < nc \right) + P_0 (C_n^c)\nonumber\\
			&\leq& \exp \left( -\frac{n \log 2\pi}{4} \right) \left( \frac{\kappa_2}{\kappa_3} \right)^{p/2} (2p+1)^p + P_0 (C_n^c) \rightarrow 0 \label{infsum}
		\end{eqnarray}
		
		\noindent
		as $n \rightarrow \infty$. By (\ref{nonquad}), (\ref{lapbound}), 
		(\ref{infsum}) and Assumption A4, it follows that 
		\begin{eqnarray*}
			& & \Pi \left( \|\bb - \hat{\bb}_{pm}\| > K'' \alpha \mid {\bf Y} \right)\\
			&\leq& \left( \frac{3\tau^2 + 3{\kappa_1^{-1} n}}{2 \pi} \right)^{p/2} \exp( 2 (K_2 + 1) n \alpha^2) \int_{\|\bu\|> K'' \alpha } \exp \left( - \frac{n \alpha c \|\bu\|}{4} \right) d \bu\\
			&\leq& \left( \frac{3\tau^2 + 3{\kappa_1^{-1} n}}{2 \pi} \right)^{p/2} \exp( 2 (K_2 + 1) n \alpha^2) \exp \left( - \frac{nc K'' \alpha^2}{8} \right) 
			\int_{\|\bu\|> K'' \alpha } \exp \left( - \frac{n \alpha c \|\bu\|}{8} \right) d \bu\\
			&\leq& \left( \frac{3\tau^2 + 3{\kappa_1^{-1} n}}{2 \pi} \right)^{p/2} \exp( 2 (K_2 + 1) n \alpha^2) \exp \left( - \frac{nc K'' \alpha^2}{8} \right) 
			\int_{\|\bu\|> K'' \alpha } \exp \left( - \frac{n \alpha c \sum_{i=1}^p |u_i|}{8 \sqrt{p}} \right) d \bu\\
			&\leq& \left( \frac{3\tau^2 + 3{\kappa_1^{-1} n}}{2 \pi} \right)^{p/2} \exp( 2 (K_2 + 1) n \alpha^2) \exp \left( - \frac{nc K'' \alpha^2}{8} \right) 
			\left( \frac{16 \sqrt{p}}{n \alpha c} \right)^p. 
		\end{eqnarray*}
		
		\noindent
		on an event with $P_0$-probability converging to $1$. By 
		Assumptions A1 and A5, it follows that 
		\begin{equation} \label{ibound}
			E_0 \left[ \Pi \left( \|\bb - \hat{\bb}_{pm}\| > K'' \alpha \mid {\bf Y} 
			\right) \right] \rightarrow 0 
		\end{equation}
		
		\noindent
		for a suitably large constant $K''$ as $n \rightarrow \infty$. In light 
		of (\ref{ibound}), to prove the desired result, it is enough to show that 
		$$
		\Pi \left( \|\bb - \bb_0\| > M^* \delta_n, \; \|\bb - \bb_0\| \leq 
		(K''-1) \alpha \mid {\bf Y} \right) 
		$$
		
		\noindent
		converges in $P_0$-probability to zero as $n \rightarrow \infty$. Note 
		that 
		\begin{eqnarray}
			& & \Pi \left( \|\bb - \bb_0\| > M^* \delta_n, \; \|\bb - \bb_0\| \leq 
			(K''-1) \alpha \mid {\bf Y} \right) \nonumber\\
			&=& \frac{\int_{\|\bu\| \leq \tilde{K} \alpha, \|\bu\| > M^* \delta_n} \exp(-n\alpha Q_\alpha ({\bb}_0 + \bu)) d \bu}{\int_{\mathbb{R}^p} \exp(-n\alpha Q_\alpha ({\bb})) d \bb} \nonumber\\
			&=& \frac{\int_{\|\bu\| \leq \tilde{K} \alpha, \|\bu\| > M^* \delta_n} \exp(-n\alpha \left\{ Q_\alpha ({\bb}_{0} + \bu) - Q_\alpha ({\bb}_{0}) \right\}) d \bu}{\int_{\mathbb{R}^p} \exp(-n\alpha \left\{ Q_\alpha (\hat{\bb}_{pm} + \bu) - Q_\alpha (\hat{\bb}_{pm}) \right\}) d \bu} \times \exp \left( n \alpha \left( Q_\alpha (\bb_0) - Q_\alpha (\hat{\bb}_{pm} \right) \right) \nonumber\\
			& & \label{phood}
		\end{eqnarray}
		
		\noindent
		where $\tilde{K} = K'' - 1$. For any vector $\bu$, the vector 
		$\tilde{\bu}$ denotes $\bu/\|\bu\|$. For every $\bu$ such that $\|\bu\| 
		\leq \tilde{K} \alpha, \|\bu\| > M^* \delta_n$, (\ref{taylor}) (without the 
		$\delta_n$ term) and (\ref{taylor1}), along with (\ref{lterm}) and 
		Assumption A5 imply that on an event with $P_0$-probability converging 
		to one 
		\begin{eqnarray}
			Q_\alpha (\bb_0 + \bu) - Q_\alpha (\bb_0) 
			&\geq& \frac{1}{n \alpha} \sum_{i=1}^n \ell_{SPH,\alpha}' (\epsilon_i) 
			\xb_i^T \bu - \frac{2 \tau^2}{n \alpha} \|\bu\| \|\bb_0\| + \frac{\tau^2}{n 
				\alpha} \bu^T \bu + \nonumber\\
			& & \frac{\sqrt{1+\alpha^{-2}}}{2 n \alpha} \sum_{i=1}^n \left( 1 + 
			2\alpha^{-2} \epsilon_i^2 + 2 \alpha^{-2} (\xb_i^T \bu)^2 
			\right)^{-3/2} (\xb_i^T \bu)^2 \nonumber\\
			&=& \frac{\|\bu\|}{n \alpha} \sum_{i=1}^n \ell_{SPH,\alpha}' (\epsilon_i) 
			\xb_i^T \tilde{\bu} - \frac{2 \tau^2}{n \alpha} \|\bu\| \|\bb_0\| + \frac{\tau^2}{n 
				\alpha} \bu^T \bu + \nonumber\\
			& &  \frac{\sqrt{1+\alpha^{-2}} \|\bu\|^2}{2 n \alpha} \sum_{i=1}^n \left( 1 
			+ 2\alpha^{-2} \epsilon_i^2 + 2 \alpha^{-2} \|\bu\|^2 (\xb_i^T 
			\tilde{\bu})^2 \right)^{-3/2} (\xb_i^T \tilde{\bu})^2 \nonumber\\
			&\geq& -C M^* \delta_n \sqrt{\frac{p}{n}} + \frac{\tau^2}{n 
				\alpha} \bu^T \bu + \nonumber\\
			& & + \frac{\sqrt{1+\alpha^{-2}} \|\bu\|^2}{2 n \alpha} 
			\sum_{i=1}^n \left( 1 + 2 \epsilon_i^2 + 2 \tilde{K}^2 (\xb_i^T 
			\tilde{\bu})^2 \right)^{-3/2} (\xb_i^T \tilde{\bu})^2 \label{lb1}
		\end{eqnarray}
		
		\noindent
		for an appropriate constant $C$. Now, by the exact same argument starting 
		from the end of (\ref{lterm}) to (\ref{qterm}) (adjusting for 
		relevant constants in the definition of $Z_i (\bu)$), it follows that 
		\begin{equation} \label{bfn}
			P_0 \left( \inf_{\bu: \|\bu\|=1} \frac{1}{n} \sum_{i=1}^n \left( 1 + 2 \epsilon_i^2 + 2 
			\tilde{K}^2 (\xb_i^T \bu)^2 \right)^{-3/2} (\xb_i^T 
			\bu)^2 > \tilde{M} \right) \rightarrow 1 
		\end{equation}
		
		\noindent
		as $n \rightarrow \infty$ for an appropriate constant $\tilde{M}$. It follows by (\ref{phood}), (\ref{lb1}), (\ref{ts}), (\ref{ts1}) and (\ref{bfn}) that 
		\begin{eqnarray*}
			& & \Pi \left( \|\bb - \bb_0\| > M^* \delta_n, \; \|\bb - \bb_0\| \leq 
			(K''-1) \alpha \mid {\bf Y} \right)\\
			&\leq& \exp \left( C M^* \delta_n \alpha \sqrt{np} + \frac{\sqrt{1+\alpha^{-2}}}{2} 
			\|\sum_{i=1}^n \xb_i \xb_i^T \| \|\hat{\bb}_{pm} - \bb_0\|^2 + \tau^2 
			\|\hat{\bb}_{pm} - \bb_0\|^2 \right) \times\\
			& & \frac{\int_{\|\bu\|> M^* \delta_n} \exp \left( - \frac{n \tilde{M} \sqrt{1+\alpha^{-2}}}{2} \bu^T \bu - \tau^2 \bu^T \bu \right)}{\int_{\mathbb{R}^p} \exp \left( - n \sqrt{1+\alpha^{-2}} \kappa_1 \bu^T \bu - \tau^2 \sqrt{1+\alpha^{-2}} \bu^T \bu \right) d \bu}
		\end{eqnarray*}
		
		\noindent
		on an event whose $P_0$-probability converges to one as $n \rightarrow 
		\infty$. It follows by (\ref{covbound}) and Theorem \ref{thm:postmean} 
		that 
		\begin{eqnarray*}
			& & \Pi \left( \|\bb - \bb_0\| > M^* \delta_n, \; \|\bb - \bb_0\| \leq 
			(K''-1) \alpha \mid {\bf Y} \right)\\
			&\leq& \exp \left( C M^* \delta_n \alpha \sqrt{np} + 2 n \kappa_1 \delta_n^2 + \tau^2 \delta_n^2 - \frac{n \tilde{M} (M^*)^2}{4} \delta_n^2 - \frac{\tau^2 (M^*)^2}{2} \delta_n^2 \right) \times\\
			& & \frac{\int_{\|\bu\|> M^* \delta_n} \exp \left( - \frac{n \tilde{M} \sqrt{1+\alpha^{-2}}}{2} \bu^T \bu - \tau^2 \bu^T \bu \right)}{\int_{\mathbb{R}^p} \exp \left( - n \sqrt{1+\alpha^{-2}} \kappa_1 \bu^T \bu - \tau^2 \sqrt{1+\alpha^{-2}} \bu^T \bu \right) d \bu}\\
			&\leq& \exp \left( C M^* \delta_n \alpha \sqrt{np} + 2 n \kappa_1 \delta_n^2 + \tau^2 \delta_n^2 - \frac{n \tilde{M} (M^*)^2}{4} \delta_n^2 - \frac{\tau^2 (M^*)^2}{2} \delta_n^2 \right) \left( \frac{n \tilde{M} + 2 \tau^2}{5 n \kappa_1 + 5 \tau^2} \right)^{-p/2}\\
			&\leq& \exp \left( C M^* \delta_n \alpha \sqrt{np} + 2 n \kappa_1 \delta_n^2 + \tau^2 \delta_n^2 - \frac{n \tilde{M} (M^*)^2}{4} \delta_n^2 - \frac{\tau^2 (M^*)^2}{2} \delta_n^2 + \frac{p}{2} \log \kappa_3 \right), 
		\end{eqnarray*}
		
		\noindent
		on an event whose $P_0$-probability converges to one as $n \rightarrow 
		\infty$, where $\kappa_3 = 5\kappa_1/\tilde{M}+5/2$. Since 
		$\delta_n \alpha \sqrt{np} = o(n \delta_n^2)$ and $p = o(n \delta_n^2)$, 
		choosing $M^* = 4 \max \left( 1, 2 \kappa_1, \frac{C+1}{\tilde{M}} 
		\right)$ ensures that 
		$$
		\Pi \left( \|\bb - \bb_0\| > M^* \delta_n, \; \|\bb - 
		\bb_0\| \leq (K''-1) \alpha \mid {\bf Y} \right)
		$$
		
		\noindent
		converges to zero in $P_0$-probability as $n \rightarrow \infty$. \hfill$\Box$

		\subsection{Proof of Theorem \ref{thm:high-dim-consistency}}
		
		\noindent
		Let ${\bf s}$ be any element of $\{0,1\}^p$ which satisfies $|{\bf s}| \leq n/(\log (\max(n,p)))^{1+\delta} + |{\bf s}_0|$. Using the same arguments that led to (\ref{covbound}), but replacing 
		$\xb_i$ by $\xb_{i,s}$, $\sqrt{p/n}$ by $\sqrt{\frac{|s| \log p}{n}}$, 
		$\Gamma_n (0)$ by $(\Gamma_n (0))_{ss}$, we get 
		\begin{equation} \label{covboundsubset}
			P_0 \left( \left\| \frac{1}{n} \sum_{i=1}^n \xb_{i,s} \xb_{i,s}^T - (\Gamma_n (0))_{ss} \right\| > \frac{10 \kappa_2}{\sqrt{c}} \sqrt{\frac{|{\bf s}| \log p}{n}} \right) \leq 2 \exp \left( -|{\bf s}| \log p (25 - 2 \log(21)) \right) 
		\end{equation}
		
		\noindent
		Let 
		$$
		D_n := \cap_{{\bf s} \in \{0,1\}^p: \; {\bf s} \neq {\bf 0}, |{\bf s}| \leq n/(\log (\max(n,p)))^{1+\delta} + |{\bf s}_0|} \left\{ \left\| \frac{1}{n} \sum_{i=1}^n \xb_{i,s} \xb_{i,s}^T - (\Gamma_n (0))_{ss} \right\| \leq \frac{10 \kappa_2}{\sqrt{c}} \sqrt{\frac{|{\bf s}| \log p}{n}} \right\}. 
		$$
		
		\noindent
		It follows by (\ref{covboundsubset}) that 
		\begin{eqnarray*}
			P(D_n) 
			&\geq& 1 - \sum_{{\bf s} \in \{0,1\}^p: \; {\bf s} \neq {\bf 0}, s|{\bf s}| \leq n/(\log (\max(n,p)))^{1+\delta} + |{\bf s}_0|} 2 \exp \left( -|{\bf s}| \log p (25 - 2 \log(21)) \right)\\
			&\geq& 1 - \sum_{k=1}^\infty {{p}\choose{k}} 2 \exp \left( -k \log p (25 - 2 \log(21)) \right)\\
			&\geq& 1 - 2 \sum_{k=1}^\infty p^k p^{-3k}\\
			&=& 1 - \frac{p^{-2}}{1 - p^{-2}} \rightarrow 1 
		\end{eqnarray*}
		
		\noindent
		as $n \rightarrow \infty$. We now derive bounds for the ratio of the 
		posterior probability assigned to a given sparsity pattern ${\bf s}$ and 
		the posterior probability assigned to the true sparsity pattern 
		${\bf s}_0$ under different cases. 
		
		\medskip
		
		\noindent
		{\bf Case I: ${\bf s}$ is a `superset' of ${\bf s}_0$ with $|{\bf s}| \leq n/(\log (\max(n,p)))^{1+\delta}$}. Let ${\bf s} \in 
		\{0,1\}^p$ be such that ${\bf s}_0 \subset {\bf s}$. Hence $s_j = 1$ 
		whenever $s_{0j} = 1$. Prior to examining the ratio in (\ref{post:ratio}), we need to establish consistency of the restricted posterior mode for 
		$\bb$ under the sparsity constraint imposed by ${\bf s}$. This posterior mode is denoted by $\hat{\bb}_{pm,s}$. The proof goes along the same lines as the proof of Theorem \ref{thm:postmean}, with some key changes that we highlight. Similar to the proof of Theorem \ref{thm:postmean}, with $\delta_{n,s} := M^{**} \alpha \sqrt{\frac{|{\bf s}| \log p}{n}}$ (for an appropriately chosen $M^{**}$ independent of $n$ and ${\bf s}$), we aim to establish that 
		$$
		P_0 \left( \inf_{{\bf u} \in \mathbb{R}^{|s|}: \|{\bf u}\| = 1} Q_\alpha 
		(\bb_{0,s} + \delta_{n,s} \bu) > Q_\alpha (\bb_{0,s}) \right) \rightarrow 1 
		$$
		
		\noindent
		as $n \rightarrow \infty$. Since ${\bf s}_0 \subset {\bf s}$, it follows 
		that for every $1 \leq i \leq n$
		$$
		\epsilon_i = y_i - \xb_i^T \bb_0 = y_i - \xb_{i,s_0}^T \bb_{0,s_0} = y_i - \xb_{i,s}^T 
		\bb_{0,s} \; \mbox{ and } \; \|\bb_{0}\| = \|\bb_{0,s_0}\| = \|\bb_{0,s}\|. 
		$$
		
		\noindent
		Using this fact along with similar arguments leading up to equation 
		(\ref{taylor1}), we obtain 
		\begin{eqnarray}
			& & Q_\alpha (\bb_{0,s} + \delta_{n,s} \bu) - Q_\alpha (\bb_{0,s}) 
			\nonumber\\
			&\geq& \frac{\delta_{n,s}}{n \alpha} \sum_{i=1}^n \ell_{SPH,\alpha}' (\epsilon_i) 
			\xb_{i,s}^T \bu - \frac{2 \tau^2 \delta_{n,s}}{n \alpha} \|\bb_{0}\| +\nonumber\\
			& & \frac{\sqrt{1+\alpha^{-2}} \delta_{n,s}^2}{2 n \alpha} \sum_{i=1}^n \left( 1 + 
			2\alpha^{-2} \epsilon_i^2 + 2 \delta_{n,s}^2 \alpha^{-2} (\xb_{i,s}^T \bu)^2 
			\right)^{-3/2} (\xb_{i,s}^T \bu)^2. \label{taylor1:ss}
		\end{eqnarray}
		
		\noindent
		Again, repeating the exact same arguments between (\ref{taylor1}) and (\ref{lterm}) replacing $\xb_i$ by $\xb_{i,s}$, $\sqrt{p/n}$ by $\sqrt{\frac{|s| \log p}{n}}$, $\Gamma_n (k)$ by $(\Gamma_n (k))_s$, and $21^p$ by $21^{|s|}$, we get 
		\begin{eqnarray}
			& & P_0 \left( \sup_{{\bf u}: \|{\bf u}\| = 1} \left| \frac{1}{n \alpha \sqrt{1+\alpha^{-2}}} \sum_{i=1}^n \ell_{SPH,\alpha}' (\epsilon_i) \xb_i^T \bu \right| > K_1 \sqrt{\frac{|s|\log p}{n}} \right) \nonumber\\
			&=& \exp \left( - \left\{ \frac{81K_1^2}{400 \kappa_2} - \log 21 \right\} |s| \log p \right) \rightarrow 0 \mbox{ as } n \rightarrow \infty 
			\label{lterm:ss}
		\end{eqnarray}
		
		\noindent
		if $K_1$ is chosen to be sufficiently large. Now fix $\bu \in \mathbb{R}^{|s|}$ with 
		$\|\bu\| = 1$ and define 
		$$
		Z_{i,s} (\bu) := \left( 1 + \epsilon_i^2 + \frac{(\xb_{i,s}^T \bu)^2}{\kappa_1 \bu^T 
			(\Gamma_n (0))_{ss} \bu} \right)^{-3/2} \frac{(\xb_{i,s}^T \bu)^2}{\bu^T (\Gamma_n (0))_{ss} 
			\bu} \; \; \forall 1 \leq i \leq n. 
		$$
		
		\noindent
		It follows by Assumptions A2 and A3 that $\{Z_{i,s} (\bu)\}_{i=1}^n$ 
		are i.i.d. random variables and are uniformly bounded by $\kappa_1$. 
		Note that $G_s (\bu) := \xb_{1,s}^T \bu/\sqrt{\bu^T (\Gamma_n (0))_{ss} \bu}$ has 
		a standard normal distribution and is independent of $\epsilon_1$, and
		$E_0 [Z_{1,s} (\bu)] = E_0 [Z_1 (\bu)] = M_1$ (see proof of Theorem \ref{thm:postmean}). 
		Also, by the definition of the function $g$ in Assumption A3, it follows that 
		$g(\epsilon_i) = E[Z_{i,s} (\bu) \mid {\boldsymbol \epsilon}]$ (and 
		$E_0 [Z_{i,s} (\bu)] = E[g(\epsilon_i)]$ by tower property). Now, the entire argument from equation (\ref{union}) to (\ref{bound}), can essentially be repeated verbatim (with $\xb_i$ replaced by $\xb_{i,s}$ and $\Gamma_n (\cdot)$ replaced by $(\Gamma_n (\cdot))_s$), leading to the conclusion that 
		\begin{eqnarray}
			P_0 \left( \frac{1}{n} \sum_{i=1}^n \left( 1 + \epsilon_i^2 + (\xb_{i,s}^T 
			\bu)^2 \right)^{-3/2} (\xb_{i,s}^T \bu)^2 < \frac{\kappa_1 M_1}{2} \right)
			\leq 2 \exp \left( -\min(M_2, M_3) n \right). \label{bound_ss}
		\end{eqnarray}
		
		\noindent
		Again, by \cite[Theorem 5.2]{Vershynin:2011}, there exists a subset $\tilde{S}_{1/\max{n,p}}$ of $\{{\bf u} \in \mathbb{R}^{|s|}: \|{\bu} = 1\}$ with the property that $|\tilde{S}_{1/\max(n,p)}| \leq (2\max(n,p)+1)^{|{\bf s}|}$, and that for any $\bu \in \mathbb{R}^{|{\bf s}|}$ with $\|\bu\| = 1$, there exists $\tilde{\bf w}_{(\bu)} \in \tilde{S}_{1/\max(n,p)}$ such that $\|\bu - \tilde{\bf w}_{(\bu)}\| \leq 
		\frac{2}{\max(n,p)}$. Again, the entire argument from equation (\ref{derivative}) to (\ref{qdiff}), can essentially be repeated verbatim (with $\xb_i$ replaced by $\xb_{i,s}$ and $\Theta_n$ replaced by $(\Theta_n)_s$), leading to the conclusion that 
		\begin{eqnarray*}
			& & \inf_{{\bf u} \in \mathbb{R}^{|{\bf s}|}: \|{\bf u}\| = 1} \frac{1}{n} \sum_{i=1}^n \left( 1 + \epsilon_i^2 + (\xb_i^T \bu)^2 \right)^{-3/2} (\xb_i^T \bu)^2\\
			&\geq& \min_{{\bf w} \in \tilde{S}_{1/\max(p,n)}} \frac{1}{n} \sum_{i=1}^n \left( 1 + \epsilon_i^2 + (\xb_{i,s}^T {\bf w})^2 \right)^{-3/2} (\xb_{i,s}^T {\bf w})^2 - \frac{20 \sqrt{|{\bf s}| \kappa_2}}{{\max(n,p)}}\\
			&\geq& \min_{{\bf w} \in \tilde{S}_{1/\max(p,n)}} \frac{1}{n} \sum_{i=1}^n \left( 1 + \epsilon_i^2 + (\xb_{i,s}^T {\bf w})^2 \right)^{-3/2} (\xb_{i,s}^T {\bf w})^2 - \frac{20 \sqrt{\kappa_2}}{\sqrt{n}}
		\end{eqnarray*}
		
		\noindent
		on an event with $P_0$-probability converging to one. In particular, for appropriately chosen constants $K_1$ and $M^{**}$ (independent of ${\bf s}$), and for $n$ large enough to satisfy $\sqrt{n} \kappa_1 M_1 > 80 \kappa_2$ and $\min(M_2, M_3) (\log \max(n,p))^{1+\delta} > 2 \log (2n+1)$, we obtain 
		\begin{eqnarray}
			& & P_0 \left( \inf_{{\bf u} \in \mathbb{R}^{|s|}: \|{\bf u}\| = 1} Q_\alpha 
			(\bb_{0,s} + \delta_{n,s} \bu) > Q_\alpha (\bb_{0,s}) \right) 
			\nonumber\\
			&\geq& 1 - \left( 2 \exp \left( -\frac{2n|s|}{3} \right) + \exp \left( - \left\{ \frac{81K_1^2}{400 \kappa_2} - \log 21 \right\} |{\bf s}| \log p \right) \right) - \nonumber\\
			& & 2 \times \exp \left( -\min(M_2, M_3) n + |{\bf s}| \log(2|{\bf s}|+1) \right) \nonumber\\
			&\geq& 1 - \left( 2 \exp \left( -\frac{2n|s|}{3} \right) + \exp \left( - 3 |{\bf s}| \log p \right) +  2 \times \exp \left( -\frac{\min(M_2, M_3) n}{2} \right) \right) \label{setbound}
		\end{eqnarray}
		
		\noindent
		Let $C_{n,s} := \{ \inf_{{\bf u} \in \mathbb{R}^{|s|}: \|{\bf u}\| = 1} Q_\alpha (\bb_{0,s} + \delta_{n,s} \bu) > Q_\alpha (\bb_{0,s})\}$. It follows that $\|\hat{\bb}_{pm,s} - \bb_0\| \leq \delta_{n,s}$ on the event $C_{n,s}$. Now, by second order Taylor series expansion and the fact that $0 \leq \ell''(y) \leq 1$, it follows that for every $\bu \in \mathbb{R}^{|{\bf s}|}$ 
		\begin{equation} \label{ts:lower}
			n \alpha \left( Q_\alpha (\hat{\bb}_{pm,s} + \bu) - 
			Q_\alpha (\hat{\bb}_{pm,s}) \right) \geq \tau^2 \bu^T \bu,  
		\end{equation}
		
		\noindent
		and for every $\bv \in \mathbb{R}^{|s_0|}$
		\begin{eqnarray} \label{ts:upper}
			& & n \alpha \left( Q_\alpha (\hat{\bb}_{pm, s_0} + \bv) - 
			Q_\alpha (\hat{\bb}_{pm, s_0}) \right) \nonumber\\
			&\leq& \sqrt{1+\alpha^{-2}} \bv^T \frac{\sum_{i=1}^n \xb_{i,s_0} \xb_{i,s_0}^T}{2} \bv + \tau^2 \bv^T \bv \nonumber\\
			&\leq& \kappa_1^{-1} \bv^T \bv + \tau^2 \bv^T \bv 
		\end{eqnarray}
		
		\noindent
		on the event $D_n$ defined at the beginning of this proof when $n$ is large enough so that 
		$$
		\frac{\sqrt{1+\alpha^{-2}}}{2 - \sqrt{1+\alpha^{-2}}} \frac{10 \kappa_2}{\sqrt{c}} \left( \frac{1}{(\log \max(n,p))^{\delta/2}} + \sqrt{\frac{|{\bf s}_0| \log p}{n}} \right) < \frac{1}{\kappa_1}. 
		$$
		
		\noindent
		Note that due to Assumption B1, the LHS of the above inequality converges to zero as $n \rightarrow \infty$; hence, this inequality eventually holds for all $n$ above a relevant threshold. Combining (\ref{post:ratio}), (\ref{ts:lower}) and (\ref{ts:upper}), we 
		now get 
		\begin{eqnarray}
			\frac{\Pi \left( {\bf s} \mid {\bf Y} \right)}{\Pi \left( {\bf s}_0 \mid {\bf Y} \right)} 
			&\leq& \left( \frac{q \tau}{(1-q)} \right)^{|{\bf s}|-|{\bf s}_0|} \frac{(\tau^2 + \kappa_1^{-1})^{|s_0|/2}}{\tau^{|s|}} \exp \left( n \alpha \left( Q_\alpha (\hat{\bb}_{pm, s_0}) - Q_\alpha (\hat{\bb}_{pm, s}) \right) \right) \nonumber\\
			&=& \left( \frac{q}{(1-q)} \right)^{|{\bf s}|-|{\bf s}_0|} \left( 
			1 + \frac{1}{\kappa_1 \tau^2} \right)^{|s_0|} \exp \left( n \alpha \left( Q_\alpha (\hat{\bb}_{pm, s_0}) - Q_\alpha (\hat{\bb}_{pm, s}) \right) \right). \label{ratiobound}
		\end{eqnarray}
		
		\noindent
		Again, noting that ${\bf s}$ is a superset of ${\bf s}_0$, and by repeating the arguments between (\ref{gbound}) and (\ref{covbound}) with appropriate changes, we get 
		\begin{eqnarray*}
			n \alpha (Q_\alpha (\hat{\bb}_{pm, s_0}) - Q_\alpha (\hat{\bb}_{pm, s})) 
			&\leq& (\kappa_1^{-1} + \tau^2) \|\hat{\bb}_{fill, pm, s_0} - \hat{\bb}_{fill, pm, s}\|^2\\
			&\leq& (M^{**})^2 \alpha^2 (\kappa_1^{-1} + \tau^2) \frac{(|{\bf s}| + |{\bf s}_0|) \log p}{n} 
		\end{eqnarray*}
		
		\noindent
		on the event $C_{n,s} \cap C_{n, s_0} \cap D_n$. Note that 
		$$
		\frac{|{\bf s}| + |{\bf s}_0|}{|{\bf s}| - |{\bf s}_0|} = 1 + \frac{2|{\bf s}_0|}{|{\bf s}| - |{\bf s}_0|} \leq 1 + 2 |{\bf s}_0|. 
		$$
		
		\noindent
		Let $N_0$ be such that $\alpha^\delta = \alpha_n^\delta > 4(1 + 2 |{\bf s}_0|)$ for $n > N_0$. Then 
		$$
		(|{\bf s}| + |{\bf s}_0|) \alpha^2 \log p \leq 0.25 \left( |{\bf s}| - 
		|{\bf s}_0| \right) \alpha^{2+\delta} \log p
		$$
		
		\noindent
		for $n > N_0$. It follows by (\ref{ratiobound}) and the definition of $q$ 
		that on $C_{n,s} \cap C_{n, s_0} \cap D_n$
		\begin{equation} \label{superbound}
			\frac{\Pi_{SS} \left( {\bf s} \mid {\bf Y} \right)}{\Pi_{SS} \left( {\bf s}_0 \mid 
				{\bf Y} \right)} \leq K_0 q^{\frac{|{\bf s}| - |{\bf s}_0|}{2}}
		\end{equation}
		
		\noindent
		for large enough $n$ (cutoff not depending on ${\bf s}$) and an appropriate constant $K_0$ (not depending on $n$ and ${\bf s}$). 
		
		\medskip
		
		\noindent
		{\bf Case II: ${\bf s}$ is a `subset' of ${\bf s}_0$}. Let ${\bf s} \in 
		\{0,1\}^p$ be such that ${\bf s} \subset {\bf s}_0$. Note that under the true model $P_0$, we have 
		\begin{eqnarray*}
			y_i 
			&=& \xb_i^T \bb_0 + \epsilon_i\\
			&=& \xb_{i,s}^T \bb_{0,s} + \xb_{i,s_0 \setminus s}^T \bb_{0,s_0 
				\setminus s} + \epsilon_i\\
			&=& \xb_{i,s}^T \left( \bb_{0,s} + (\Gamma_n (0))_{ss} (\Gamma_n (0))_{s,
				s_0 \setminus s} \bb_{0, s_0 \setminus s} \right) +\\
			& & \left (\xb_{i,s_0 \setminus s} - (\Gamma_n (0))_{s_0 \setminus s, s} (\Gamma_n (0))_{ss} \xb_{i,s} \right)^T \bb_{0, s_0 \setminus s} + \epsilon_i\\
			&=& \xb_{i,s}^T \tilde{\bb}_{0,s} + \tilde{\epsilon}_{i,s} 
		\end{eqnarray*}
		
		\noindent
		where 
		$$
		\tilde{\bb}_{0,s} := \bb_{0,s} + (\Gamma_n (0))_{ss} (\Gamma_n 
		(0))_{s,s_0 \setminus s} \bb_{0, s_0 \setminus s}
		$$
		
		\noindent
		and 
		$$
		\tilde{\epsilon}_{i,s} := \left (\xb_{i,s_0 \setminus s} - (\Gamma_n (0))_{s_0 \setminus s, s} (\Gamma_n (0))_{ss} \xb_{i,s} \right)^T \bb_{0, s_0 \setminus s} + \epsilon_i. 
		$$
		
		\noindent
		Note that by construction $\tilde{\epsilon}_{i,s}$ is independent of 
		$\xb_{i,s}$. For any ${\bf u} \in \mathbb{R}^{|s|}$ with $\|{\bf u}\| = 1$, define the random variables 
		$$
		\tilde{Z}_{i,s} (\bu) := \left( 1 + \tilde{\epsilon}_{i,s}^2 + \frac{(\xb_{i,s}^T \bu)^2}{\kappa_1 \bu^T (\Gamma_n (0))_{ss} \bu} \right)^{-3/2} 
		\frac{(\xb_{i,s}^T \bu)^2}{\bu^T (\Gamma_n (0))_{ss} \bu} \; \; \forall 1 \leq i \leq n. 
		$$
		
		\noindent
		Now, note that 
		\begin{eqnarray*}
			& & \left| \sum_{i=1}^n Z_i (\bu) - n E_0 [Z_1 (\bu)] \right|\\
			&\leq& \left| \sum_{i=1}^n Z_i (\bu) - \sum_{i=1}^n g(\tilde{\epsilon}_i) \right| + \left| \sum_{i=1}^n g(\epsilon_i) - n E_0 [Z_1 (\bu)] \right|\\
			&\leq& \left| \sum_{i=1}^n Z_i (\bu) - \sum_{i=1}^n g(\tilde{\epsilon}_i) \right| + \left| \sum_{i=1}^n g(\tilde{\epsilon}_i) - \sum_{i=1}^n E_0 [g(\tilde{\epsilon}_i) \mid {\boldsymbol \epsilon}] \right| + \left| \sum_{i=1}^n E_0 [g(\tilde{\epsilon}_i) \mid {\boldsymbol \epsilon}] - 
			n E_0 [Z_1 (\bu)] \right|, 
		\end{eqnarray*}
		
		\noindent
		where $g(\tilde{\epsilon}_i) = E_0 \left[ Z_i (\bu) \mid 
		\tilde{\boldsymbol \epsilon} \right]$. Using the independence of 
		$\tilde{\epsilon}_{i,s}$ and $\xb_{i,s}$, and observing that 
		$\tilde{Z}_{i,s} (\bu)$ is a uniformly bounded function of $\xb_{i,s}^T \bu$ (conditional on $\tilde{\epsilon}_i$), a parallel argument to the one right after equation 
		(\ref{union}) leads to the bound 
		$$
		V(\sum_{i=1}^n \tilde{Z}_{i,s} ({\bf u}) \mid \tilde{\boldsymbol \epsilon}) \leq 4n \kappa_1 \kappa_2. 
		$$
		
		\noindent
		Similarly, using independence of $\epsilon_i$ and $\xb_i$, and observing that $g(\tilde{\epsilon}_i)$ is a uniformly bounded function of $\left (\xb_{i,s_0 \setminus s} - (\Gamma_n (0))_{s_0 \setminus s, s} (\Gamma_n (0))_{ss} \xb_{i,s} \right)^T \bb_{0, s_0 \setminus s}$ (conditional on $\epsilon_i$), it can be  shown that $n^{-1} V(\sum_{i=1}^n g(\tilde{\epsilon}_i) \mid {\boldsymbol \epsilon})$ is uniformly bounded (in ${\boldsymbol \epsilon}$ and ${\bf s}$). Finally, Assumption B3 can be used to show that $n^{-1} V(\sum_{i=1}^n E_0 [g(\tilde{\epsilon}_i) \mid {\boldsymbol \epsilon}])$ is uniformly bounded (in 
		${\bf s}$). The above facts can be leveraged to repeat the arguments in the proof of Theorem \ref{thm:postmean} with straightforward changes/adjustments to conclude that there exists a constant $M^{***}$ (not depending on ${\bf s}$) such that 
		$$
		\|\hat{\bb}_{pm,s} - \tilde{\bb}_{0,s}\| \leq M^{***} \alpha \sqrt{\frac{|{\bf s}|\log p}{n}} 
		$$
		
		\noindent
		on a set $C_{n,s}$ with $P_0 (C_{n,s}) \rightarrow 1$ as $n 
		\rightarrow \infty$. Let $\bv \in \mathbb{R}^{|s_0|}$ be such that $\hat{\bb}_{pm, s_0} + \bv$ corresponds to the filled version of $\hat{\bb}_{pm, s}$ in $\mathbb{R}^{|s_0|}$ (with zeros appended in relevant places). It follows that for large enough $n$, there exists a constant $K^*$ such that $\|\bv\| \leq K^*$ on $C_{n,s}$. By a second order Taylor series expansion around the restricted mode $\hat{\bb}_{pm, s_0}$, we get 
		\begin{eqnarray*}
			& & \alpha \left( Q_\alpha (\hat{\bb}_{pm, s_0} + \bv) - 
			Q_\alpha (\hat{\bb}_{pm, s_0}) \right)\\
			&\geq& \frac{\|\bv\|^2}{2 n} \sum_{i=1}^n \left( 1 + 
			2 \epsilon_i^2 + 2 \|\bv\|^2 (\xb_{i,s_0}^T \tilde{\bv})^2 
			\right)^{-3/2} (\xb_{i,s_0}^T \tilde{\bv})^2 - \frac{2 \tau^2}{n} \|\bv\| \|\bb_0\|\\
			&\geq& \frac{\|\bv\|^2}{2 n} \sum_{i=1}^n \left( 1 + 
			2 \epsilon_i^2 + 2 (K^*)^2 (\xb_{i,s_0}^T \tilde{\bv})^2 
			\right)^{-3/2} (\xb_{i,s_0}^T \tilde{\bv})^2 - \frac{2 \tau^2}{n} K^* \|\bb_0\| 
		\end{eqnarray*}
		
		\noindent
		with $\tilde{\bv} = \bv/\|\bv\|.$ By a similar argument as the one leading to (\ref{bfn}), there exists a constant $\bar{M}$ such that 
		\begin{equation} \label{bfn:new}
			P_0 \left( \inf_{\bv: \|\bv\|=1} \frac{1}{n} \sum_{i=1}^n \left( 1 + 2 \epsilon_i^2 + 2 (K^*)^2 (\xb_{i,s_0}^T \bv)^2 
			\right)^{-3/2} (\xb_{i,s_0}^T \bv)^2 > \bar{M} \right) \rightarrow 1 
		\end{equation}
		
		\noindent
		Note that the bound in (\ref{ratiobound}) holds for any ${\bf s} \in 
		\{0,1\}^p$. Also, by construction of $\bv$, it follows that $\|v\|^2 \geq 
		(|{\bf s}_0| - |{\bf s}|) S^2$, where $S = \min_{1 \leq i \leq |s_0|} |\beta_{s_0, i}|$. Combining everything, we get 
		\begin{eqnarray}
			\frac{\Pi_{SS} \left( {\bf s} \mid {\bf Y} \right)}{\Pi_{SS} \left( {\bf s}_0 \mid {\bf Y} \right)} 
			&\leq& \left( \frac{q}{(1-q)} \right)^{|{\bf s}|-|{\bf s}_0|} \left( 
			1 + \frac{1}{\kappa_1 \tau^2} \right)^{|s_0|} \exp \left( n \alpha \left( Q_\alpha (\hat{\bb}_{pm, s_0}) - Q_\alpha (\hat{\bb}_{pm, s}) \right) \right) \nonumber\\
			&\leq& K_1 q^{|{\bf s}|-|{\bf s}_0|} \exp \left( - 0.25 n (|{\bf s}_0| - |{\bf s}|) \bar{M} S^2 \right) \nonumber\\
			&\leq& K_1 \exp \left( - 0.125 n (|{\bf s}_0| - |{\bf s}|) \bar{M} S^2 \right) \label{subbound}
		\end{eqnarray}
		
		\noindent
		for large enough $n$ (cutoff not depending on ${\bf s}$) on a set, say 
		$\tilde{C}_{n,s}$, with $P_0$-probability converging to $1$ as $n 
		\rightarrow \infty$. Here $K_1$ is a constant which does not depend on $n$ 
		or ${\bf s}$. The last inequality follows from Assumption B4. 
		
		\medskip
		
		\noindent
		{\bf Case III: ${\bf s}$ satisfies ${\bf s} \not\subset {\bf s}_0$, ${\bf s}_0 \not\subset {\bf s}$, $|{\bf s}| \leq n/(\log (\max(n,p)))^{1+\delta}$ and $|{\bf s}| > |{\bf s}_0|$}. Let $\tilde{\bf s} := {\bf s} \cup {\bf s}_0$. Note that $\tilde{\bf s}$ is a superset of ${\bf s}_0$. By repeating the arguments in Case I up to equation (\ref{setbound}) verbatim, and noting $|\tilde{\bf s}| \leq n/(\log (\max(n,p)))^{1+\delta} + |s_0| = o(n/\log n)$, there exists a set $C_{n,s}$ such that 
		\begin{eqnarray*}
			P_0 (C_{n,s}) 
			&\geq& 1 - \left( 2 \exp \left( -\frac{2n|\tilde{\bf s}|}{3} \right) + \exp \left( - 3 |\tilde{\bf s}| \log p \right) +  2 \times \exp \left( -\frac{\min(M_2, M_3) n}{2} \right) \right)\\
			&\geq& 1 - \left( 2 \exp \left( -\frac{2n|{\bf s}|}{3} \right) + \exp \left( - 3 |{\bf s}| \log p \right) +  2 \times \exp \left( -\frac{\min(M_2, M_3) n}{2} \right) \right), 
		\end{eqnarray*}
		
		\noindent
		for large enough $n$ (cutoff not depending on ${\bf s}$), and 
		$\|\hat{\bb}_{pm,\tilde{s}} - \bb_0\| \leq \delta_{n,\tilde{s}}$ on 
		$C_{n,s}$. It follows that 
		\begin{eqnarray*}
			n \alpha (Q_\alpha (\hat{\bb}_{pm, s_0}) - Q_\alpha (\hat{\bb}_{pm, \tilde{s}})) 
			&\leq& (\kappa_1^{-1} + \tau^2) \|\hat{\bb}_{fill, pm, s_0} - \hat{\bb}_{fill, pm, \tilde{s}}\|^2\\
			&\leq& (M^{**})^2 \alpha^2 (\kappa_1^{-1} + \tau^2) \frac{(|\tilde{\bf s}| + |{\bf s}_0|) \log p}{n}\\
			&\leq& (M^{**})^2 \alpha^2 (\kappa_1^{-1} + \tau^2) \frac{(|{\bf s}| + 2 |{\bf s}_0|) \log p}{n}
		\end{eqnarray*}
		
		\noindent
		on the event $C_{n,s} \cap C_{n, s_0} \cap D_n$ for large enough $n$ (cutoff not depending on ${\bf s}$). Note that 
		$$
		\frac{|{\bf s}| + 2|{\bf s}_0|}{|{\bf s}| - |{\bf s}_0|} = 1 + \frac{3|{\bf s}_0|}{|{\bf s}| - |{\bf s}_0|} \leq 1 + 3 |{\bf s}_0|. 
		$$
		
		\noindent
		Let $N_0^*$ be such that $\alpha^\delta = \alpha_n^\delta > 4 |{\bf s}_0|(1 + 3 |{\bf s}_0|)$ for $n > N_0^*$. Then 
		$$
		(|{\bf s}| + 2 |{\bf s}_0|) \alpha^2 \log p \leq \frac{1}{4|{\bf s}_0|} \left( |{\bf s}| - |{\bf s}_0| \right) \alpha^{2+\delta} \log p
		$$
		
		\noindent
		for $n > N_0^*$. Let $d({\bf s}, {\bf s}_0) = |{\bf s} \cap {\bf s}_0^c| + |{\bf s}_0 \cap {\bf s}^c|$ denote the number of disagreements 
		between ${\bf s}$ and ${\bf s}_0$. Since $|{\bf s}| - |{\bf s}_0| \geq 1$ and $|{\bf s}_0| \geq 1$, we get 
		\begin{eqnarray*}
			d({\bf s}, {\bf s}_0) 
			&=& |{\bf s} \cap {\bf s}_0^c| + |{\bf s}_0 \cap {\bf s}^c|\\
			&=& |{\bf s} \cap {\bf s}_0^c| - |{\bf s}_0 \cap {\bf s}^c| + 2|{\bf s}_0 \cap {\bf s}^c|\\
			&=& |{\bf s}| - |{\bf s}_0| + 2|{\bf s}_0 \cap {\bf s}^c|\\
			&\leq& |{\bf s}| - |{\bf s}_0| + 2 |{\bf s}_0| (|{\bf s}| - |{\bf s}_0|)\\
			&\leq& 3 |{\bf s}_0| (|{\bf s}| - |{\bf s}_0|). 
		\end{eqnarray*}
		
		\noindent
		It follows by (\ref{ratiobound}) and the definition of $q$ that on $C_{n,s} \cap C_{n, s_0} \cap D_n$
		\begin{equation} \label{bigbound}
			\frac{\Pi_{SS} \left( {\bf s} \mid {\bf Y} \right)}{\Pi_{SS} \left( {\bf s}_0 \mid 
				{\bf Y} \right)} \leq K_0^* \left( q^{1/|{\bf s}_0|} \right)^{\frac{|{\bf s}_0|(|{\bf s}| - |{\bf s}_0|)}{2}} \leq K_0^* \left( q^{1/(6|{\bf s}_0|)} \right)^{d({\bf s}, {\bf s}_0)} 
		\end{equation}
		
		\noindent
		for large enough $n$ (cutoff not depending on ${\bf s}$) and an appropriate constant $K_0^*$ (not depending on ${\bf s}$ as well). 
		
		\medskip
		
		\noindent
		{\bf Case IV: ${\bf s}$ satisfies ${\bf s} \not\subset {\bf s}_0$, 
			${\bf s}_0 \not\subset {\bf s}$, $|{\bf s}| \leq n/(\log 
			(\max(n,p)))^{1+\delta}$ and $|{\bf s}| \leq |{\bf s}_0|$}. Let 
		$\bar{\bf s} := {\bf s} \cap {\bf s}_0$. Note that $\bar{\bf s}$ is a 
		subset of both ${\bf s}$ and ${\bf s}_0$. It follows by (\ref{ratiobound}) 
		that 
		\begin{eqnarray}
			\frac{\Pi_{SS} \left( {\bf s} \mid {\bf Y} \right)}{\Pi_{SS} \left( {\bf s}_0 \mid {\bf Y} \right)} 
			&\leq& \left( \frac{q}{(1-q)} \right)^{|{\bf s}|-|{\bf s}_0|} \left( 
			1 + \frac{1}{\kappa_1 \tau^2} \right)^{|s_0|} \exp \left( n \alpha \left( Q_\alpha (\hat{\bb}_{pm, s_0}) - Q_\alpha (\hat{\bb}_{pm, s}) \right) \right) \nonumber\\
			&=& \left( \frac{q}{(1-q)} \right)^{|{\bf s}|-|{\bf s}_0|} \left( 
			1 + \frac{1}{\kappa_1 \tau^2} \right)^{|s_0|} \exp \left( n \alpha \left( Q_\alpha (\hat{\bb}_{pm, s_0}) - Q_\alpha ({\bb}_{0, s_0}) \right) \right) \nonumber\\
			& & \times \exp \left( n \alpha \left( Q_\alpha ({\bb}_{0, s_0}) - Q_\alpha (\hat{\bb}_{pm, s}) \right) \right) \nonumber\\
			&\leq& \left( \frac{q}{(1-q)} \right)^{|{\bf s}|-|{\bf s}_0|} \left( 
			1 + \frac{1}{\kappa_1 \tau^2} \right)^{|s_0|} \exp \left( n \alpha \left( Q_\alpha ({\bb}_{0, s_0}) - Q_\alpha (\hat{\bb}_{pm, s}) \right) \right). \label{mixedbound}
		\end{eqnarray}
		
		\noindent
		Let 
		$$
		{\bf s}^* := {\bf s}_0 \cup {\bf s} = {\bf s}_0 \uplus ({\bf s} \setminus 
		\bar{\bf s}) = {\bf s} \uplus ({\bf s}_0 \setminus \bar{\bf s}). 
		$$
		
		\noindent
		Let $\hat{\bb}_{pm, s, fill(s^*)}$ denote the $s^*$-dimensional vector 
		obtained by appending relevant zeros to $\hat{\bb}_{pm, s}$. Noting that 
		$|{\bf s}^*| \leq 2|{\bf s}_0|$, and repeating the analysis in Case I 
		(replacing ${\bf s}$ by ${\bf s^*}$), we get 
		\begin{eqnarray*}
			& & Q_\alpha (\hat{\bb}_{pm, s}) - Q_\alpha ({\bb}_{0, s_0})\\
			&=& Q_\alpha (\hat{\bb}_{pm, s, fill(s^*)}) - Q_\alpha ({\bb}_{0, s^*})\\
			&\geq& \|\hat{\bb}_{pm, s, fill(s^*)} - {\bb}_{0, s^*}\|^2 \frac{\kappa_1 M_1}{8 \alpha} - \|\hat{\bb}_{pm, s, fill(s^*)} - {\bb}_{0, s^*}\| \left( K_1 \sqrt{\frac{|{\bf s}^*| \log p}{n}} + \frac{2 \tau^2 \|\bb_0\|}{n \alpha} \right)
		\end{eqnarray*}
		
		\noindent
		on an event with $P_0$-probability converging to $1$. Also, note that 
		\begin{eqnarray*}
			y_i 
			&=& \xb_i^T \bb_0 + \epsilon_i\\
			&=& \xb_{i,s^*}^T \bb_{0, s^*} + \epsilon_i\\
			&=& \xb_{i,s}^T \bb_{0,s} + \xb_{i,s^* \setminus s}^T \bb_{0,s^* 
				\setminus s} + \epsilon_i\\
			&=& \xb_{i,s}^T \left( \bb_{0,s} + (\Gamma_n (0))_{ss}^{-1} 
			(\Gamma_n (0))_{s,s^* \setminus s} \bb_{0, s^* \setminus s} \right) + 
			\left( \xb_{i,s^* \setminus s} - (\Gamma_n (0))_{s^* \setminus s, s} (\Gamma_n (0))_{ss}^{-1} \xb_{i,s} \right)^T \bb_{0, s^* \setminus s} + \epsilon_i\\
			&=& \xb_{i,s}^T \left( \bb_{0,s} + (\Gamma_n (0))_{ss}^{-1} 
			(\Gamma_n (0))_{s,s_0 \setminus \bar{s}} \bb_{0, s_0 \setminus \bar{s}} 
			\right) + \left( \xb_{i,s_0 \setminus \bar{s}} - (\Gamma_n (0))_{s_0 \setminus \bar{s}, s} (\Gamma_n (0))_{ss}^{-1} \xb_{i,s} \right)^T \bb_{0, s_0 \setminus \bar{s}} + \epsilon_i\\
			&=& \xb_{i,s}^T \tilde{\bb}_{0,s} + \tilde{\epsilon}_{i,s} 
		\end{eqnarray*}

		\noindent
		where 
		$$
		\tilde{\bb}_{0,s} := \bb_{0,s} + (\Gamma_n (0))_{ss}^{-1} 
		(\Gamma_n (0))_{s,s_0 \setminus \bar{s}} \bb_{0, s_0 \setminus \bar{s}}
		$$
		
		\noindent
		and 
		$$
		\tilde{\epsilon}_{i,s} := \left( \xb_{i,s_0 \setminus \bar{s}} - (\Gamma_n (0))_{s_0 \setminus \bar{s}, s} (\Gamma_n (0))_{ss}^{-1} \xb_{i,s} \right)^T \bb_{0, s_0 \setminus \bar{s}} + \epsilon_i. 
		$$
		
		\noindent
		By repeating the arguments in Case II (with ${\bf s}$ replaced by 
		$\bar{\bf s}$) up to equation (\ref{subbound}), we get 
		$$
		\|\hat{\bb}_{pm,s} - \tilde{\bb}_{0,s}\| \leq M^{***} \alpha \sqrt{\frac{|{\bf s}|\log p}{n}} 
		$$
		
		\noindent
		on an event with $P_0$-probability converging to one as $n \rightarrow 
		\infty$. Since the true model ${\bf s}_0$ does not vary with $n$, and 
		$|{\bf s}^*| \leq 2 |{\bf s}_0|$, it follows that 
		\begin{eqnarray*}
			& & Q_\alpha (\hat{\bb}_{pm, s, fill(s^*)}) - Q_\alpha ({\bb}_{0, s^*})\\
			&\geq& \|\hat{\bb}_{pm, s, fill(s^*)} - {\bb}_{0, s^*}\|^2 \frac{\kappa_1 M_1}{8 \alpha} - \\
			& & \|\hat{\bb}_{pm, s, fill(s^*)} - {\bb}_{0, s^*}\| \left( K_1 \sqrt{\frac{(1+\alpha^{-2})|{\bf s}^*| \log p}{n}} + \frac{2 \tau^2 \|\bb_0\|}{n \alpha} \right)\\
			&\geq& \left( \|\bb_{0,s} - \hat{\bb}_{pm,s}\|^2 + \|\bb_{0,s_0 \setminus \bar{s}}\|^2 \right) \frac{\kappa_1 M_1}{8 \alpha} - \left( \|\bb_{0,s} - \tilde{\bb}_{0,s}\| + \|\bb_{0,s_0 \setminus \bar{s}}\| + M^{***} \alpha \sqrt{\frac{2|{\bf s}_0|\log p}{n}} \right) \times\\
			& & \left( K_1 \sqrt{\frac{2(1+\alpha^{-2})|{\bf s}_0| \log p}{n}} + \frac{2 \tau^2 \|\bb_0\|}{n \alpha} \right)\\
			&\geq& \frac{(|{\bf s}_0| - |\bar{\bf s}|) S^2 \kappa_1 M_1}{16 \alpha}
		\end{eqnarray*}
		
		\noindent
		for large enough $n$ (cutoff not depending on ${\bf s}$), on an event with 
		$P_0$-probability converging to one as $n \rightarrow \infty$. Using 
		(\ref{mixedbound}), we conclude that 
		\begin{eqnarray}
			\frac{\Pi_{SS} \left( {\bf s} \mid {\bf Y} \right)}{\Pi_{SS} \left( {\bf s}_0 \mid {\bf Y} \right)} 
			&\leq& \bar{K_1} q^{|{\bf s}| - |{\bf s}_0|} \exp \left( - \frac{(|{\bf s}_0| - |\bar{\bf s}|) n S^2 \kappa_1 M_1}{16} \right) \nonumber\\
			&\leq& \bar{K_1} q^{- |{\bf s}_0|} \exp \left( - \frac{n S^2 \kappa_1 M_1}{16} \right) \nonumber\\
			&\leq& \bar{K}_1 \exp \left( - \frac{n S^2 \kappa_1 M_1}{32} \right)\\
			&\leq& \bar{K}_1 \left( \exp \left( - \frac{n S^2 \kappa_1 M_1}{64 
				|{\bf s}_0|} \right) \right)^{d({\bf s}, {\bf s}_0)} 
			\label{smallbound}
		\end{eqnarray}
		
		\noindent
		for large enough $n$ (cutoff not depending on ${\bf s}$) on a set with 
		$P_0$-probability converging to $1$ as $n \rightarrow \infty$ and where the constant
		$\bar{K}_1$ does not depend on $n$ or ${\bf s}$. The 
		second to last inequality follows from Assumptions B1 and B4, and the last 
		inequality uses $d({\bf s}, {\bf s}_0) \leq 2 |{\bf s}_0|$. 
		
		\medskip
		
		\noindent
		We now gather the results from all the four scenarios above to establish
		strong selection consistency. Note that 
		\begin{eqnarray*}
			& & \sum_{{\bf s}: |{\bf s}| > |{\bf s}_0|, |{\bf s}| \leq n/(\log(\max(n,p)))^{1+\delta}} P_0 (C_{n,s}^c)\\
			&\leq& \sum_{{\bf s}: |{\bf s}| > |{\bf s}_0|, |{\bf s}| \leq n/(\log(\max(n,p)))^{1+\delta}} \left( 2 \exp \left( -\frac{2n|{\bf s}|}{3} \right) + \exp \left( - 3 |{\bf s}| \log p \right) + 2 \exp \left( -\frac{\min(M_2, M_3) n}{2} \right) \right)\\
			&\leq& \sum_{j=1}^\infty 2 p^j \exp \left( -\frac{2nj}{3} \right) + \sum_{j=1}^\infty p^j \exp \left( - 3 j \log p \right) + \\
			& & 2 \exp \left( 
			n/(\log(\max(n,p)))^{\delta} + \log p - \frac{\min(M_2, M_3) n}{2} \right)\\
			&\leq& \frac{p \exp \left( -\frac{2n}{3} \right)}{1 - p \exp \left( -\frac{2n}{3} \right)} + \frac{1}{p^2 - 1} + 2 \exp \left( 
			n/(\log(\max(n,p)))^{\delta} + \log p - \frac{\min(M_2, M_3) n}{2} 
			\right) \rightarrow 0 
		\end{eqnarray*}
		
		\noindent
		as $n \rightarrow \infty$. Note that the number of sparsity patterns 
		satisfying the conditions in Case II and Case IV are uniformly bounded in $n$ (since the indices in ${\bf s}_0$ which are one do not change with $n$). It follows that the inequalities in (\ref{superbound}), (\ref{subbound}), (\ref{bigbound}) and (\ref{smallbound}) hold jointly on a common event whose $P_0$-probability converges to $1$ as $n \rightarrow \infty$. On this common set, denoted by $\tilde{C}_n$, we have that for every ${\bf s} \neq {\bf s}_0$ with $|{\bf s}| \leq n/(\log(\max(n,p)))^{1+\delta}$ 
		$$
		\frac{\Pi_{SS} \left( {\bf s} \mid {\bf Y} \right)}{\Pi_{SS} \left( {\bf s}_0 \mid {\bf Y} \right)} \leq K^{**} f_n^{d({\bf s}, {\bf s}_0)} 
		$$
		
		\noindent
		where 
		$$
		f_n = \min \left( q_n^{1/2}, q_n^{1/(6 |{\bf s}_0|)}, \exp \left( - 0.125 
		n \bar{M} S^2 \right), \exp \left( - \frac{n S^2 \kappa_1 M_1}{64 
			|{\bf s}_0|} \right) \right)
		$$
		
		\noindent
		and $K^{**}$ is a constant not depending on ${\bf s}$ or on $n$. By 
		Assumptions B1 and B4, it follows that $pf_n \rightarrow 0$ as $n \rightarrow \infty$. Hence, 
		\begin{eqnarray*}
			& & \sum_{{\bf s}: {\bf s} \neq {\bf s}_0, |{\bf s}| \leq n/(\log(\max(n,p)))^{1+\delta}} \frac{\Pi \left( {\bf s} \mid {\bf Y} \right)}{\Pi \left( {\bf s}_0 \mid {\bf Y} \right)}\\
			&\leq& K^{**} \sum_{{\bf s}: {\bf s} \neq {\bf s}_0, |{\bf s}| \leq n/(\log(\max(n,p)))^{1+\delta}} f_n^{d({\bf s}, {\bf s}_0)}\\
			&\leq& K^{**} \sum_{j=1}^p \sum_{{\bf s}: d({\bf s}, {\bf s}_0) = j} 
			f_n^{d({\bf s}, {\bf s}_0)}\\
			&\leq& K^{**} \sum_{j=1}^p  (pf_n)^j\\
			&\leq& K^{**} \frac{pf_n}{1 - pf_n} \rightarrow 0 
		\end{eqnarray*}
		
		\noindent
		as $n \rightarrow \infty$. \hfill$\Box$

		\section{Additional Details on Simulation Experiments}
		
		\subsection{Simuation settings}
		
		The following tables present detailed descriptions of the extensive simulation settings (choices of $n$, $p$, error ($\epsilon$) and predictor ($x$) correlation, and error distributions) considered in our experiments. 
		
		\begin{longtable}[t]{lrlll}
			\caption{\label{tab:settings-extremely heavy-normal}Simulation settings for scenarios with data generated from extremely heavy-tailed error distributions and models fitted with a ridge prior on the regression parameters.}\\
			\toprule
			Setting & $n$ & $p$ & Correlation & Error Distribution\\
			\midrule
			\endfirsthead
			\caption[]{Simulation settings for scenarios with data generated from extremely heavy-tailed error distributions and models fitted with a ridge prior on the regression parameters. \textit{(continued)}}\\
			\toprule
			Setting & $n$ & $p$ & Correlation & Error Distribution\\
			\midrule
			\endhead
			
			\endfoot
			\bottomrule
			\endlastfoot
			Setting-1 & 50 & 10 & $x$: 0; $\varepsilon$: 0 & discrete mix $\mathcal{N}(0, 1)$ and $\mathcal{U}(-10^{10}, 10^{10})$ $(90\%; 10\%)$\\
			Setting-2 & 100 & 10 & $x$: 0; $\varepsilon$: 0 & discrete mix $\mathcal{N}(0, 1)$ and $\mathcal{U}(-10^{10}, 10^{10})$ $(90\%; 10\%)$\\
			Setting-3 & 200 & 10 & $x$: 0; $\varepsilon$: 0 & discrete mix $\mathcal{N}(0, 1)$ and $\mathcal{U}(-10^{10}, 10^{10})$ $(90\%; 10\%)$\\
			Setting-4 & 500 & 10 & $x$: 0; $\varepsilon$: 0 & discrete mix $\mathcal{N}(0, 1)$ and $\mathcal{U}(-10^{10}, 10^{10})$ $(90\%; 10\%)$\\
			Setting-5 & 1,000 & 10 & $x$: 0; $\varepsilon$: 0 & discrete mix $\mathcal{N}(0, 1)$ and $\mathcal{U}(-10^{10}, 10^{10})$ $(90\%; 10\%)$\\
			\addlinespace
			Setting-6 & 2,000 & 10 & $x$: 0; $\varepsilon$: 0 & discrete mix $\mathcal{N}(0, 1)$ and $\mathcal{U}(-10^{10}, 10^{10})$ $(90\%; 10\%)$\\
			Setting-7 & 5,000 & 10 & $x$: 0; $\varepsilon$: 0 & discrete mix $\mathcal{N}(0, 1)$ and $\mathcal{U}(-10^{10}, 10^{10})$ $(90\%; 10\%)$\\
			Setting-8 & 10,000 & 10 & $x$: 0; $\varepsilon$: 0 & discrete mix $\mathcal{N}(0, 1)$ and $\mathcal{U}(-10^{10}, 10^{10})$ $(90\%; 10\%)$\\
			Setting-9 & 20,000 & 10 & $x$: 0; $\varepsilon$: 0 & discrete mix $\mathcal{N}(0, 1)$ and $\mathcal{U}(-10^{10}, 10^{10})$ $(90\%; 10\%)$\\
			Setting-10 & 50 & 10 & $x$: 0; $\varepsilon$: 0 & discrete mix $\mathcal{N}(0, 1)$ and $\mathcal{U}(-10^{10}, 10^{10})$ $(50\%; 50\%)$\\
			\addlinespace
			Setting-11 & 100 & 10 & $x$: 0; $\varepsilon$: 0 & discrete mix $\mathcal{N}(0, 1)$ and $\mathcal{U}(-10^{10}, 10^{10})$ $(50\%; 50\%)$\\
			Setting-12 & 200 & 10 & $x$: 0; $\varepsilon$: 0 & discrete mix $\mathcal{N}(0, 1)$ and $\mathcal{U}(-10^{10}, 10^{10})$ $(50\%; 50\%)$\\
			Setting-13 & 500 & 10 & $x$: 0; $\varepsilon$: 0 & discrete mix $\mathcal{N}(0, 1)$ and $\mathcal{U}(-10^{10}, 10^{10})$ $(50\%; 50\%)$\\
			Setting-14 & 1,000 & 10 & $x$: 0; $\varepsilon$: 0 & discrete mix $\mathcal{N}(0, 1)$ and $\mathcal{U}(-10^{10}, 10^{10})$ $(50\%; 50\%)$\\
			Setting-15 & 2,000 & 10 & $x$: 0; $\varepsilon$: 0 & discrete mix $\mathcal{N}(0, 1)$ and $\mathcal{U}(-10^{10}, 10^{10})$ $(50\%; 50\%)$\\
			\addlinespace
			Setting-16 & 5,000 & 10 & $x$: 0; $\varepsilon$: 0 & discrete mix $\mathcal{N}(0, 1)$ and $\mathcal{U}(-10^{10}, 10^{10})$ $(50\%; 50\%)$\\
			Setting-17 & 10,000 & 10 & $x$: 0; $\varepsilon$: 0 & discrete mix $\mathcal{N}(0, 1)$ and $\mathcal{U}(-10^{10}, 10^{10})$ $(50\%; 50\%)$\\
			Setting-18 & 20,000 & 10 & $x$: 0; $\varepsilon$: 0 & discrete mix $\mathcal{N}(0, 1)$ and $\mathcal{U}(-10^{10}, 10^{10})$ $(50\%; 50\%)$\\*
		\end{longtable}

		\begin{longtable}[t]{lllll}
			\caption{\label{tab:settings-heavy-normal}Simulation settings for scenarios with data generated from heavy-tailed error distributions and models fitted with a ridge prior on the regression parameters.}\\
			\toprule
			Setting & $n$ & $p$ & Correlation & Error Distribution\\
			\midrule
			\endfirsthead
			\caption[]{Simulation settings for scenarios with data generated from heavy-tailed error distributions and models fitted with a ridge prior on the regression parameters. \textit{(continued)}}\\
			\toprule
			Setting & $n$ & $p$ & Correlation & Error Distribution\\
			\midrule
			\endhead
			
			\endfoot
			\bottomrule
			\endlastfoot
			Setting-1 & 100 & 20 & $x$: 0.2; $\varepsilon$: 0.3 & discrete mix $\mathcal{N}(0, 1)$ and $\mathcal{N}(0, 10^2)$ $(90\%; 10\%)$\\
			Setting-2 & 100 & 50 & $x$: 0.2; $\varepsilon$: 0.3 & discrete mix $\mathcal{N}(0, 1)$ and $\mathcal{N}(0, 10^2)$ $(90\%; 10\%)$\\
			Setting-3 & 100 & 75 & $x$: 0.2; $\varepsilon$: 0.3 & discrete mix $\mathcal{N}(0, 1)$ and $\mathcal{N}(0, 10^2)$ $(90\%; 10\%)$\\
			Setting-4 & 250 & 50 & $x$: 0.2; $\varepsilon$: 0.3 & discrete mix $\mathcal{N}(0, 1)$ and $\mathcal{N}(0, 10^2)$ $(90\%; 10\%)$\\
			Setting-5 & 250 & 125 & $x$: 0.2; $\varepsilon$: 0.3 & discrete mix $\mathcal{N}(0, 1)$ and $\mathcal{N}(0, 10^2)$ $(90\%; 10\%)$\\
			\addlinespace
			Setting-6 & 250 & 187 & $x$: 0.2; $\varepsilon$: 0.3 & discrete mix $\mathcal{N}(0, 1)$ and $\mathcal{N}(0, 10^2)$ $(90\%; 10\%)$\\
			Setting-7 & 100 & 20 & $x$: 0.2; $\varepsilon$: 0.3 & continuous $t$ with df=1\\
			Setting-8 & 100 & 50 & $x$: 0.2; $\varepsilon$: 0.3 & continuous $t$ with df=1\\
			Setting-9 & 100 & 75 & $x$: 0.2; $\varepsilon$: 0.3 & continuous $t$ with df=1\\
			Setting-10 & 250 & 50 & $x$: 0.2; $\varepsilon$: 0.3 & continuous $t$ with df=1\\
			\addlinespace
			Setting-11 & 250 & 125 & $x$: 0.2; $\varepsilon$: 0.3 & continuous $t$ with df=1\\
			Setting-12 & 250 & 187 & $x$: 0.2; $\varepsilon$: 0.3 & continuous $t$ with df=1\\
			Setting-13 & 100 & 20 & $x$: 0.2; $\varepsilon$: 0.3 & continuous $t$ with df=2\\
			Setting-14 & 100 & 50 & $x$: 0.2; $\varepsilon$: 0.3 & continuous $t$ with df=2\\
			Setting-15 & 100 & 75 & $x$: 0.2; $\varepsilon$: 0.3 & continuous $t$ with df=2\\
			\addlinespace
			Setting-16 & 250 & 50 & $x$: 0.2; $\varepsilon$: 0.3 & continuous $t$ with df=2\\
			Setting-17 & 250 & 125 & $x$: 0.2; $\varepsilon$: 0.3 & continuous $t$ with df=2\\
			Setting-18 & 250 & 187 & $x$: 0.2; $\varepsilon$: 0.3 & continuous $t$ with df=2\\
			Setting-19 & 100 & 20 & $x$: 0.4; $\varepsilon$: 0.6 & continuous $t$ with df=1\\
			Setting-20 & 100 & 50 & $x$: 0.4; $\varepsilon$: 0.6 & continuous $t$ with df=1\\
			\addlinespace
			Setting-21 & 100 & 75 & $x$: 0.4; $\varepsilon$: 0.6 & continuous $t$ with df=1\\
			Setting-22 & 250 & 50 & $x$: 0.4; $\varepsilon$: 0.6 & continuous $t$ with df=1\\
			Setting-23 & 250 & 125 & $x$: 0.4; $\varepsilon$: 0.6 & continuous $t$ with df=1\\
			Setting-24 & 250 & 187 & $x$: 0.4; $\varepsilon$: 0.6 & continuous $t$ with df=1\\
			Setting-25 & 100 & 20 & $x$: 0.4; $\varepsilon$: 0.6 & continuous $t$ with df=2\\
			\addlinespace
			Setting-26 & 100 & 50 & $x$: 0.4; $\varepsilon$: 0.6 & continuous $t$ with df=2\\
			Setting-27 & 100 & 75 & $x$: 0.4; $\varepsilon$: 0.6 & continuous $t$ with df=2\\
			Setting-28 & 250 & 50 & $x$: 0.4; $\varepsilon$: 0.6 & continuous $t$ with df=2\\
			Setting-29 & 250 & 125 & $x$: 0.4; $\varepsilon$: 0.6 & continuous $t$ with df=2\\
			Setting-30 & 250 & 187 & $x$: 0.4; $\varepsilon$: 0.6 & continuous $t$ with df=2\\
			\addlinespace
			Setting-31 & 200 & 20 & $x$: 0.2; $\varepsilon$: 0.3 & discrete mix $\mathcal{N}(0, 1)$ and $\mathcal{N}(0, 10^2)$ $(90\%; 10\%)$\\
			Setting-32 & 200 & 50 & $x$: 0.2; $\varepsilon$: 0.3 & discrete mix $\mathcal{N}(0, 1)$ and $\mathcal{N}(0, 10^2)$ $(90\%; 10\%)$\\
			Setting-33 & 200 & 75 & $x$: 0.2; $\varepsilon$: 0.3 & discrete mix $\mathcal{N}(0, 1)$ and $\mathcal{N}(0, 10^2)$ $(90\%; 10\%)$\\
			Setting-34 & 500 & 50 & $x$: 0.2; $\varepsilon$: 0.3 & discrete mix $\mathcal{N}(0, 1)$ and $\mathcal{N}(0, 10^2)$ $(90\%; 10\%)$\\
			Setting-35 & 500 & 125 & $x$: 0.2; $\varepsilon$: 0.3 & discrete mix $\mathcal{N}(0, 1)$ and $\mathcal{N}(0, 10^2)$ $(90\%; 10\%)$\\
			\addlinespace
			Setting-36 & 500 & 187 & $x$: 0.2; $\varepsilon$: 0.3 & discrete mix $\mathcal{N}(0, 1)$ and $\mathcal{N}(0, 10^2)$ $(90\%; 10\%)$\\
			Setting-37 & 200 & 20 & $x$: 0.2; $\varepsilon$: 0.3 & continuous $t$ with df=1\\
			Setting-38 & 200 & 50 & $x$: 0.2; $\varepsilon$: 0.3 & continuous $t$ with df=1\\
			Setting-39 & 200 & 75 & $x$: 0.2; $\varepsilon$: 0.3 & continuous $t$ with df=1\\
			Setting-40 & 500 & 50 & $x$: 0.2; $\varepsilon$: 0.3 & continuous $t$ with df=1\\
			\addlinespace
			Setting-41 & 500 & 125 & $x$: 0.2; $\varepsilon$: 0.3 & continuous $t$ with df=1\\
			Setting-42 & 500 & 187 & $x$: 0.2; $\varepsilon$: 0.3 & continuous $t$ with df=1\\
			Setting-43 & 200 & 20 & $x$: 0.2; $\varepsilon$: 0.3 & continuous $t$ with df=2\\
			Setting-44 & 200 & 50 & $x$: 0.2; $\varepsilon$: 0.3 & continuous $t$ with df=2\\
			Setting-45 & 200 & 75 & $x$: 0.2; $\varepsilon$: 0.3 & continuous $t$ with df=2\\
			\addlinespace
			Setting-46 & 500 & 50 & $x$: 0.2; $\varepsilon$: 0.3 & continuous $t$ with df=2\\
			Setting-47 & 500 & 125 & $x$: 0.2; $\varepsilon$: 0.3 & continuous $t$ with df=2\\
			Setting-48 & 500 & 187 & $x$: 0.2; $\varepsilon$: 0.3 & continuous $t$ with df=2\\
			Setting-49 & 200 & 20 & $x$: 0.4; $\varepsilon$: 0.6 & discrete mix $\mathcal{N}(0, 1)$ and $\mathcal{N}(0, 10^2)$ $(90\%; 10\%)$\\
			Setting-50 & 200 & 50 & $x$: 0.4; $\varepsilon$: 0.6 & discrete mix $\mathcal{N}(0, 1)$ and $\mathcal{N}(0, 10^2)$ $(90\%; 10\%)$\\
			\addlinespace
			Setting-51 & 200 & 75 & $x$: 0.4; $\varepsilon$: 0.6 & discrete mix $\mathcal{N}(0, 1)$ and $\mathcal{N}(0, 10^2)$ $(90\%; 10\%)$\\
			Setting-52 & 500 & 50 & $x$: 0.4; $\varepsilon$: 0.6 & discrete mix $\mathcal{N}(0, 1)$ and $\mathcal{N}(0, 10^2)$ $(90\%; 10\%)$\\
			Setting-53 & 500 & 125 & $x$: 0.4; $\varepsilon$: 0.6 & discrete mix $\mathcal{N}(0, 1)$ and $\mathcal{N}(0, 10^2)$ $(90\%; 10\%)$\\
			Setting-54 & 500 & 187 & $x$: 0.4; $\varepsilon$: 0.6 & discrete mix $\mathcal{N}(0, 1)$ and $\mathcal{N}(0, 10^2)$ $(90\%; 10\%)$\\
			Setting-55 & 200 & 20 & $x$: 0.4; $\varepsilon$: 0.6 & continuous $t$ with df=1\\
			\addlinespace
			Setting-56 & 200 & 50 & $x$: 0.4; $\varepsilon$: 0.6 & continuous $t$ with df=1\\
			Setting-57 & 200 & 75 & $x$: 0.4; $\varepsilon$: 0.6 & continuous $t$ with df=1\\
			Setting-58 & 500 & 50 & $x$: 0.4; $\varepsilon$: 0.6 & continuous $t$ with df=1\\
			Setting-59 & 500 & 125 & $x$: 0.4; $\varepsilon$: 0.6 & continuous $t$ with df=1\\
			Setting-60 & 500 & 187 & $x$: 0.4; $\varepsilon$: 0.6 & continuous $t$ with df=1\\
			\addlinespace
			Setting-61 & 200 & 20 & $x$: 0.4; $\varepsilon$: 0.6 & continuous $t$ with df=2\\
			Setting-62 & 200 & 50 & $x$: 0.4; $\varepsilon$: 0.6 & continuous $t$ with df=2\\
			Setting-63 & 200 & 75 & $x$: 0.4; $\varepsilon$: 0.6 & continuous $t$ with df=2\\
			Setting-64 & 500 & 50 & $x$: 0.4; $\varepsilon$: 0.6 & continuous $t$ with df=2\\
			Setting-65 & 500 & 125 & $x$: 0.4; $\varepsilon$: 0.6 & continuous $t$ with df=2\\
			\addlinespace
			Setting-66 & 500 & 187 & $x$: 0.4; $\varepsilon$: 0.6 & continuous $t$ with df=2\\
			Setting-67 & 50 & 10 & $x$: 0; $\varepsilon$: 0 & discrete mix $\mathcal{N}(0, 1)$ and $\mathcal{C}(0, 10)$ $(90\%; 10\%)$\\
			Setting-68 & 100 & 10 & $x$: 0; $\varepsilon$: 0 & discrete mix $\mathcal{N}(0, 1)$ and $\mathcal{C}(0, 10)$ $(90\%; 10\%)$\\
			Setting-69 & 200 & 10 & $x$: 0; $\varepsilon$: 0 & discrete mix $\mathcal{N}(0, 1)$ and $\mathcal{C}(0, 10)$ $(90\%; 10\%)$\\
			Setting-70 & 500 & 10 & $x$: 0; $\varepsilon$: 0 & discrete mix $\mathcal{N}(0, 1)$ and $\mathcal{C}(0, 10)$ $(90\%; 10\%)$\\
			\addlinespace
			Setting-71 & 1,000 & 10 & $x$: 0; $\varepsilon$: 0 & discrete mix $\mathcal{N}(0, 1)$ and $\mathcal{C}(0, 10)$ $(90\%; 10\%)$\\
			Setting-72 & 2,000 & 10 & $x$: 0; $\varepsilon$: 0 & discrete mix $\mathcal{N}(0, 1)$ and $\mathcal{C}(0, 10)$ $(90\%; 10\%)$\\
			Setting-73 & 5,000 & 10 & $x$: 0; $\varepsilon$: 0 & discrete mix $\mathcal{N}(0, 1)$ and $\mathcal{C}(0, 10)$ $(90\%; 10\%)$\\
			Setting-74 & 10,000 & 10 & $x$: 0; $\varepsilon$: 0 & discrete mix $\mathcal{N}(0, 1)$ and $\mathcal{C}(0, 10)$ $(90\%; 10\%)$\\
			Setting-75 & 20,000 & 10 & $x$: 0; $\varepsilon$: 0 & discrete mix $\mathcal{N}(0, 1)$ and $\mathcal{C}(0, 10)$ $(90\%; 10\%)$\\
			\addlinespace
			Setting-76 & 50 & 10 & $x$: 0; $\varepsilon$: 0 & continuous $t$ with df=1\\
			Setting-77 & 100 & 10 & $x$: 0; $\varepsilon$: 0 & continuous $t$ with df=1\\
			Setting-78 & 200 & 10 & $x$: 0; $\varepsilon$: 0 & continuous $t$ with df=1\\
			Setting-79 & 500 & 10 & $x$: 0; $\varepsilon$: 0 & continuous $t$ with df=1\\
			Setting-80 & 1,000 & 10 & $x$: 0; $\varepsilon$: 0 & continuous $t$ with df=1\\
			\addlinespace
			Setting-81 & 2,000 & 10 & $x$: 0; $\varepsilon$: 0 & continuous $t$ with df=1\\
			Setting-82 & 5,000 & 10 & $x$: 0; $\varepsilon$: 0 & continuous $t$ with df=1\\
			Setting-83 & 10,000 & 10 & $x$: 0; $\varepsilon$: 0 & continuous $t$ with df=1\\
			Setting-84 & 20,000 & 10 & $x$: 0; $\varepsilon$: 0 & continuous $t$ with df=1\\
			Setting-85 & 50 & 10 & $x$: 0; $\varepsilon$: 0 & continuous $t$ with df=2\\
			\addlinespace
			Setting-86 & 100 & 10 & $x$: 0; $\varepsilon$: 0 & continuous $t$ with df=2\\
			Setting-87 & 200 & 10 & $x$: 0; $\varepsilon$: 0 & continuous $t$ with df=2\\
			Setting-88 & 500 & 10 & $x$: 0; $\varepsilon$: 0 & continuous $t$ with df=2\\
			Setting-89 & 1,000 & 10 & $x$: 0; $\varepsilon$: 0 & continuous $t$ with df=2\\
			Setting-90 & 2,000 & 10 & $x$: 0; $\varepsilon$: 0 & continuous $t$ with df=2\\
			\addlinespace
			Setting-91 & 5,000 & 10 & $x$: 0; $\varepsilon$: 0 & continuous $t$ with df=2\\
			Setting-92 & 10,000 & 10 & $x$: 0; $\varepsilon$: 0 & continuous $t$ with df=2\\
			Setting-93 & 20,000 & 10 & $x$: 0; $\varepsilon$: 0 & continuous $t$ with df=2\\*
		\end{longtable}

		\begin{longtable}[t]{lllll}
			\caption{\label{tab:settings-moderate-normal}Simulation settings for scenarios with data generated from moderate-tailed error distributions and models fitted with a ridge prior on the regression parameters.}\\
			\toprule
			Setting & $n$ & $p$ & Correlation & Error Distribution\\
			\midrule
			\endfirsthead
			\caption[]{Simulation settings for scenarios with data generated from moderate-tailed error distributions and models fitted with a ridge prior on the regression parameters. \textit{(continued)}}\\
			\toprule
			Setting & $n$ & $p$ & Correlation & Error Distribution\\
			\midrule
			\endhead
			
			\endfoot
			\bottomrule
			\endlastfoot
			Setting-1 & 100 & 20 & $x$: 0.2; $\varepsilon$: 0.3 & discrete mix $\mathcal{N}(0, 1)$ and $\mathcal{N}(0, 10^2)$ $(99\%; 1\%)$\\
			Setting-2 & 100 & 50 & $x$: 0.2; $\varepsilon$: 0.3 & discrete mix $\mathcal{N}(0, 1)$ and $\mathcal{N}(0, 10^2)$ $(99\%; 1\%)$\\
			Setting-3 & 100 & 75 & $x$: 0.2; $\varepsilon$: 0.3 & discrete mix $\mathcal{N}(0, 1)$ and $\mathcal{N}(0, 10^2)$ $(99\%; 1\%)$\\
			Setting-4 & 250 & 50 & $x$: 0.2; $\varepsilon$: 0.3 & discrete mix $\mathcal{N}(0, 1)$ and $\mathcal{N}(0, 10^2)$ $(99\%; 1\%)$\\
			Setting-5 & 250 & 125 & $x$: 0.2; $\varepsilon$: 0.3 & discrete mix $\mathcal{N}(0, 1)$ and $\mathcal{N}(0, 10^2)$ $(99\%; 1\%)$\\
			\addlinespace
			Setting-6 & 250 & 187 & $x$: 0.2; $\varepsilon$: 0.3 & discrete mix $\mathcal{N}(0, 1)$ and $\mathcal{N}(0, 10^2)$ $(99\%; 1\%)$\\
			Setting-7 & 100 & 20 & $x$: 0.2; $\varepsilon$: 0.3 & discrete mix $\mathcal{N}(0, 1)$ and $\mathcal{N}(0, 10^2)$ $(95\%; 5\%)$\\
			Setting-8 & 100 & 50 & $x$: 0.2; $\varepsilon$: 0.3 & discrete mix $\mathcal{N}(0, 1)$ and $\mathcal{N}(0, 10^2)$ $(95\%; 5\%)$\\
			Setting-9 & 100 & 75 & $x$: 0.2; $\varepsilon$: 0.3 & discrete mix $\mathcal{N}(0, 1)$ and $\mathcal{N}(0, 10^2)$ $(95\%; 5\%)$\\
			Setting-10 & 250 & 50 & $x$: 0.2; $\varepsilon$: 0.3 & discrete mix $\mathcal{N}(0, 1)$ and $\mathcal{N}(0, 10^2)$ $(95\%; 5\%)$\\
			\addlinespace
			Setting-11 & 250 & 125 & $x$: 0.2; $\varepsilon$: 0.3 & discrete mix $\mathcal{N}(0, 1)$ and $\mathcal{N}(0, 10^2)$ $(95\%; 5\%)$\\
			Setting-12 & 250 & 187 & $x$: 0.2; $\varepsilon$: 0.3 & discrete mix $\mathcal{N}(0, 1)$ and $\mathcal{N}(0, 10^2)$ $(95\%; 5\%)$\\
			Setting-13 & 100 & 20 & $x$: 0.2; $\varepsilon$: 0.3 & continuous $t$ with df=4\\
			Setting-14 & 100 & 50 & $x$: 0.2; $\varepsilon$: 0.3 & continuous $t$ with df=4\\
			Setting-15 & 100 & 75 & $x$: 0.2; $\varepsilon$: 0.3 & continuous $t$ with df=4\\
			\addlinespace
			Setting-16 & 250 & 50 & $x$: 0.2; $\varepsilon$: 0.3 & continuous $t$ with df=4\\
			Setting-17 & 250 & 125 & $x$: 0.2; $\varepsilon$: 0.3 & continuous $t$ with df=4\\
			Setting-18 & 250 & 187 & $x$: 0.2; $\varepsilon$: 0.3 & continuous $t$ with df=4\\
			Setting-19 & 100 & 20 & $x$: 0.2; $\varepsilon$: 0.3 & continuous $t$ with df=8\\
			Setting-20 & 100 & 50 & $x$: 0.2; $\varepsilon$: 0.3 & continuous $t$ with df=8\\
			\addlinespace
			Setting-21 & 100 & 75 & $x$: 0.2; $\varepsilon$: 0.3 & continuous $t$ with df=8\\
			Setting-22 & 250 & 50 & $x$: 0.2; $\varepsilon$: 0.3 & continuous $t$ with df=8\\
			Setting-23 & 250 & 125 & $x$: 0.2; $\varepsilon$: 0.3 & continuous $t$ with df=8\\
			Setting-24 & 250 & 187 & $x$: 0.2; $\varepsilon$: 0.3 & continuous $t$ with df=8\\
			Setting-25 & 100 & 20 & $x$: 0.4; $\varepsilon$: 0.6 & discrete mix $\mathcal{N}(0, 1)$ and $\mathcal{N}(0, 10^2)$ $(95\%; 5\%)$\\
			\addlinespace
			Setting-26 & 100 & 50 & $x$: 0.4; $\varepsilon$: 0.6 & discrete mix $\mathcal{N}(0, 1)$ and $\mathcal{N}(0, 10^2)$ $(95\%; 5\%)$\\
			Setting-27 & 100 & 75 & $x$: 0.4; $\varepsilon$: 0.6 & discrete mix $\mathcal{N}(0, 1)$ and $\mathcal{N}(0, 10^2)$ $(95\%; 5\%)$\\
			Setting-28 & 250 & 50 & $x$: 0.4; $\varepsilon$: 0.6 & discrete mix $\mathcal{N}(0, 1)$ and $\mathcal{N}(0, 10^2)$ $(95\%; 5\%)$\\
			Setting-29 & 250 & 125 & $x$: 0.4; $\varepsilon$: 0.6 & discrete mix $\mathcal{N}(0, 1)$ and $\mathcal{N}(0, 10^2)$ $(95\%; 5\%)$\\
			Setting-30 & 250 & 187 & $x$: 0.4; $\varepsilon$: 0.6 & discrete mix $\mathcal{N}(0, 1)$ and $\mathcal{N}(0, 10^2)$ $(95\%; 5\%)$\\
			\addlinespace
			Setting-31 & 100 & 20 & $x$: 0.4; $\varepsilon$: 0.6 & continuous $t$ with df=4\\
			Setting-32 & 100 & 50 & $x$: 0.4; $\varepsilon$: 0.6 & continuous $t$ with df=4\\
			Setting-33 & 100 & 75 & $x$: 0.4; $\varepsilon$: 0.6 & continuous $t$ with df=4\\
			Setting-34 & 250 & 50 & $x$: 0.4; $\varepsilon$: 0.6 & continuous $t$ with df=4\\
			Setting-35 & 250 & 125 & $x$: 0.4; $\varepsilon$: 0.6 & continuous $t$ with df=4\\
			\addlinespace
			Setting-36 & 250 & 187 & $x$: 0.4; $\varepsilon$: 0.6 & continuous $t$ with df=4\\
			Setting-37 & 100 & 20 & $x$: 0.4; $\varepsilon$: 0.6 & continuous $t$ with df=8\\
			Setting-38 & 100 & 50 & $x$: 0.4; $\varepsilon$: 0.6 & continuous $t$ with df=8\\
			Setting-39 & 100 & 75 & $x$: 0.4; $\varepsilon$: 0.6 & continuous $t$ with df=8\\
			Setting-40 & 250 & 50 & $x$: 0.4; $\varepsilon$: 0.6 & continuous $t$ with df=8\\
			\addlinespace
			Setting-41 & 250 & 125 & $x$: 0.4; $\varepsilon$: 0.6 & continuous $t$ with df=8\\
			Setting-42 & 250 & 187 & $x$: 0.4; $\varepsilon$: 0.6 & continuous $t$ with df=8\\
			Setting-43 & 200 & 20 & $x$: 0.2; $\varepsilon$: 0.3 & discrete mix $\mathcal{N}(0, 1)$ and $\mathcal{N}(0, 10^2)$ $(99\%; 1\%)$\\
			Setting-44 & 200 & 50 & $x$: 0.2; $\varepsilon$: 0.3 & discrete mix $\mathcal{N}(0, 1)$ and $\mathcal{N}(0, 10^2)$ $(99\%; 1\%)$\\
			Setting-45 & 200 & 75 & $x$: 0.2; $\varepsilon$: 0.3 & discrete mix $\mathcal{N}(0, 1)$ and $\mathcal{N}(0, 10^2)$ $(99\%; 1\%)$\\
			\addlinespace
			Setting-46 & 500 & 50 & $x$: 0.2; $\varepsilon$: 0.3 & discrete mix $\mathcal{N}(0, 1)$ and $\mathcal{N}(0, 10^2)$ $(99\%; 1\%)$\\
			Setting-47 & 500 & 125 & $x$: 0.2; $\varepsilon$: 0.3 & discrete mix $\mathcal{N}(0, 1)$ and $\mathcal{N}(0, 10^2)$ $(99\%; 1\%)$\\
			Setting-48 & 500 & 187 & $x$: 0.2; $\varepsilon$: 0.3 & discrete mix $\mathcal{N}(0, 1)$ and $\mathcal{N}(0, 10^2)$ $(99\%; 1\%)$\\
			Setting-49 & 200 & 20 & $x$: 0.2; $\varepsilon$: 0.3 & discrete mix $\mathcal{N}(0, 1)$ and $\mathcal{N}(0, 10^2)$ $(95\%; 5\%)$\\
			Setting-50 & 200 & 50 & $x$: 0.2; $\varepsilon$: 0.3 & discrete mix $\mathcal{N}(0, 1)$ and $\mathcal{N}(0, 10^2)$ $(95\%; 5\%)$\\
			\addlinespace
			Setting-51 & 200 & 75 & $x$: 0.2; $\varepsilon$: 0.3 & discrete mix $\mathcal{N}(0, 1)$ and $\mathcal{N}(0, 10^2)$ $(95\%; 5\%)$\\
			Setting-52 & 500 & 50 & $x$: 0.2; $\varepsilon$: 0.3 & discrete mix $\mathcal{N}(0, 1)$ and $\mathcal{N}(0, 10^2)$ $(95\%; 5\%)$\\
			Setting-53 & 500 & 125 & $x$: 0.2; $\varepsilon$: 0.3 & discrete mix $\mathcal{N}(0, 1)$ and $\mathcal{N}(0, 10^2)$ $(95\%; 5\%)$\\
			Setting-54 & 500 & 187 & $x$: 0.2; $\varepsilon$: 0.3 & discrete mix $\mathcal{N}(0, 1)$ and $\mathcal{N}(0, 10^2)$ $(95\%; 5\%)$\\
			Setting-55 & 200 & 20 & $x$: 0.2; $\varepsilon$: 0.3 & continuous $t$ with df=4\\
			\addlinespace
			Setting-56 & 200 & 50 & $x$: 0.2; $\varepsilon$: 0.3 & continuous $t$ with df=4\\
			Setting-57 & 200 & 75 & $x$: 0.2; $\varepsilon$: 0.3 & continuous $t$ with df=4\\
			Setting-58 & 500 & 50 & $x$: 0.2; $\varepsilon$: 0.3 & continuous $t$ with df=4\\
			Setting-59 & 500 & 125 & $x$: 0.2; $\varepsilon$: 0.3 & continuous $t$ with df=4\\
			Setting-60 & 500 & 187 & $x$: 0.2; $\varepsilon$: 0.3 & continuous $t$ with df=4\\
			\addlinespace
			Setting-61 & 200 & 20 & $x$: 0.2; $\varepsilon$: 0.3 & continuous $t$ with df=8\\
			Setting-62 & 200 & 50 & $x$: 0.2; $\varepsilon$: 0.3 & continuous $t$ with df=8\\
			Setting-63 & 200 & 75 & $x$: 0.2; $\varepsilon$: 0.3 & continuous $t$ with df=8\\
			Setting-64 & 500 & 50 & $x$: 0.2; $\varepsilon$: 0.3 & continuous $t$ with df=8\\
			Setting-65 & 500 & 125 & $x$: 0.2; $\varepsilon$: 0.3 & continuous $t$ with df=8\\
			\addlinespace
			Setting-66 & 500 & 187 & $x$: 0.2; $\varepsilon$: 0.3 & continuous $t$ with df=8\\
			Setting-67 & 200 & 20 & $x$: 0.4; $\varepsilon$: 0.6 & discrete mix $\mathcal{N}(0, 1)$ and $\mathcal{N}(0, 10^2)$ $(99\%; 1\%)$\\
			Setting-68 & 200 & 50 & $x$: 0.4; $\varepsilon$: 0.6 & discrete mix $\mathcal{N}(0, 1)$ and $\mathcal{N}(0, 10^2)$ $(99\%; 1\%)$\\
			Setting-69 & 200 & 75 & $x$: 0.4; $\varepsilon$: 0.6 & discrete mix $\mathcal{N}(0, 1)$ and $\mathcal{N}(0, 10^2)$ $(99\%; 1\%)$\\
			Setting-70 & 500 & 50 & $x$: 0.4; $\varepsilon$: 0.6 & discrete mix $\mathcal{N}(0, 1)$ and $\mathcal{N}(0, 10^2)$ $(99\%; 1\%)$\\
			\addlinespace
			Setting-71 & 500 & 125 & $x$: 0.4; $\varepsilon$: 0.6 & discrete mix $\mathcal{N}(0, 1)$ and $\mathcal{N}(0, 10^2)$ $(99\%; 1\%)$\\
			Setting-72 & 500 & 187 & $x$: 0.4; $\varepsilon$: 0.6 & discrete mix $\mathcal{N}(0, 1)$ and $\mathcal{N}(0, 10^2)$ $(99\%; 1\%)$\\
			Setting-73 & 200 & 20 & $x$: 0.4; $\varepsilon$: 0.6 & discrete mix $\mathcal{N}(0, 1)$ and $\mathcal{N}(0, 10^2)$ $(95\%; 5\%)$\\
			Setting-74 & 200 & 50 & $x$: 0.4; $\varepsilon$: 0.6 & discrete mix $\mathcal{N}(0, 1)$ and $\mathcal{N}(0, 10^2)$ $(95\%; 5\%)$\\
			Setting-75 & 200 & 75 & $x$: 0.4; $\varepsilon$: 0.6 & discrete mix $\mathcal{N}(0, 1)$ and $\mathcal{N}(0, 10^2)$ $(95\%; 5\%)$\\
			\addlinespace
			Setting-76 & 500 & 50 & $x$: 0.4; $\varepsilon$: 0.6 & discrete mix $\mathcal{N}(0, 1)$ and $\mathcal{N}(0, 10^2)$ $(95\%; 5\%)$\\
			Setting-77 & 500 & 125 & $x$: 0.4; $\varepsilon$: 0.6 & discrete mix $\mathcal{N}(0, 1)$ and $\mathcal{N}(0, 10^2)$ $(95\%; 5\%)$\\
			Setting-78 & 500 & 187 & $x$: 0.4; $\varepsilon$: 0.6 & discrete mix $\mathcal{N}(0, 1)$ and $\mathcal{N}(0, 10^2)$ $(95\%; 5\%)$\\
			Setting-79 & 200 & 20 & $x$: 0.4; $\varepsilon$: 0.6 & continuous $t$ with df=4\\
			Setting-80 & 200 & 50 & $x$: 0.4; $\varepsilon$: 0.6 & continuous $t$ with df=4\\
			\addlinespace
			Setting-81 & 200 & 75 & $x$: 0.4; $\varepsilon$: 0.6 & continuous $t$ with df=4\\
			Setting-82 & 500 & 50 & $x$: 0.4; $\varepsilon$: 0.6 & continuous $t$ with df=4\\
			Setting-83 & 500 & 125 & $x$: 0.4; $\varepsilon$: 0.6 & continuous $t$ with df=4\\
			Setting-84 & 500 & 187 & $x$: 0.4; $\varepsilon$: 0.6 & continuous $t$ with df=4\\
			Setting-85 & 200 & 20 & $x$: 0.4; $\varepsilon$: 0.6 & continuous $t$ with df=8\\
			\addlinespace
			Setting-86 & 200 & 50 & $x$: 0.4; $\varepsilon$: 0.6 & continuous $t$ with df=8\\
			Setting-87 & 200 & 75 & $x$: 0.4; $\varepsilon$: 0.6 & continuous $t$ with df=8\\
			Setting-88 & 500 & 50 & $x$: 0.4; $\varepsilon$: 0.6 & continuous $t$ with df=8\\
			Setting-89 & 500 & 125 & $x$: 0.4; $\varepsilon$: 0.6 & continuous $t$ with df=8\\
			Setting-90 & 500 & 187 & $x$: 0.4; $\varepsilon$: 0.6 & continuous $t$ with df=8\\
			\addlinespace
			Setting-91 & 50 & 10 & $x$: 0; $\varepsilon$: 0 & discrete mix $\mathcal{N}(0, 1)$ and $\mathcal{C}(0, 10)$ $(99\%; 1\%)$\\
			Setting-92 & 100 & 10 & $x$: 0; $\varepsilon$: 0 & discrete mix $\mathcal{N}(0, 1)$ and $\mathcal{C}(0, 10)$ $(99\%; 1\%)$\\
			Setting-93 & 200 & 10 & $x$: 0; $\varepsilon$: 0 & discrete mix $\mathcal{N}(0, 1)$ and $\mathcal{C}(0, 10)$ $(99\%; 1\%)$\\
			Setting-94 & 500 & 10 & $x$: 0; $\varepsilon$: 0 & discrete mix $\mathcal{N}(0, 1)$ and $\mathcal{C}(0, 10)$ $(99\%; 1\%)$\\
			Setting-95 & 1,000 & 10 & $x$: 0; $\varepsilon$: 0 & discrete mix $\mathcal{N}(0, 1)$ and $\mathcal{C}(0, 10)$ $(99\%; 1\%)$\\
			\addlinespace
			Setting-96 & 2,000 & 10 & $x$: 0; $\varepsilon$: 0 & discrete mix $\mathcal{N}(0, 1)$ and $\mathcal{C}(0, 10)$ $(99\%; 1\%)$\\
			Setting-97 & 5,000 & 10 & $x$: 0; $\varepsilon$: 0 & discrete mix $\mathcal{N}(0, 1)$ and $\mathcal{C}(0, 10)$ $(99\%; 1\%)$\\
			Setting-98 & 10,000 & 10 & $x$: 0; $\varepsilon$: 0 & discrete mix $\mathcal{N}(0, 1)$ and $\mathcal{C}(0, 10)$ $(99\%; 1\%)$\\
			Setting-99 & 20,000 & 10 & $x$: 0; $\varepsilon$: 0 & discrete mix $\mathcal{N}(0, 1)$ and $\mathcal{C}(0, 10)$ $(99\%; 1\%)$\\
			Setting-100 & 50 & 10 & $x$: 0; $\varepsilon$: 0 & discrete mix $\mathcal{N}(0, 1)$ and $\mathcal{C}(0, 10)$ $(95\%; 5\%)$\\
			\addlinespace
			Setting-101 & 100 & 10 & $x$: 0; $\varepsilon$: 0 & discrete mix $\mathcal{N}(0, 1)$ and $\mathcal{C}(0, 10)$ $(95\%; 5\%)$\\
			Setting-102 & 200 & 10 & $x$: 0; $\varepsilon$: 0 & discrete mix $\mathcal{N}(0, 1)$ and $\mathcal{C}(0, 10)$ $(95\%; 5\%)$\\
			Setting-103 & 500 & 10 & $x$: 0; $\varepsilon$: 0 & discrete mix $\mathcal{N}(0, 1)$ and $\mathcal{C}(0, 10)$ $(95\%; 5\%)$\\
			Setting-104 & 1,000 & 10 & $x$: 0; $\varepsilon$: 0 & discrete mix $\mathcal{N}(0, 1)$ and $\mathcal{C}(0, 10)$ $(95\%; 5\%)$\\
			Setting-105 & 2,000 & 10 & $x$: 0; $\varepsilon$: 0 & discrete mix $\mathcal{N}(0, 1)$ and $\mathcal{C}(0, 10)$ $(95\%; 5\%)$\\
			\addlinespace
			Setting-106 & 5,000 & 10 & $x$: 0; $\varepsilon$: 0 & discrete mix $\mathcal{N}(0, 1)$ and $\mathcal{C}(0, 10)$ $(95\%; 5\%)$\\
			Setting-107 & 10,000 & 10 & $x$: 0; $\varepsilon$: 0 & discrete mix $\mathcal{N}(0, 1)$ and $\mathcal{C}(0, 10)$ $(95\%; 5\%)$\\
			Setting-108 & 20,000 & 10 & $x$: 0; $\varepsilon$: 0 & discrete mix $\mathcal{N}(0, 1)$ and $\mathcal{C}(0, 10)$ $(95\%; 5\%)$\\
			Setting-109 & 50 & 10 & $x$: 0; $\varepsilon$: 0 & continuous $t$ with df=4\\
			Setting-110 & 100 & 10 & $x$: 0; $\varepsilon$: 0 & continuous $t$ with df=4\\
			\addlinespace
			Setting-111 & 200 & 10 & $x$: 0; $\varepsilon$: 0 & continuous $t$ with df=4\\
			Setting-112 & 500 & 10 & $x$: 0; $\varepsilon$: 0 & continuous $t$ with df=4\\
			Setting-113 & 1,000 & 10 & $x$: 0; $\varepsilon$: 0 & continuous $t$ with df=4\\
			Setting-114 & 2,000 & 10 & $x$: 0; $\varepsilon$: 0 & continuous $t$ with df=4\\
			Setting-115 & 5,000 & 10 & $x$: 0; $\varepsilon$: 0 & continuous $t$ with df=4\\
			\addlinespace
			Setting-116 & 10,000 & 10 & $x$: 0; $\varepsilon$: 0 & continuous $t$ with df=4\\
			Setting-117 & 20,000 & 10 & $x$: 0; $\varepsilon$: 0 & continuous $t$ with df=4\\
			Setting-118 & 50 & 10 & $x$: 0; $\varepsilon$: 0 & continuous $t$ with df=8\\
			Setting-119 & 100 & 10 & $x$: 0; $\varepsilon$: 0 & continuous $t$ with df=8\\
			Setting-120 & 200 & 10 & $x$: 0; $\varepsilon$: 0 & continuous $t$ with df=8\\
			\addlinespace
			Setting-121 & 500 & 10 & $x$: 0; $\varepsilon$: 0 & continuous $t$ with df=8\\
			Setting-122 & 1,000 & 10 & $x$: 0; $\varepsilon$: 0 & continuous $t$ with df=8\\
			Setting-123 & 2,000 & 10 & $x$: 0; $\varepsilon$: 0 & continuous $t$ with df=8\\
			Setting-124 & 5,000 & 10 & $x$: 0; $\varepsilon$: 0 & continuous $t$ with df=8\\
			Setting-125 & 10,000 & 10 & $x$: 0; $\varepsilon$: 0 & continuous $t$ with df=8\\
			\addlinespace
			Setting-126 & 20,000 & 10 & $x$: 0; $\varepsilon$: 0 & continuous $t$ with df=8\\*
		\end{longtable}

		\begin{longtable}[t]{lllll}
			\caption{\label{tab:settings-thin-normal}Simulation settings for scenarios with data generated from thin-tailed error distributions and models fitted with a ridge prior on the regression parameters.}\\
			\toprule
			Setting & $n$ & $p$ & Correlation & Error Distribution\\
			\midrule
			\endfirsthead
			\caption[]{Simulation settings for scenarios with data generated from thin-tailed error distributions and models fitted with a ridge prior on the regression parameters. \textit{(continued)}}\\
			\toprule
			Setting & $n$ & $p$ & Correlation & Error Distribution\\
			\midrule
			\endhead
			
			\endfoot
			\bottomrule
			\endlastfoot
			Setting-1 & 100 & 20 & $x$: 0.2; $\varepsilon$: 0.3 & continuous $\mathcal{N}(0, 1)$\\
			Setting-2 & 100 & 50 & $x$: 0.2; $\varepsilon$: 0.3 & continuous $\mathcal{N}(0, 1)$\\
			Setting-3 & 100 & 75 & $x$: 0.2; $\varepsilon$: 0.3 & continuous $\mathcal{N}(0, 1)$\\
			Setting-4 & 250 & 50 & $x$: 0.2; $\varepsilon$: 0.3 & continuous $\mathcal{N}(0, 1)$\\
			Setting-5 & 250 & 125 & $x$: 0.2; $\varepsilon$: 0.3 & continuous $\mathcal{N}(0, 1)$\\
			\addlinespace
			Setting-6 & 250 & 187 & $x$: 0.2; $\varepsilon$: 0.3 & continuous $\mathcal{N}(0, 1)$\\
			Setting-7 & 100 & 20 & $x$: 0.4; $\varepsilon$: 0.6 & continuous $\mathcal{N}(0, 1)$\\
			Setting-8 & 100 & 50 & $x$: 0.4; $\varepsilon$: 0.6 & continuous $\mathcal{N}(0, 1)$\\
			Setting-9 & 100 & 75 & $x$: 0.4; $\varepsilon$: 0.6 & continuous $\mathcal{N}(0, 1)$\\
			Setting-10 & 250 & 50 & $x$: 0.4; $\varepsilon$: 0.6 & continuous $\mathcal{N}(0, 1)$\\
			\addlinespace
			Setting-11 & 250 & 125 & $x$: 0.4; $\varepsilon$: 0.6 & continuous $\mathcal{N}(0, 1)$\\
			Setting-12 & 250 & 187 & $x$: 0.4; $\varepsilon$: 0.6 & continuous $\mathcal{N}(0, 1)$\\
			Setting-13 & 200 & 20 & $x$: 0.2; $\varepsilon$: 0.3 & continuous $\mathcal{N}(0, 1)$\\
			Setting-14 & 200 & 50 & $x$: 0.2; $\varepsilon$: 0.3 & continuous $\mathcal{N}(0, 1)$\\
			Setting-15 & 200 & 75 & $x$: 0.2; $\varepsilon$: 0.3 & continuous $\mathcal{N}(0, 1)$\\
			\addlinespace
			Setting-16 & 500 & 50 & $x$: 0.2; $\varepsilon$: 0.3 & continuous $\mathcal{N}(0, 1)$\\
			Setting-17 & 500 & 125 & $x$: 0.2; $\varepsilon$: 0.3 & continuous $\mathcal{N}(0, 1)$\\
			Setting-18 & 500 & 187 & $x$: 0.2; $\varepsilon$: 0.3 & continuous $\mathcal{N}(0, 1)$\\
			Setting-19 & 200 & 20 & $x$: 0.4; $\varepsilon$: 0.6 & continuous $\mathcal{N}(0, 1)$\\
			Setting-20 & 200 & 50 & $x$: 0.4; $\varepsilon$: 0.6 & continuous $\mathcal{N}(0, 1)$\\
			\addlinespace
			Setting-21 & 200 & 75 & $x$: 0.4; $\varepsilon$: 0.6 & continuous $\mathcal{N}(0, 1)$\\
			Setting-22 & 500 & 50 & $x$: 0.4; $\varepsilon$: 0.6 & continuous $\mathcal{N}(0, 1)$\\
			Setting-23 & 500 & 125 & $x$: 0.4; $\varepsilon$: 0.6 & continuous $\mathcal{N}(0, 1)$\\
			Setting-24 & 500 & 187 & $x$: 0.4; $\varepsilon$: 0.6 & continuous $\mathcal{N}(0, 1)$\\
			Setting-25 & 50 & 10 & $x$: 0; $\varepsilon$: 0 & continuous $\mathcal{N}(0, 1)$\\
			\addlinespace
			Setting-26 & 100 & 10 & $x$: 0; $\varepsilon$: 0 & continuous $\mathcal{N}(0, 1)$\\
			Setting-27 & 200 & 10 & $x$: 0; $\varepsilon$: 0 & continuous $\mathcal{N}(0, 1)$\\
			Setting-28 & 500 & 10 & $x$: 0; $\varepsilon$: 0 & continuous $\mathcal{N}(0, 1)$\\
			Setting-29 & 1,000 & 10 & $x$: 0; $\varepsilon$: 0 & continuous $\mathcal{N}(0, 1)$\\
			Setting-30 & 2,000 & 10 & $x$: 0; $\varepsilon$: 0 & continuous $\mathcal{N}(0, 1)$\\
			\addlinespace
			Setting-31 & 5,000 & 10 & $x$: 0; $\varepsilon$: 0 & continuous $\mathcal{N}(0, 1)$\\
			Setting-32 & 10,000 & 10 & $x$: 0; $\varepsilon$: 0 & continuous $\mathcal{N}(0, 1)$\\
			Setting-33 & 20,000 & 10 & $x$: 0; $\varepsilon$: 0 & continuous $\mathcal{N}(0, 1)$\\*
		\end{longtable}

		\begin{longtable}[t]{lllll}
			\caption{\label{tab:settings-heavy-spikeslab}Simulation settings for scenarios with data generated from heavy-tailed error distributions and models fitted with a spike and slab prior on the regression parameters.}\\
			\toprule
			Setting & $n$ & $p$ & Correlation & Error Distribution\\
			\midrule
			\endfirsthead
			\caption[]{Simulation settings for scenarios with data generated from heavy-tailed error distributions and models fitted with a spike and slab prior on the regression parameters. \textit{(continued)}}\\
			\toprule
			Setting & $n$ & $p$ & Correlation & Error Distribution\\
			\midrule
			\endhead
			
			\endfoot
			\bottomrule
			\endlastfoot
			Setting-1 & 75 & 100 & $x$: 0; $\varepsilon$: 0 & discrete mix $\mathcal{N}(0, 1)$ and $\mathcal{C}(0, 10)$ $(90\%; 10\%)$\\
			Setting-2 & 75 & 200 & $x$: 0; $\varepsilon$: 0 & discrete mix $\mathcal{N}(0, 1)$ and $\mathcal{C}(0, 10)$ $(90\%; 10\%)$\\
			Setting-3 & 75 & 250 & $x$: 0; $\varepsilon$: 0 & discrete mix $\mathcal{N}(0, 1)$ and $\mathcal{C}(0, 10)$ $(90\%; 10\%)$\\
			Setting-4 & 75 & 100 & $x$: 0; $\varepsilon$: 0 & continuous $t$ with df=1\\
			Setting-5 & 75 & 200 & $x$: 0; $\varepsilon$: 0 & continuous $t$ with df=1\\
			\addlinespace
			Setting-6 & 75 & 250 & $x$: 0; $\varepsilon$: 0 & continuous $t$ with df=1\\
			Setting-7 & 75 & 100 & $x$: 0; $\varepsilon$: 0 & continuous $t$ with df=2\\
			Setting-8 & 75 & 200 & $x$: 0; $\varepsilon$: 0 & continuous $t$ with df=2\\
			Setting-9 & 75 & 250 & $x$: 0; $\varepsilon$: 0 & continuous $t$ with df=2\\
			Setting-10 & 75 & 100 & $x$: 0.2; $\varepsilon$: 0.3 & discrete mix $\mathcal{N}(0, 1)$ and $\mathcal{C}(0, 10)$ $(90\%; 10\%)$\\
			\addlinespace
			Setting-11 & 75 & 200 & $x$: 0.2; $\varepsilon$: 0.3 & discrete mix $\mathcal{N}(0, 1)$ and $\mathcal{C}(0, 10)$ $(90\%; 10\%)$\\
			Setting-12 & 75 & 250 & $x$: 0.2; $\varepsilon$: 0.3 & discrete mix $\mathcal{N}(0, 1)$ and $\mathcal{C}(0, 10)$ $(90\%; 10\%)$\\
			Setting-13 & 100 & 100 & $x$: 0.2; $\varepsilon$: 0.3 & discrete mix $\mathcal{N}(0, 1)$ and $\mathcal{C}(0, 10)$ $(90\%; 10\%)$\\
			Setting-14 & 100 & 200 & $x$: 0.2; $\varepsilon$: 0.3 & discrete mix $\mathcal{N}(0, 1)$ and $\mathcal{C}(0, 10)$ $(90\%; 10\%)$\\
			Setting-15 & 100 & 250 & $x$: 0.2; $\varepsilon$: 0.3 & discrete mix $\mathcal{N}(0, 1)$ and $\mathcal{C}(0, 10)$ $(90\%; 10\%)$\\
			\addlinespace
			Setting-16 & 75 & 100 & $x$: 0.2; $\varepsilon$: 0.3 & continuous $t$ with df=1\\
			Setting-17 & 75 & 200 & $x$: 0.2; $\varepsilon$: 0.3 & continuous $t$ with df=1\\
			Setting-18 & 75 & 250 & $x$: 0.2; $\varepsilon$: 0.3 & continuous $t$ with df=1\\
			Setting-19 & 100 & 100 & $x$: 0.2; $\varepsilon$: 0.3 & continuous $t$ with df=1\\
			Setting-20 & 100 & 200 & $x$: 0.2; $\varepsilon$: 0.3 & continuous $t$ with df=1\\
			\addlinespace
			Setting-21 & 100 & 250 & $x$: 0.2; $\varepsilon$: 0.3 & continuous $t$ with df=1\\
			Setting-22 & 75 & 100 & $x$: 0.2; $\varepsilon$: 0.3 & continuous $t$ with df=2\\
			Setting-23 & 75 & 200 & $x$: 0.2; $\varepsilon$: 0.3 & continuous $t$ with df=2\\
			Setting-24 & 75 & 250 & $x$: 0.2; $\varepsilon$: 0.3 & continuous $t$ with df=2\\
			Setting-25 & 100 & 100 & $x$: 0.2; $\varepsilon$: 0.3 & continuous $t$ with df=2\\
			\addlinespace
			Setting-26 & 100 & 200 & $x$: 0.2; $\varepsilon$: 0.3 & continuous $t$ with df=2\\
			Setting-27 & 100 & 250 & $x$: 0.2; $\varepsilon$: 0.3 & continuous $t$ with df=2\\
			Setting-28 & 75 & 100 & $x$: 0.4; $\varepsilon$: 0.6 & discrete mix $\mathcal{N}(0, 1)$ and $\mathcal{C}(0, 10)$ $(90\%; 10\%)$\\
			Setting-29 & 75 & 200 & $x$: 0.4; $\varepsilon$: 0.6 & discrete mix $\mathcal{N}(0, 1)$ and $\mathcal{C}(0, 10)$ $(90\%; 10\%)$\\
			Setting-30 & 75 & 250 & $x$: 0.4; $\varepsilon$: 0.6 & discrete mix $\mathcal{N}(0, 1)$ and $\mathcal{C}(0, 10)$ $(90\%; 10\%)$\\
			\addlinespace
			Setting-31 & 100 & 100 & $x$: 0.4; $\varepsilon$: 0.6 & discrete mix $\mathcal{N}(0, 1)$ and $\mathcal{C}(0, 10)$ $(90\%; 10\%)$\\
			Setting-32 & 100 & 200 & $x$: 0.4; $\varepsilon$: 0.6 & discrete mix $\mathcal{N}(0, 1)$ and $\mathcal{C}(0, 10)$ $(90\%; 10\%)$\\
			Setting-33 & 100 & 250 & $x$: 0.4; $\varepsilon$: 0.6 & discrete mix $\mathcal{N}(0, 1)$ and $\mathcal{C}(0, 10)$ $(90\%; 10\%)$\\
			Setting-34 & 75 & 100 & $x$: 0.4; $\varepsilon$: 0.6 & continuous $t$ with df=1\\
			Setting-35 & 75 & 200 & $x$: 0.4; $\varepsilon$: 0.6 & continuous $t$ with df=1\\
			\addlinespace
			Setting-36 & 75 & 250 & $x$: 0.4; $\varepsilon$: 0.6 & continuous $t$ with df=1\\
			Setting-37 & 100 & 100 & $x$: 0.4; $\varepsilon$: 0.6 & continuous $t$ with df=1\\
			Setting-38 & 100 & 200 & $x$: 0.4; $\varepsilon$: 0.6 & continuous $t$ with df=1\\
			Setting-39 & 100 & 250 & $x$: 0.4; $\varepsilon$: 0.6 & continuous $t$ with df=1\\
			Setting-40 & 75 & 100 & $x$: 0.4; $\varepsilon$: 0.6 & continuous $t$ with df=2\\
			\addlinespace
			Setting-41 & 75 & 200 & $x$: 0.4; $\varepsilon$: 0.6 & continuous $t$ with df=2\\
			Setting-42 & 75 & 250 & $x$: 0.4; $\varepsilon$: 0.6 & continuous $t$ with df=2\\
			Setting-43 & 100 & 100 & $x$: 0.4; $\varepsilon$: 0.6 & continuous $t$ with df=2\\
			Setting-44 & 100 & 200 & $x$: 0.4; $\varepsilon$: 0.6 & continuous $t$ with df=2\\
			Setting-45 & 100 & 250 & $x$: 0.4; $\varepsilon$: 0.6 & continuous $t$ with df=2\\*
		\end{longtable}

		\begin{longtable}[t]{lllll}
			\caption{\label{tab:settings-moderate-spikeslab}Simulation settings for scenarios with data generated from moderate-tailed error distributions and models fitted with a spike and slab prior on the regression parameters.}\\
			\toprule
			Setting & $n$ & $p$ & Correlation & Error Distribution\\
			\midrule
			\endfirsthead
			\caption[]{Simulation settings for scenarios with data generated from moderate-tailed error distributions and models fitted with a spike and slab prior on the regression parameters. \textit{(continued)}}\\
			\toprule
			Setting & $n$ & $p$ & Correlation & Error Distribution\\
			\midrule
			\endhead
			
			\endfoot
			\bottomrule
			\endlastfoot
			Setting-1 & 75 & 100 & $x$: 0; $\varepsilon$: 0 & discrete mix $\mathcal{N}(0, 1)$ and $\mathcal{C}(0, 10)$ $(99\%; 1\%)$\\
			Setting-2 & 75 & 200 & $x$: 0; $\varepsilon$: 0 & discrete mix $\mathcal{N}(0, 1)$ and $\mathcal{C}(0, 10)$ $(99\%; 1\%)$\\
			Setting-3 & 75 & 250 & $x$: 0; $\varepsilon$: 0 & discrete mix $\mathcal{N}(0, 1)$ and $\mathcal{C}(0, 10)$ $(99\%; 1\%)$\\
			Setting-4 & 75 & 100 & $x$: 0; $\varepsilon$: 0 & discrete mix $\mathcal{N}(0, 1)$ and $\mathcal{C}(0, 10)$ $(95\%; 5\%)$\\
			Setting-5 & 75 & 200 & $x$: 0; $\varepsilon$: 0 & discrete mix $\mathcal{N}(0, 1)$ and $\mathcal{C}(0, 10)$ $(95\%; 5\%)$\\
			\addlinespace
			Setting-6 & 75 & 250 & $x$: 0; $\varepsilon$: 0 & discrete mix $\mathcal{N}(0, 1)$ and $\mathcal{C}(0, 10)$ $(95\%; 5\%)$\\
			Setting-7 & 75 & 100 & $x$: 0; $\varepsilon$: 0 & continuous $t$ with df=4\\
			Setting-8 & 75 & 200 & $x$: 0; $\varepsilon$: 0 & continuous $t$ with df=4\\
			Setting-9 & 75 & 250 & $x$: 0; $\varepsilon$: 0 & continuous $t$ with df=4\\
			Setting-10 & 75 & 100 & $x$: 0; $\varepsilon$: 0 & continuous $t$ with df=8\\
			\addlinespace
			Setting-11 & 75 & 200 & $x$: 0; $\varepsilon$: 0 & continuous $t$ with df=8\\
			Setting-12 & 75 & 250 & $x$: 0; $\varepsilon$: 0 & continuous $t$ with df=8\\
			Setting-13 & 75 & 100 & $x$: 0.2; $\varepsilon$: 0.3 & discrete mix $\mathcal{N}(0, 1)$ and $\mathcal{C}(0, 10)$ $(99\%; 1\%)$\\
			Setting-14 & 75 & 200 & $x$: 0.2; $\varepsilon$: 0.3 & discrete mix $\mathcal{N}(0, 1)$ and $\mathcal{C}(0, 10)$ $(99\%; 1\%)$\\
			Setting-15 & 75 & 250 & $x$: 0.2; $\varepsilon$: 0.3 & discrete mix $\mathcal{N}(0, 1)$ and $\mathcal{C}(0, 10)$ $(99\%; 1\%)$\\
			\addlinespace
			Setting-16 & 100 & 100 & $x$: 0.2; $\varepsilon$: 0.3 & discrete mix $\mathcal{N}(0, 1)$ and $\mathcal{C}(0, 10)$ $(99\%; 1\%)$\\
			Setting-17 & 100 & 200 & $x$: 0.2; $\varepsilon$: 0.3 & discrete mix $\mathcal{N}(0, 1)$ and $\mathcal{C}(0, 10)$ $(99\%; 1\%)$\\
			Setting-18 & 100 & 250 & $x$: 0.2; $\varepsilon$: 0.3 & discrete mix $\mathcal{N}(0, 1)$ and $\mathcal{C}(0, 10)$ $(99\%; 1\%)$\\
			Setting-19 & 75 & 100 & $x$: 0.2; $\varepsilon$: 0.3 & discrete mix $\mathcal{N}(0, 1)$ and $\mathcal{C}(0, 10)$ $(95\%; 5\%)$\\
			Setting-20 & 75 & 200 & $x$: 0.2; $\varepsilon$: 0.3 & discrete mix $\mathcal{N}(0, 1)$ and $\mathcal{C}(0, 10)$ $(95\%; 5\%)$\\
			\addlinespace
			Setting-21 & 75 & 250 & $x$: 0.2; $\varepsilon$: 0.3 & discrete mix $\mathcal{N}(0, 1)$ and $\mathcal{C}(0, 10)$ $(95\%; 5\%)$\\
			Setting-22 & 100 & 100 & $x$: 0.2; $\varepsilon$: 0.3 & discrete mix $\mathcal{N}(0, 1)$ and $\mathcal{C}(0, 10)$ $(95\%; 5\%)$\\
			Setting-23 & 100 & 200 & $x$: 0.2; $\varepsilon$: 0.3 & discrete mix $\mathcal{N}(0, 1)$ and $\mathcal{C}(0, 10)$ $(95\%; 5\%)$\\
			Setting-24 & 100 & 250 & $x$: 0.2; $\varepsilon$: 0.3 & discrete mix $\mathcal{N}(0, 1)$ and $\mathcal{C}(0, 10)$ $(95\%; 5\%)$\\
			Setting-25 & 75 & 100 & $x$: 0.2; $\varepsilon$: 0.3 & continuous $t$ with df=4\\
			\addlinespace
			Setting-26 & 75 & 200 & $x$: 0.2; $\varepsilon$: 0.3 & continuous $t$ with df=4\\
			Setting-27 & 75 & 250 & $x$: 0.2; $\varepsilon$: 0.3 & continuous $t$ with df=4\\
			Setting-28 & 100 & 100 & $x$: 0.2; $\varepsilon$: 0.3 & continuous $t$ with df=4\\
			Setting-29 & 100 & 200 & $x$: 0.2; $\varepsilon$: 0.3 & continuous $t$ with df=4\\
			Setting-30 & 100 & 250 & $x$: 0.2; $\varepsilon$: 0.3 & continuous $t$ with df=4\\
			\addlinespace
			Setting-31 & 75 & 100 & $x$: 0.2; $\varepsilon$: 0.3 & continuous $t$ with df=8\\
			Setting-32 & 75 & 200 & $x$: 0.2; $\varepsilon$: 0.3 & continuous $t$ with df=8\\
			Setting-33 & 75 & 250 & $x$: 0.2; $\varepsilon$: 0.3 & continuous $t$ with df=8\\
			Setting-34 & 100 & 100 & $x$: 0.2; $\varepsilon$: 0.3 & continuous $t$ with df=8\\
			Setting-35 & 100 & 200 & $x$: 0.2; $\varepsilon$: 0.3 & continuous $t$ with df=8\\
			\addlinespace
			Setting-36 & 100 & 250 & $x$: 0.2; $\varepsilon$: 0.3 & continuous $t$ with df=8\\
			Setting-37 & 75 & 100 & $x$: 0.4; $\varepsilon$: 0.6 & discrete mix $\mathcal{N}(0, 1)$ and $\mathcal{C}(0, 10)$ $(99\%; 1\%)$\\
			Setting-38 & 75 & 200 & $x$: 0.4; $\varepsilon$: 0.6 & discrete mix $\mathcal{N}(0, 1)$ and $\mathcal{C}(0, 10)$ $(99\%; 1\%)$\\
			Setting-39 & 75 & 250 & $x$: 0.4; $\varepsilon$: 0.6 & discrete mix $\mathcal{N}(0, 1)$ and $\mathcal{C}(0, 10)$ $(99\%; 1\%)$\\
			Setting-40 & 100 & 100 & $x$: 0.4; $\varepsilon$: 0.6 & discrete mix $\mathcal{N}(0, 1)$ and $\mathcal{C}(0, 10)$ $(99\%; 1\%)$\\
			\addlinespace
			Setting-41 & 100 & 200 & $x$: 0.4; $\varepsilon$: 0.6 & discrete mix $\mathcal{N}(0, 1)$ and $\mathcal{C}(0, 10)$ $(99\%; 1\%)$\\
			Setting-42 & 100 & 250 & $x$: 0.4; $\varepsilon$: 0.6 & discrete mix $\mathcal{N}(0, 1)$ and $\mathcal{C}(0, 10)$ $(99\%; 1\%)$\\
			Setting-43 & 75 & 100 & $x$: 0.4; $\varepsilon$: 0.6 & discrete mix $\mathcal{N}(0, 1)$ and $\mathcal{C}(0, 10)$ $(95\%; 5\%)$\\
			Setting-44 & 75 & 200 & $x$: 0.4; $\varepsilon$: 0.6 & discrete mix $\mathcal{N}(0, 1)$ and $\mathcal{C}(0, 10)$ $(95\%; 5\%)$\\
			Setting-45 & 75 & 250 & $x$: 0.4; $\varepsilon$: 0.6 & discrete mix $\mathcal{N}(0, 1)$ and $\mathcal{C}(0, 10)$ $(95\%; 5\%)$\\
			\addlinespace
			Setting-46 & 100 & 100 & $x$: 0.4; $\varepsilon$: 0.6 & discrete mix $\mathcal{N}(0, 1)$ and $\mathcal{C}(0, 10)$ $(95\%; 5\%)$\\
			Setting-47 & 100 & 200 & $x$: 0.4; $\varepsilon$: 0.6 & discrete mix $\mathcal{N}(0, 1)$ and $\mathcal{C}(0, 10)$ $(95\%; 5\%)$\\
			Setting-48 & 100 & 250 & $x$: 0.4; $\varepsilon$: 0.6 & discrete mix $\mathcal{N}(0, 1)$ and $\mathcal{C}(0, 10)$ $(95\%; 5\%)$\\
			Setting-49 & 75 & 100 & $x$: 0.4; $\varepsilon$: 0.6 & continuous $t$ with df=4\\
			Setting-50 & 75 & 200 & $x$: 0.4; $\varepsilon$: 0.6 & continuous $t$ with df=4\\
			\addlinespace
			Setting-51 & 75 & 250 & $x$: 0.4; $\varepsilon$: 0.6 & continuous $t$ with df=4\\
			Setting-52 & 100 & 100 & $x$: 0.4; $\varepsilon$: 0.6 & continuous $t$ with df=4\\
			Setting-53 & 100 & 200 & $x$: 0.4; $\varepsilon$: 0.6 & continuous $t$ with df=4\\
			Setting-54 & 100 & 250 & $x$: 0.4; $\varepsilon$: 0.6 & continuous $t$ with df=4\\
			Setting-55 & 75 & 100 & $x$: 0.4; $\varepsilon$: 0.6 & continuous $t$ with df=8\\
			\addlinespace
			Setting-56 & 75 & 200 & $x$: 0.4; $\varepsilon$: 0.6 & continuous $t$ with df=8\\
			Setting-57 & 75 & 250 & $x$: 0.4; $\varepsilon$: 0.6 & continuous $t$ with df=8\\
			Setting-58 & 100 & 100 & $x$: 0.4; $\varepsilon$: 0.6 & continuous $t$ with df=8\\
			Setting-59 & 100 & 200 & $x$: 0.4; $\varepsilon$: 0.6 & continuous $t$ with df=8\\
			Setting-60 & 100 & 250 & $x$: 0.4; $\varepsilon$: 0.6 & continuous $t$ with df=8\\*
		\end{longtable}

		\begin{longtable}[t]{lllll}
			\caption{\label{tab:settings-thin-spikeslab}Simulation settings for scenarios with data generated from thin-tailed error distributions and models fitted with a spike and slab prior on the regression parameters.}\\
			\toprule
			Setting & $n$ & $p$ & Correlation & Error Distribution\\
			\midrule
			\endfirsthead
			\caption[]{Simulation settings for scenarios with data generated from thin-tailed error distributions and models fitted with a spike and slab prior on the regression parameters. \textit{(continued)}}\\
			\toprule
			Setting & $n$ & $p$ & Correlation & Error Distribution\\
			\midrule
			\endhead
			
			\endfoot
			\bottomrule
			\endlastfoot
			Setting-1 & 75 & 100 & $x$: 0; $\varepsilon$: 0 & continuous $\mathcal{N}(0, 1)$\\
			Setting-2 & 75 & 200 & $x$: 0; $\varepsilon$: 0 & continuous $\mathcal{N}(0, 1)$\\
			Setting-3 & 75 & 250 & $x$: 0; $\varepsilon$: 0 & continuous $\mathcal{N}(0, 1)$\\
			Setting-4 & 75 & 100 & $x$: 0.2; $\varepsilon$: 0.3 & continuous $\mathcal{N}(0, 1)$\\
			Setting-5 & 75 & 200 & $x$: 0.2; $\varepsilon$: 0.3 & continuous $\mathcal{N}(0, 1)$\\
			\addlinespace
			Setting-6 & 75 & 250 & $x$: 0.2; $\varepsilon$: 0.3 & continuous $\mathcal{N}(0, 1)$\\
			Setting-7 & 100 & 100 & $x$: 0.2; $\varepsilon$: 0.3 & continuous $\mathcal{N}(0, 1)$\\
			Setting-8 & 100 & 200 & $x$: 0.2; $\varepsilon$: 0.3 & continuous $\mathcal{N}(0, 1)$\\
			Setting-9 & 100 & 250 & $x$: 0.2; $\varepsilon$: 0.3 & continuous $\mathcal{N}(0, 1)$\\
			Setting-10 & 75 & 100 & $x$: 0.4; $\varepsilon$: 0.6 & continuous $\mathcal{N}(0, 1)$\\
			\addlinespace
			Setting-11 & 75 & 200 & $x$: 0.4; $\varepsilon$: 0.6 & continuous $\mathcal{N}(0, 1)$\\
			Setting-12 & 75 & 250 & $x$: 0.4; $\varepsilon$: 0.6 & continuous $\mathcal{N}(0, 1)$\\
			Setting-13 & 100 & 100 & $x$: 0.4; $\varepsilon$: 0.6 & continuous $\mathcal{N}(0, 1)$\\
			Setting-14 & 100 & 200 & $x$: 0.4; $\varepsilon$: 0.6 & continuous $\mathcal{N}(0, 1)$\\
			Setting-15 & 100 & 250 & $x$: 0.4; $\varepsilon$: 0.6 & continuous $\mathcal{N}(0, 1)$\\*
		\end{longtable}

		\subsection{Further details on metrics used for assessing model performance}
		
		\subsubsection{Prediction Performance} \label{sec:detailed-prediction-perf}
		
		For prediction assessment, an independent ``test data set'' is generated for each replicated training dataset used to fit the models. The test data set maintained the same values of $n$, $p$, $\bb^{\text{true}}$, predictor and error correlation structure, and error distribution as the training data but differed in the random elements, namely $\epsilon_i$, $\bm{x}_i$, and $y_i$. Subsequently, we focus on the expected posterior predictive distribution, averaged over the data distribution, to predict $y_i^{\text{test}}$ given $x_i^{\text{test}}$ and computed the prediction (posterior) MSE $\tilde{M}_{i, \text{data}}$ and its median over replicates and coordinates (i.e., observations), analogous to estimation MSE in Section~\ref{} (see Supplement~\ref{} for detailed definitions).  
		\begin{equation}
			\label{defn:pred-post-mse}
			\tilde{M}_{i, \text{data}} = \text{prediction MSE}(i, \text{data}) = E \left[ \left( y_i^{\text{test}} - \mu - \bb^T \bm{x}_i^{\text{test}} \right)^2 \mid \text{training data} \right].
		\end{equation}
		
		Analogous to the steps involved in assessing estimation performance, the prediction MSE is first computed using posterior MCMC draws for each fitted model and every observation $i$ (i.e., $\yb$ coordinate) within each replicate of the training-test data set pair. This yields $\tilde{M}_{i, r}^{\text{model}}$ for each individual data-generating setting (see Supplementary Tables \ref{tab:settings-extremely heavy-normal}-\ref{tab:settings-thin-spikeslab}), where $r = 1, \dots, R = 200$ indexes the data replicates for the setting, and $i = 1, \cdots, n$ indexes the observation ($y$) coordinates. As a summary measure for each model in each simulation setting, we then compute the median posterior MSE, defined as  
		\[
		\median_{i=1, \dots, n} \left( \median_{r=1, \dots, R} \tilde{M}_{i, r}^{\text{model}} \right).
		\]
		
	\end{document}